\documentclass[a4paper,11pt]{article}

\usepackage{amsmath, amsthm, amssymb, amsfonts}
\usepackage{mathtools}
\usepackage{hyperref}
\usepackage{graphicx}
\usepackage{dsfont}
\usepackage{fullpage}
\usepackage{color}
\usepackage{soul} 
\usepackage[title]{appendix}
\usepackage{tikz}
\usetikzlibrary{patterns}
\usetikzlibrary{shapes.multipart}
\usetikzlibrary{arrows}
\usepackage{appendix}
\theoremstyle{definition}

\definecolor{labelkey}{cmyk}{.4,.2,0,0}

\newcommand{\be}{\begin{equation}}
\newcommand{\ee}{\end{equation}}
\newcommand{\bea}{\begin{eqnarray}}
\newcommand{\eea}{\end{eqnarray}}
\newcommand{\nn}{\nonumber}

\usepackage{titlesec}
\titleformat{\section}{\large\bf}{\thesection}{1em}{}
\titleformat{\subsection}[runin]{\bf}{\thesubsection}{1em}{}

\usepackage{authblk}
\makeatletter
\newcommand\appendix@section[1]{%
  \refstepcounter{section}%
  \orig@section*{Appendix \@Alph\c@section: #1}%
  \addcontentsline{toc}{section}{Appendix \@Alph\c@section: #1}%
}
\let\orig@section\section
\g@addto@macro\appendix{\let\section\appendix@section}
\makeatother

\title{\bf \large Quench dynamics of noninteracting fermions with a delta impurity}

\author[1]{Gabriel Gouraud}
\author[1]{Pierre Le Doussal}
\affil[1]{\normalsize Laboratoire de Physique de l'\'Ecole Normale Sup\'erieure, ENS, Universit\'e PSL, CNRS, Sorbonne Universit\'e, Universit\'e de Paris, 75005 Paris, France}
\author[2]{Gr{\'e}gory Schehr }
\affil[2]{Sorbonne Universit\'e, Laboratoire de Physique Th\'eorique et Hautes Energies, CNRS UMR 7589, 4 Place Jussieu, 75252 Paris Cedex 05, France}
\date{}

\begin{document}

\maketitle

\begin{abstract}
We study the out-of-equilibrium dynamics of noninteracting fermions in one dimension and in continuum space, in the presence of a delta impurity potential at the origin
whose strength $g$ is varied at time $t=0$. The system is prepared in its ground state with $g=g_0=+\infty$, with two different densities and Fermi wave-vectors $k_L$ and $k_R$ 
on the two half-spaces $x>0$ and $x<0$ respectively. It then evolves for $t>0$ as an isolated system, with a finite impurity strength $g$. 
We compute exactly the time dependent density and current. For a fixed position $x$ and in the large time limit $t \to \infty$, the system reaches a non-equilibrium stationary state (NESS). We obtain analytically the correlation kernel, density, particle current, and energy current in the NESS, and characterize their relaxation,
which is algebraic in time. {In particular, in the NESS, we show that, away from the impurity, the particle density displays oscillations which are the non-equilibrium analog of the Friedel oscillations}. 
In the regime of ``rays'', $x/t=\xi$ fixed with $x, t \to \infty$, we compute the same quantities
and observe the emergence of two light cones, associated to the Fermi velocities $k_L$ and $k_R$ in the initial state.
Interestingly, we find non trivial quantum correlations between two opposite rays with velocities $\xi$ and $-\xi$ which
we compute explicitly. We extend to a continuum setting and to a correlated initial state 
the analytical methods developed in a recent work of Ljubotina, Sotiriadis and Prosen, in the context of a discrete fermionic chain with an impurity. 
We also generalize our results to an initial state at finite temperature, recovering, via explicit calculations, some predictions of conformal field theory
in the low energy limit. 
\end{abstract}
%\noindent{\it Keywords\/}: 

% \pacs{72.20.-i, 71.23.An, 71.23.-k}
%\note{do we need the PACS numbers ?}

\newpage

{\pagestyle{plain}
 \tableofcontents
\cleardoublepage}

\section{Introduction}

There is a recent revived interest in noninteracting fermions in continuum inhomegeneous settings
as analytically tractable models for studying equilibrium and out-of-equilibrium quantum correlations. 
For one-dimensional fermions at equilibrium in an external potential, there are interesting connections to random matrix theory (for a review see \cite{us_review_JPA}). 
These relations allow to compute the density, the full counting statistics and
the entanglement entropy in a variety of potentials \cite{us_review_PRA,us_FCS} using the tools of determinantal point processes \cite{Borodin,Joh} in agreement with other approaches using inhomogeneous bosonization
\cite{Dubail_stat}. In particular, the correlation functions can be expressed
as determinants built from a central object called the kernel \cite{mehta,forrester}. In the limit of a large number of
fermions the kernel in the bulk of the Fermi gas becomes universal at the scale of the inter-particle distance $k_F^{-1}$, where 
$k_F$ is the local Fermi wave-vector. It is given, 
at zero temperature, by the celebrated sine kernel \cite{mehta}.
This universal behavior holds for smooth potentials \cite{us_review_JPA} but breaks down for more singular potentials
which vary over a region ${\cal D}$ of size $O(k_F^{-1})$. In this case, the kernel differs from the sine kernel and has been calculated, for instance, for the hard wall \cite{Calabrese_hwall,us_hardwall_epl,us_hardwall_jstat,CMO2018}, the step potential \cite{us_step}, as well as for static impurities modeled by a delta potential \cite{DLMS2021}. 
Far from the singular region ${\cal D}$ the kernel exhibits Friedel type oscillations \cite{Friedel,Fuchs1,Fuchs2}, which have been characterized using reflection and
transmission coefficients through~${\cal D}$ \cite{us_step,KohnSham}.

It was shown that the bulk universality for smooth potentials can be extended to equilibrium dynamics \cite{us_eq_dyn} (in terms of the so-called extended sine kernel
\cite{moerbeke}).
It is natural to ask similar questions in the case of non-equilibrium dynamics. In the case of fermions trapped in a confining potential it was found for 
some time-dependent potentials (such as the harmonic oscillator or the $1/x^2$ potential) and some initial conditions, 
that the kernel keeps its equilibrium form up to time dependent factors \cite{Gangardt,Gritsev,Dubail_dyn,DLSM2019}. 
For more general potentials and initial conditions the situation is more complicated~\cite{DLSM2019,Kulkarni,RCDD21}. 
The large time limit of the kernel can be obtained from the so-called diagonal ensemble and coincides with the prediction from the generalized Gibbs ensemble (GGE) \cite{EF2016}. However the large time limit
of the multi-point correlations exhibits a more complicated behavior~\cite{DLSM2019}.

One can also consider the out of equilibrium dynamics in the absence of a confining potential. A seminal example is the Landauer-B\"uttiker theory for transport between two reservoirs \cite{Buttiker_review, CJM2005}. Another important question, much studied in the context of Luttinger liquids, is to characterize the transport that takes place when two $1d$ systems are connected via a contact \cite{KaneFisher1992,FendleySaleur1996,BernardDoyonViti2015,Mazza2016,GiamarchiRuggieroQuenchLL2021,GiamarchiCalabreseQuenchLL2021}.
This problem has attracted a renewed interest in the context of quantum quenches where the system is translationally invariant but the initial condition is
inhomogeneous. Important questions concern the large time behavior of the system, in particular the local convergence to a non equilibrium steady state (NESS), characterized by stationary currents, density profiles and counting statistics. Another question concerns non-stationary fluctuations and moving fronts. These questions have been addressed using conformal field theory (CFT)
\cite{SotiriadisCardy2008,BernardDoyonEnergy2012,DeLucaVitiBernardDoyon2013}, generalized hydrodynamics \cite{GHDDoyonYoshimura,GHDBertiniDeNardis},
as well as exact solutions of free fermions either on a lattice
\cite{AntalXX1999,Krapivsky2008,DubailViti2016,EislerRacz2013,EislerMaislinger2016,PerfettoGambassi2017,Kormos2017,
Sasamoto2019} or in the continuum \cite{Calabrese_release,ColluraKarevski2014}. 

It is natural to ask how these results for noninteracting fermions will be modified in the presence of a (non-confining) external potential.
The simplest example is the delta function impurity model in the continuum which was solved at equilibrium in \cite{DLMS2021,Friedel,Fuchs1,Fuchs2}.
In this paper we explore the out-of-equilibrium counterpart of this model, starting from an inhomogeneous initial condition. 
It is inspired by a recent work of Ljubotina, Sotiriadis and Prosen \cite{Prosen2018} where a similar model with an impurity
was studied for a discrete fermionic chain. In that work the initial state was chosen fully decorrelated. On the other
hand it is known that fermionic correlations can affect the transport, see e.~g.~\cite{alex}. In the present work, we consider an initial
state that consists in two independent Fermi gases on semi-infinite lines $\mathbb{R}^\pm$ separated by an infinite wall which are each in their ground state, and hence 
both exhibit the sine kernel bulk correlations with different Fermi wavevectors $k_R > k_L$. They are joined at $t=0$ and evolve freely in the presence of a delta impurity at the origin. We find that the system reaches a NESS at large time. We calculate exactly the asymptotic kernel in the NESS, from
which we obtain the asymptotic particle and energy current as well as the density profile. These quantities depend only on $k_R, k_L$ and $g$ the strength of the impurity. 
In the case where $k_L=k_R$ and when the impurity is repulsive $g>0$, one finds that the NESS coincides with the ground state of the system on the infinite line, although the initial state is far from equilibrium. In this case one thus recovers the equilibrium kernel obtained in \cite{DLMS2021}. 
This can be understood since all the excitations move to infinity. On the contrary when $g<0$, the Hamiltonian possesses a bound state and one
finds that the NESS always differs from the ground state (even when $k_L = k_R$). In this case, the mean occupation number of the bound state does not converge to unity at large time. 
Our methods also allow us to study the relaxation towards the NESS, which we find to be algebraic in time.
We also compute the kernel in the vicinity of space-time ``rays'' in the $(x,t)$ plane
with fixed velocities $\xi = x/t$. It exhibits a change of behaviors on the two
light cones $\xi=k_R$ and $\xi=k_L$. Interestingly we find nontrivial correlations between points
belonging to rays with opposite velocities $\pm \xi$, an effect not discussed in \cite{Prosen2018}, although we expect this effect to be also present in the discrete setting. An important difference with Ref. \cite{Prosen2018} is the existence of two light cones which is due
to the presence of local correlations in the initial state which are different on both sides of the impurity. Outside these light cones
one recovers the sine kernel, but inside the light cones (including in the NESS for $\xi=0$) the asymptotic currents are non zero 
and the kernel is different from the sine kernel. 

Our results are then extended to an initial condition with
two different nonzero temperatures $T_L$ and $T_R$. In the low temperature and weak impurity strength
limit and equal Fermi wavevectors $k_L=k_R$, we recover, via explicit computations, the result for the energy current in the NESS obtained from CFT given in \cite{BernardDoyonViti2015}. 
Finally we show that the asymptotic kernel, hence the correlations, along rays at fixed velocity $\xi=x/t$ can be recovered by a semi-classical argument. 
However this is not the case for correlations between points belonging to rays with opposite velocities $\pm \xi$. 
Also, the semi-classical method does not allow to recover the kernel inside the NESS. 
Finally our main results are checked versus numerical evaluation of the exact starting formulas. 

Note that other recent works have addressed different effects of an impurity on the dynamics. In particular, {full counting statistics and} entanglement growth in fermionic chains \cite{EislerDefect2012,EislerDefect2020,Gamayun1,SKCD2021},
interactions localized at a contact \cite{DeLucaDefect2021}, dynamics starting from a homogeneous state~\cite{BertiniFagottiDefect2016} {and moving and time dependent defects}
\cite{Gamayun2,deLucaMovingDefect1,ColluraDeLucaDWMelting,DeLucaMoving Entropy} have been studied.

The paper is organized as follows. In Section \ref{sec:model} we define the model, the initial conditions
and the observables that we study here. In Section \ref{sec:main} we give a detailed and pedagogical presentation of our main results.
In Section \ref{sec:exact} we derive the expression of the time dependent kernel in a hard box of finite size $[-\ell/2,\ell/2]$.
In Section \ref{sec:thermo} we obtain the finite time kernel in the limit $\ell \to +\infty$. To this aim we
have adapted a trick based on contour integrals used in \cite{Prosen2018} (see also \cite{CEF12, Sotiriadis}). {As we show in that section, the 
study of the limit $\ell \to +\infty$ turns out to be more involved in the present continuum setting with a correlated initial condition}.
In Section \ref{sec:largetime} we obtain the large time limit of the kernel in the regime $x=O(1)$
which characterizes the NESS. From it we compute the density and currents. 
In Section \ref{sec:light} we study the large time limit of the kernel along rays. In Section \ref{sec:wigner} we apply a semi -classical method which allows us to 
recover some of the above exact results in the regime of rays. Finally, we conclude in Section \ref{sec:conclu}. The appendices contain many of the technical details of the calculations.

\section{Model, quench protocol and observables} \label{sec:model} 

\subsection{Model}

$\quad$ \\

We consider $N$ noninteracting fermions in one dimension and in the presence of a delta impurity at the origin described by the
single particle Hamiltonian $\hat H_g$
\be  \label{def_H}
\hat H_g=-\frac{1}{2}\partial_x^2+g\delta(x) \;.
\ee 
We work here in units where the fermion mass is unity and $\hbar=1$. In these units, $g$ denotes the strength of the impurity, which can be repulsive ($g>0$) or attractive ($g<0$). To fully specify the model we define it on the interval $[-\ell/2,\ell/2]$ with a hard wall boundary condition (i.e., vanishing wave-function at $x= \pm \ell/2$). We will be eventually interested in the problem on the full line obtained by taking the limit $\ell \to +\infty$, with fixed fermion densities. 

\subsection{Initial state} \label{sec:initial} 

$\quad$ \\

The system is prepared at $t=0$ in the ground state of the many body Hamiltonian with $g=+\infty$, associated to the single particle Hamiltonian
$\hat H_{\infty}$. This corresponds to imposing a hard-wall at $x=0$, so that the system is cut into two independent halves $x>0$ and $x<0$ at $t=0$.
We will denote by $N_L$ and $N_R$ respectively the number of fermions in the left ($x<0$) and right ($x>0$) halves. Let us introduce $\phi^L_n(x)$ and $\phi^R_n(x)$, with $n = 1, 2, \cdots$, the normalized eigenfunctions of the single particle Hamiltonian $\hat H_{\infty}$
\be \label{def_eigen}
\phi^L_n(x) = \Theta(-x)  \sqrt{\frac{4}{\ell}} \sin(k_n \, x) \quad , \quad \phi^R_n(x) = \Theta(x)  \sqrt{\frac{4}{\ell}} \sin(k_n\,x) \;,
\ee 
where $\Theta(x)$ denotes the Heaviside theta function and 
\be
k_n = \frac{2\pi n}{\ell} \;.
\ee 
The corresponding energy levels are $\epsilon_n= \frac{1}{2} k_n^2 = \frac{2 \pi^2 n^2}{\ell^2}$ which are doubly degenerate ($L, R$).

The ground state many-body wave function is a Slater determinant built from the eigenstates $\phi^L_n(x)$ and $\phi^R_{n'}(x)$ with $1 \leq n \leq N_L$ and 
$1 \leq n' \leq N_R$. The $m$-point correlation function (see below for a precise definition) can be expressed as a $m \times m$ determinant built from the so-called correlation kernel 
\begin{equation}
K_0(x,x')=K_{L}(x,x')+K_{R}(x,x') \;, \label{K0}
\end{equation}  
where
\bea 
&&K_{L}(x,x') = \Theta(-x) \Theta(-x') \sum_{n=1}^{N_L} \frac{4}{\ell}\sin\left(\frac{2\pi n x}{\ell}\right)\sin\left(\frac{2\pi n x'}{\ell}\right) \;, \label{KL} \\
&&K_{R}(x,x') = \Theta(x) \Theta(x') \sum_{n'=1}^{N_R} \frac{4}{\ell}\sin\left(\frac{2\pi n' x}{\ell}\right)\sin\left(\frac{2\pi n' x'}{\ell}\right)  \label{KR} \;.
\eea
In particular, the mean fermion density is given by $\rho(x,t=0) = K_0(x,x) = \rho_L(x) + \rho_R(x)$ with $\rho_{L/R}(x) = K_{L/R}(x,x)$, where
$\rho_L(x)$ and $\rho_R(x)$ denote respectively the average fermion density to the left and to the right of the origin.
We also define the Fermi momenta of the left and right half-spaces associated to the initial condition
\be  \label{def_kLR}
k_L = k_{N_L} = \frac{2 \pi N_L}{\ell} \quad , \quad k_R = k_{N_R} = \frac{2 \pi N_R}{\ell}
\ee 
and the corresponding Fermi energies 
\be \label{def_muLR}
\mu_L = \frac{k_L^2}{2} \quad , \quad \mu_R = \frac{k_R^2}{2} \;,
\ee 
which will be useful in the following. 

\subsection{Dynamical evolution and observables}\label{sec:dyn}

$\quad$ \\

We now consider the time evolution for $t>0$ described by the single particle Hamiltonian $\hat H_g$ (\ref{def_H}) with a finite strengtgh $g$ of the impurity at $x=0$. We denote $\psi_n^L(x,t)$ the solution
of the Schr\"odinger equation $i \partial_t \psi_n^L(x,t) = \hat H_g \psi_n^L(x,t)$ with initial condition $\psi_n^L(x,t=0)= \phi_n^L(x)$, and similarly $\psi_n^R(x,t)$ with $\psi_n^R(x,t=0)= \phi_n^R(x)$ where $\phi_n^{L/R}(x)$ are given in Eq. (\ref{def_eigen}). Under this evolution, the time dependent many-body wave function for the $N=N_R+N_L$ fermions,
$\Psi(x_1,\dots,x_N;t)$, remains a Slater determinant at all times,
built from the time-dependent wave-functions $\psi_n^L(x,t)$ and $\psi_{n'}^R(x,t)$ with $1 \leq n \leq N_L$ and 
$1 \leq n' \leq N_R$. The observables of interest are the time dependent $m$-point correlations defined as 
\be 
R_m(x_1,\dots,x_m;t) = \frac{N!}{(N-m)!} \int dx_{m+1} \dots dx_N |\Psi(x_1,\dots,x_N;t)|^2 \;. \label{def_correl}
\ee 
Using standard manipulations, $R_m(x_1,\dots;x_m,t)$ can be written as (see e.g. \cite{DLSM2019})
\be
R_m(x_1,\dots,x_m;t) =  \det_{1 \leq i,j \leq m} K(x_i,x_j,t) \;, \label{def_correlt}
\ee 
where $K(x_i,x_j;t)$ is the time dependent correlation kernel. In the present case it reads
\be 
K(x,x',t) =  K_L(x,x',t)+K_R(x,x',t) \label{def_Kernel_t}
\ee 
where
\be 
K_L(x,x',t) = \sum_{n=1}^{N_L} \psi_n^{L*}(x,t) \psi_n^L(x',t)  \quad , \quad K_R(x,x',t) = \sum_{n=1}^{N_R} \psi_n^{R*}(x,t) \psi_n^R(x',t) \;. 
\ee 
Of particular interest is the time-dependent density $\rho(x,t)$ given by 
\bea\label{def_rhoxt}
\rho(x,t) = K(x,x,t)\;.
\eea
Another important observable is {the total particle current} defined as
\bea
&& J(x,t) = J_L(x,t) + J_R(x,t) \\
&& J_{L/R}(x,t) = \frac{1}{2 i} \sum_{n=1}^{N_{L/R}}  ( \psi^{L/R *}_n(x,t) \partial_x \psi^{L/R}_n(x,t) - \psi^{L/R}_n(x,t) \partial_x \psi^{L/R*}_n(x,t) ) \;,\nn
\eea 
which can be rewritten in terms of the kernel as
\be  \label{Jdef} 
J(x,t) = \frac{1}{2 i} (\partial_{x'} - \partial_{x} ) K(x,x';t)|_{x'=x} = {\rm Im} K_{01}(x,x;t) \;,
\ee 
where $K_{01}(x,y,t)= \partial_y K(x,y,t)$.
From the Schr\"odinger equation, the current $J(x,t)$ satisfies the fermion number conservation equation
\be \label{conservation}
\partial_t \rho(x,t) + \partial_x J(x,t) = 0 \;.
\ee 
\noindent 

In this paper we compute the time evolution of the density $\rho(x,t)$, of the kernel $K(x,x',t)$
and of the current $J(x,t)$. Since we are interested in the large time behavior of the bulk of the system, 
we will take the limit $\ell \to +\infty$ before taking $t \to + \infty$. 
More precisely we will take the limit $\ell \to +\infty$, $N_{L/R} \to +\infty$ with fixed $k_L$ and $k_R$, i.e. with fixed mean densities 
\be \label{def_rhoLR}
\rho_L = \frac{2 N_L}{\ell} = \frac{k_L}{\pi} \quad, \quad \rho_R = \frac{2 N_R}{\ell} = \frac{k_R}{\pi} 
\ee 
or equivalently with fixed Fermi energies $\mu_L, \mu_R$ (see Eq. (\ref{def_muLR})).

\section{Main results} \label{sec:main}

In this section we present our main results. The first one is the expression for the kernel
$K(x,x',t)$ in the thermodynamic limit $\ell \to +\infty$ at any fixed time $t$. It is a lengthy although fully explicit
expression which is given in \eqref{K_ABCD} as a sum of terms which are given respectively in 
\eqref{ref_sine}, \eqref{res33}, \eqref{Bsum}, \eqref{BR_offdiag_large} and \eqref{BR_diag_large}. 
From this expression one can read the time dependent density from \eqref{def_rhoxt} and 
the current from \eqref{Jdef}. 

The subsequent results concern the large time behavior obtained from the kernel, once the thermodynamic limit is taken. 
{We have found that there are actually} two different scaling regimes. The first one is the NESS where $x,x' =O(1)$ 
and the second one is the regime of rays where both $x,x'=O(t)$. Since the analytical computations 
of the asymptotic behaviors are quite tricky, we have carefully checked numerically
our main predictions, which are shown in Figs. \ref{densityprofile}, \ref{currentprofile} and \ref{densitycurrentprofile} below.

Finally, we extend our study to an initial state at finite temperature, with two different temperatures $T_L$ and $T_R$
to the left and to the right of the origin. We also obtain the heat current as a function of these two temperatures. 

As we will see below some of these results (but not all) can also be obtained from a heuristic semi-classical method
which relies on the momentum dependent transmission and reflection coefficients $T(k)$ and $R(k)$, which
for a delta function impurity are given by
\be 
\label{RTformula} 
T(k) = \frac{k^2}{k^2 + g^2} \quad , \quad 
R(k) = 1 - T(k) = \frac{g^2}{k^2+g^2} \;.
\ee
The results presented here are derived from first principles, starting from from an exact expansion over the eigenfunctions of many-body system. 

\subsection{Non equilibrium stationary state (NESS)}

$\quad$\vspace*{0.5cm}

The first regime corresponds to fixed spatial positions $x,x'$ with $t \to +\infty$. 
In this case the kernel, the density and the particle current reach a stationary limit which we compute explicitly, namely
\be 
\rho(x,t) \to \rho_{\infty}(x)  \quad , \quad J(x,t) \to J_{\infty} \quad , \quad K(x,x';t) \to K_\infty(x,x') \;.
\ee 
Note that from the fermion number conservation in Eq. (\ref{conservation}) the current is constant in space in the large time limit. From the symmetry of the problem under the change $x \to -x$, these observables satisfy the following relations
\be  \label{symmetry} 
K_\infty(x,x')|_{k_L,k_R} = K_\infty(-x,-x')|_{k_R,k_L} \,, \rho_{\infty}(x)|_{k_L,k_R} = \rho_\infty(-x)|_{k_R,k_L}\,, J_\infty|_{k_R,k_L} = - J_\infty|_{k_L,k_R} \;.
\ee

\subsubsection{The case of a repulsive impurity ($g>0$)} 
\vspace*{0.5cm}

We first give the results in the case of a repulsive impurity $g>0$, and consider the case $g<0$ in the next section.

\begin{figure}[t]
\centering
\includegraphics[width=0.55\linewidth]{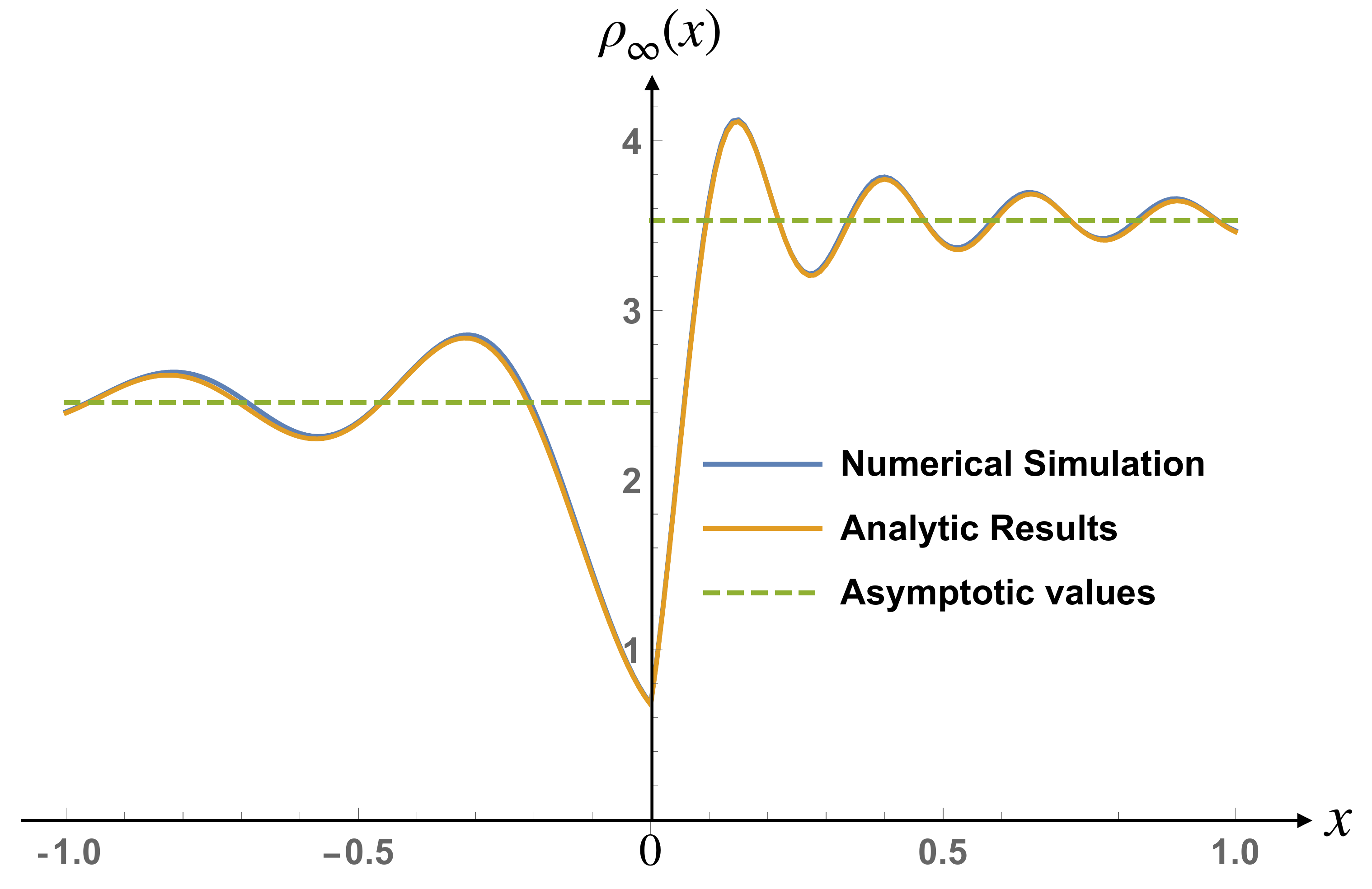}
\caption{Plot of the mean fermion density $\rho_{\infty}(x)$ (in orange) as a function of $x$ in the presence of a repulsive impurity ($g>0$) in the non equilibrium stationary state (NESS) given by Eq. (\ref{densitysteadystate>0}) for $\rho_R=4$, $\rho_L=2$ and $g=10$. {The oscillating behavior of $\rho_\infty(x)$ is the non-equilibrium analog of the Friedel oscillations.} The asymptotic values 
for $\rho_\infty(\pm \infty)$ given by \eqref{rho_asympt}
are indicated as horizontal dashed lines. These analytical results are compared with a numerical evaluation (blue) of $\rho(x,t)$ evaluated from the exact formula in \eqref{exact_K} with $x=x'$ for large time $t=\frac{\ell}{4\pi \rho_R}$ and system size $\ell=50$.
The agreement is excellent. Although large, this time has to be small enough so that the fastest fermions traveling at speed $k_R$ have not been reflected by the boundaries at $x=\pm \frac{\ell}{2}$. This is achieved if $k_R t<\frac{\ell}{2}$, i.e., if $t< t_\ell=\frac{\ell}{2\pi \rho_R}$.}
\label{densityprofile}
\end{figure}

\begin{figure}[t]
\centering
  \includegraphics[width=\linewidth]{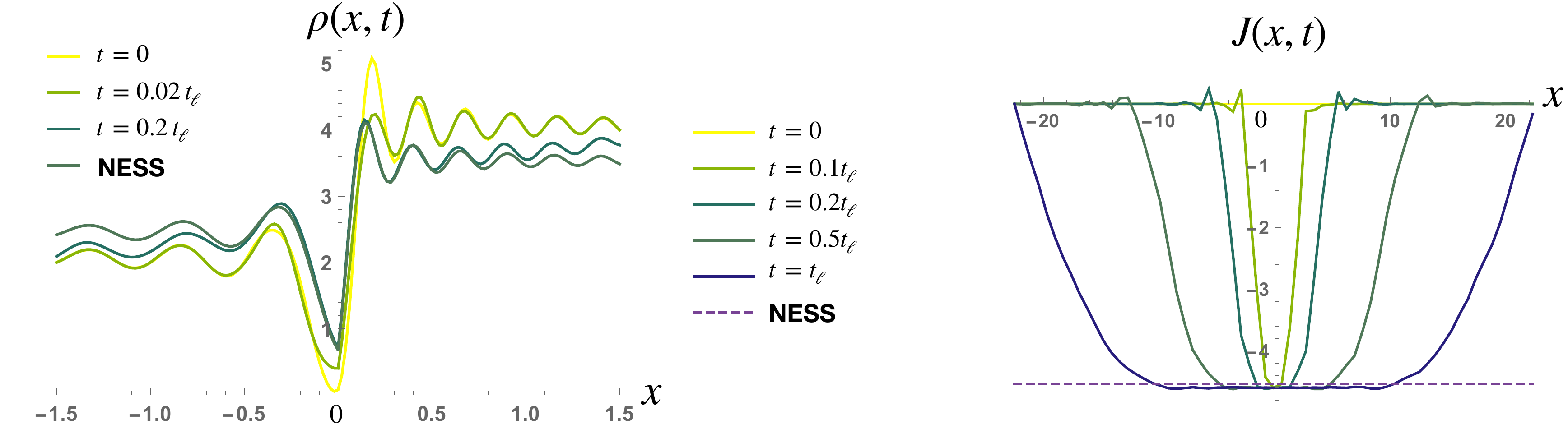}
\caption{
Plots showing the relaxation towards the non-equilibrium steady state (NESS). Left panel: plot of $\rho(x,t)$ (obtained from Eq. (\ref{exact_K} for $x=x'$) vs $x$ for various times $t$ compared to $t_\ell =\frac{\ell}{2\pi \rho_R}$, the time at which the fastest fermions reach the boundaries at $x = \pm \frac{\ell}{2}$
with $\ell=15$.
The density in the steady state $\rho_\infty(x)$ is given in Eq. (\ref{densitysteadystate>0}) and is also plotted in Fig.~\ref{densityprofile}. The parameter used are $\rho_R=4$, $\rho_L=2$, $g=10$.
We can see waves propagating away until the steady state value is reached. Right panel: plot of $J(x,t)$ obtained from Eq. (\ref{exact_K}) together with Eq. (\ref{Jdef}) vs $x$ for $\ell=45$. The dashed line indicates the exact value of $J_\infty$ in the NESS, given in Eq. (\ref{current_ness}).}
\label{dynamic-relaxation}
\end{figure}

\vspace*{0.5cm}
\noindent{\bf Density}. For the density we find, for $x>0$ 
\vspace*{0.5cm}
\begin{align}\label{densitysteadystate>0}
\rho_{\infty}(x>0)&=\frac{k_R}{\pi}  - \int_{k_L}^{k_R}\frac{dk}{2\pi}\frac{k^2}{g^2+k^2}+\frac{g}{\pi}\int_0^{k_R}dk\frac{k \sin(2k|x|)-g\cos(2kx)}{g^2+k^2}\nn\\
&=\frac{k_R}{\pi}  -  \int_{k_L}^{k_R}\frac{dk}{2\pi}\frac{k^2}{g^2+k^2} 
+\frac{g}{\pi}e^{2 g |x|}\,{\rm Im} E_1(2(g+ik_R)\,|x|) \;.
\end{align}
{Note that $\rho_\infty(x<0)$ is obtained from this expression Eq. (\ref{densitysteadystate>0}) together with the symmetry relation (\ref{symmetry}).} 
%\noindent $\bullet$ For positive $x>0$:  
%\begin{align}\label{densitysteadystate>0}
%\rho_{\infty}(x>0)&=\frac{k_R}{\pi}-\int_{k_L}^{k_R}\frac{dk}{2\pi}\frac{k^2}{g^2+k^2}+\frac{g}{\pi}\int_0^{k_R}dk\frac{k \sin(2kx)-g\cos(2kx)}{g^2+k^2}\nn\\
%&=\frac{k_R}{\pi}-\int_{k_L}^{k_R}\frac{dk}{2\pi}\frac{k^2}{g^2+k^2}+\frac{g}{\pi}e^{2 g x}\,{\rm Im} E_1(2(g+ik_R)\,x) \;.
%\end{align}
In the second line of Eq. (\ref{densitysteadystate>0}), ${\rm Im}$ denotes the imaginary part and $E_1(z) = \int_z^{+\infty} e^{-t} dt/t$ denotes the exponential integral
(see \cite{DLMS2021} for details on obtaining the second line from the first one). The function $E_1(z)$
is also denoted $\Gamma(0,z)$ in Mathematica,  the incomplete gamma function of index $0$, and the contour of integration defining it should not cross the negative real axis. 
%\vspace*{0.5cm}
%\noindent $\bullet$ For negative $x<0$:
%\begin{align} \label{densitysteadystate<0}
%\rho_{\infty}(x<0)&=\frac{k_L}{\pi}+\int_{k_L}^{k_R}\frac{dk}{2\pi}\frac{k^2}{g^2+k^2}+\frac{g}{\pi}\int_0^{k_L}dk\frac{- k \sin(2kx)-g\cos(2kx)}{g^2+k^2}\nn\\
%& = \frac{k_L}{\pi}+\int_{k_L}^{k_R}\frac{dk}{2\pi}\frac{k^2}{g^2+k^2}+\frac{g}{\pi}e^{2 g |x|}{\rm Im} E_1(2(g+ik_L)|x|)
%\end{align}
This result for $\rho_\infty(x)$ is shown in Fig. (\ref{densityprofile}) and compared with a numerical evaluation of the exact formula 
for $\rho(x,t)$ (from \eqref{exact_K} with $x=x'$) at a relatively large time. It is also plotted in Fig.
\ref{dynamic-relaxation} for smaller values of the time. As one can see on these figures, the 
convergence to our prediction within the NESS is rather fast.

Let us now discuss a few salient features of this result. Far from the impurity, which is located at $x=0$, the stationary density profile approaches constant which is different on both sides and given by
\begin{align} 
\lim_{x\to \pm\infty}\rho_{\infty}(x)&=\frac{\rho_L+\rho_R}{2}\pm\int_{k_L}^{k_R}\frac{dk}{2\pi}\frac{g^2}{g^2+k^2}\nn\\
&=\frac{\rho_L+\rho_R}{2}\pm\frac{g}{2\pi}\left(\arctan\left(\frac{\pi \rho_R}{g}\right)-\arctan\left(\frac{\pi \rho_L}{g}\right)\right) \;.
\end{align}
These asymptotic values can also be predicted by the semi classical method (see Section \ref{sec:wigner}) and written in the equivalent form
\be 
\rho_{\infty}(+\infty)= \frac{k_R}{\pi} - \int_{k_L}^{k_R}\frac{dk}{2\pi} T(k) \quad , \quad
\rho_{\infty}(-\infty)= \frac{k_L}{\pi} + \int_{k_L}^{k_R}\frac{dk}{2\pi} T(k)  \label{rho_asympt} \;,
\ee 
in terms of the transmission coefficient $T(k)$ given in Eq. (\ref{RTformula}). The mean density is continuous at $x=0$ but exhibits a cusp, with different left and right derivatives given by
\be \label{rholimit} 
\rho'_{\infty}(0^+) =  \int_0^{k_R} \frac{dk}{\pi} \frac{2 k^2 g}{k^2 + g^2}  \quad , \quad 
\rho'_{\infty}(0^-) = - \int_0^{k_L} \frac{dk}{\pi} \frac{2 k^2 g}{k^2 + g^2} \;.
\ee 
At variance with the equilibrium case (see below), this cusp is asymmetric. 

Finally, the result for $\rho_\infty(x)$ can be compared to the result obtained in \cite{DLMS2021} for the mean density $\rho_{\rm eq}(x)$ of the equilibrium problem,
i.e. in the ground state with Fermi energy $\mu=\frac{k_F^2}{2}$ in the presence of a repulsive delta impurity of strength $g>0$
(see formula (60) and (138) there with $\lambda=g$) 
\bea  \label{rho_eq}
\rho_{\rm eq}(x) && = \frac{k_F}{\pi} + \frac{g}{\pi} \int_0^{k_F} dk \frac{ k \sin(2 k |x|) - g \cos(2 k x) }{ k^2 + g^2} \nn\\
&& =\frac{k_F}{\pi}+ \frac{g}{\pi} e^{2 g |x|} {\rm Im} E_1 (2 (g + i k_F) |x|) \;.
\eea 
We see that for $k_L=k_R=k_F$ the NESS density coincides with the equilibrium (ground state) density
(as discussed below this holds only for $g \geq 0$). 
This is quite interesting since the initial state in the present work is far from the ground state of the system in the presence of the impurity. 
An explanation for this property is that the components of the initial state on the excited states correspond to fermionic
waves (quasi particles) propagating towards the edges of the system. %These waves can be seen in Fig. (\ref{dynamic-relaxation}). 
If the observation time is smaller than $t_\ell=\frac{\ell}{2\pi \rho_R}$, 
such that the fastest fermions traveling at speed $k_R$ have not been reflected yet, i.e. $k_R t<\frac{\ell}{2}$, 
we expect relaxation to the equilibrium state for $k_L=k_R$ (and to the NESS for $k_L \neq k_R$). 
This will always occur if the limit $\ell \to +\infty$ is taken first. 
For a finite size system, oscillations will take place on larger time scales. 
\\

\begin{figure}[t]
\centering
\includegraphics[width = 0.5\linewidth]{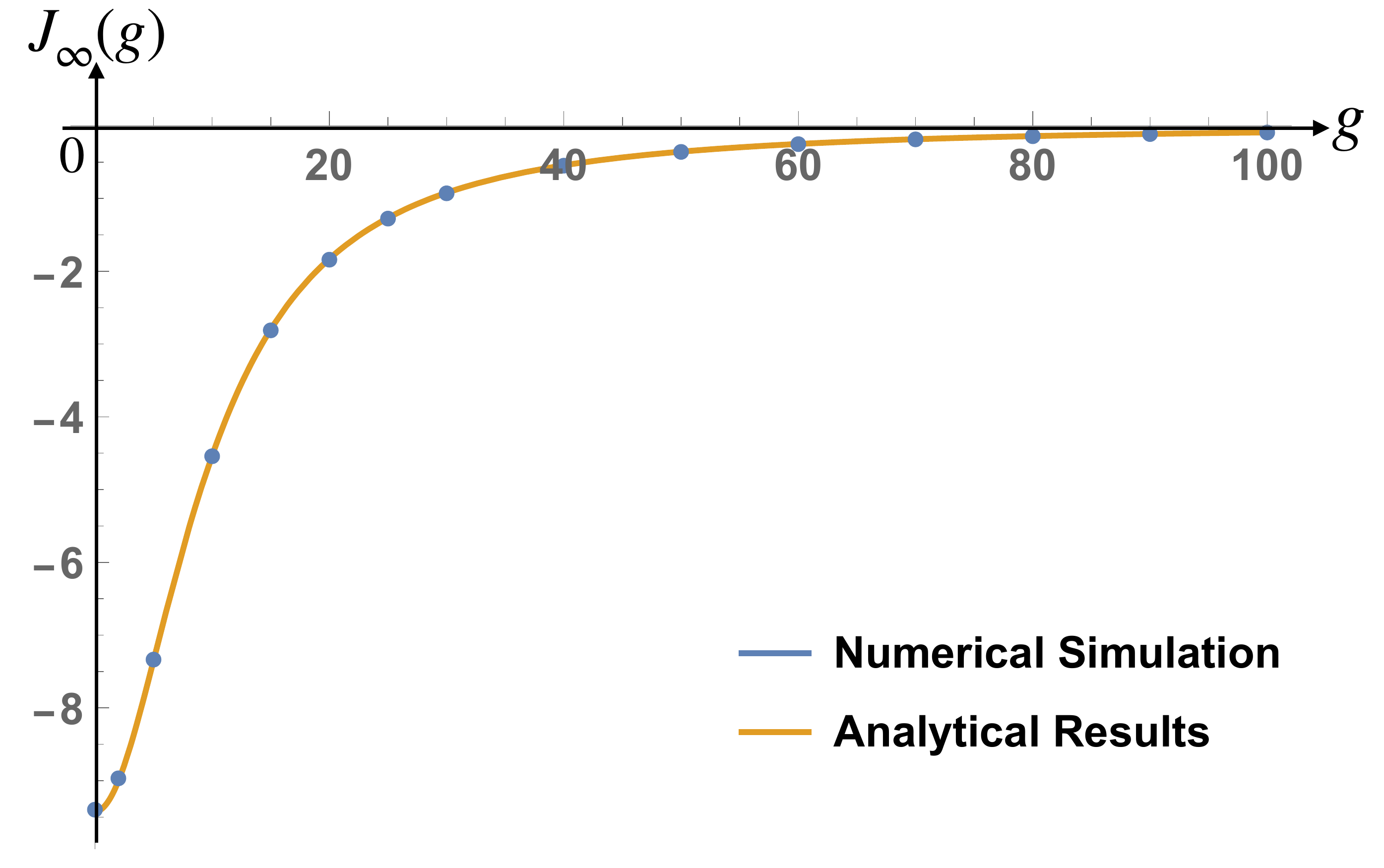}
\caption{Plot of the current $J_\infty$ (orange solid line) in the NESS as a function of $g$, the strength of the delta impurity, as given by
formula Eq. (\ref{current_ness}), with $\rho_R=4$, $\rho_L=2$. It is compared with a numerical evaluation (blue dots) of the exact expression of $J(x=0,t)$, obtained from Eq. (\ref{exact_K}) together with (\ref{Jdef}), 
for large time $t=\frac{\ell^2}{8\pi N_R}=t_{\ell}/2$ and system size $\ell=50$ for several values of~$g$.}
\label{currentprofile}
\end{figure}

\noindent{\bf Current}. In addition, in the NESS, we show that there is a non zero particle current given by 
\be \label{current_ness}
J_\infty=  -\int_{k_L}^{k_R}\frac{dk}{2\pi} \frac{k^3}{k^2 + g^2} =\frac{1}{2\pi}\left(\mu_L-\mu_R+\frac{g^2}{2}\ln\left(\frac{g^2+2\mu_R}{g^2+2\mu_L}\right)\right) \;,
\ee
which can alternatively be expressed, using the transmission coefficient $T(k)$ given in Eq. (\ref{RTformula}), as
\be 
J_\infty= -\int_{k_L}^{k_R}\frac{dk}{2\pi}k \, T(k) \;.
\ee 
This result can also be obtained by a semi-classical method, see Section \ref{sec:wigner}. The current is shown in Fig.~\ref{currentprofile} for $\rho_R=4$, $\rho_L=2$ (note that it is negative in that case). Its maximal (absolute) value is reached for $g=0$ when there is no defect. In that case one finds $J_\infty= \frac{1}{2\pi}(\mu_L-\mu_R)$ which corresponds to a unit conductance $e^2/h$.
In the limit $g \to + \infty$ it vanishes as 
\be \label{J_cond}
J_\infty = \frac{1}{2\pi g^2} (\mu_L^2 - \mu_R^2) + O\left(\frac{1}{g^4}\right) \;,
\ee 
which shows, as expected, that for a very strong defect the system is effectively cut into two almost independent halves.

%\vspace*{0.5cm}

\vspace*{0.5cm}
\noindent{\bf Kernel}. Finally we also obtain the kernel in the NESS which reads explicitly for $x,x'>0$

\bea  \label{Kpospos_intro} 
&&  K_\infty(x>0,x'>0) = \int_{0}^{k_R}\frac{dk}{\pi} \cos \left(k(x-x')\right) - 
\int_{k_L}^{k_R}\frac{dk}{2 \pi} \frac{k^2}{k^2 + g^2} e^{-i k(x-x')} \\
&+ & \int_{0}^{k_R}\frac{dk}{\pi} \frac{ g k \sin \left(k(x+x')\right) - g^2 \cos\left(k(x+x')\right) }{k^2 + g^2} \;, \nonumber 
\eea 
while for $x>0,x'<0$ it reads 
\bea \label{Kposneg_intro} 
&& K_\infty(x>0,x'<0) = \left( \int_{0}^{k_R} + \int_{0}^{k_L}  \right) \frac{dk}{2 \pi} \frac{k}{k^2 + g^2} ( k \cos (k(x-x')) + g \sin (k(x-x') ) 
\\
&& + i \int_{k_L}^{k_R} \frac{dk}{2 \pi} \frac{k}{k^2 + g^2} \left(k \sin k(x-x') - g \cos k(x-x') + g e^{- i k(x+x')} \right) \;, \nonumber 
\eea
the other cases being obtained by symmetry (see \eqref{symmetry}). For $g>0$ equivalent expressions are given in \eqref{Kernel_NESS2}. The kernel simplifies in the limit where the points $x,x' \to \infty$. In this limit, keeping $x-x'=O(1)$ one finds the asymptotic behavior
\bea  \label{Ksc10} 
&& K_\infty(x,x')  \simeq \int_0^{k_R} \frac{dk}{\pi} \cos(k(x-x')) - \int_{k_L}^{k_R} \frac{dk}{2 \pi} \frac{k^2}{k^2+g^2} e^{- i k (x-x')} \;.
\eea 
On the other hand, if $x \to +\infty$ and $x'\to - \infty$ with $x+x'=O(1)$ one finds
\bea \label{Kposneg20} 
K_\infty(x,x') \simeq  i \int_{k_L}^{k_R} \frac{dk}{2 \pi} \frac{k g}{k^2 + g^2} e^{-i k(x+x')} \;,
\eea
while the asymptotic kernel, for $x,x' \to \infty$, vanishes for generic value of $x'/x$ different from $\pm 1$. 

It is interesting to note that although the semi-classical method cannot obtain the full expressions in Eqs. 
\eqref{Kpospos_intro}, \eqref{Kposneg_intro}, it does predict the asymptotic form in \eqref{Ksc10}, see Section \ref{sec:wigner}.
However, we will see that the result \eqref{Kposneg20} cannot be predicted by this semi-classical method.

\vspace*{0.5cm}
\noindent{\bf Comparison with the GGE}. It is interesting to compare our result for the kernel $K_\infty(x,x')$ in the NESS 
to the prediction $K_{\rm GGE}(x,x')$ from the
GGE. This is done in details in the Appendix~\ref{sec:GGE}. The main result is that the kernel $K_{\rm GGE}(x,x')$
contains only "diagonal" terms (which are time independent already at finite $\ell$) and does not
lead to any current. However we show that there are non-diagonal
terms (called $C$ below) which carry current and contribute to the NESS if the limit $\ell \to \infty$ is carried first.

\subsubsection{The case of an attractive impurity ($g<0$)}  

\noindent If $g$ is negative there is an additional eigenstate of $\hat H_g$, denoted $\phi_g(x)\underset{\ell\to\infty}{\simeq}\sqrt{|g|}e^{g|x|}$ with eigenenergy $E=-\frac{g^2}{2}$, which is a bound state. All other eigenstates have positive energies and hence propagate over the whole system. One finds that the kernel $K_\infty$ for $g<0$ has the same expression as above [see Eqs.~(\ref{Kpospos_intro}) and (\ref{Kposneg_intro})], up to an additive term
denoted $\delta K_{\infty}$ which takes the form 
\bea \label{Kinf_gneg}
\delta K_{\infty}(x,x') = 2g^2 e^{g(|x|+|x'|)}\left(\int_0^{k_R}+\int_0^{k_L}\right)\frac{dk}{\pi}\frac{k^2}{(g^2+k^2)^2} 
\eea
Because of the exponential factor $e^{g(|x|+|x'|)} = e^{-|g|(|x|+|x'|)}$, this additional contribution $\delta K_{\infty}$ in the kernel is localised around $x=0$.

From the result for the kernel one obtains that for $g<0$ the density in the NESS is given for $x>0$
\begin{align}\label{densitysteadystate<0}
\rho_{\infty}(x>0)&=\frac{k_R}{\pi} - \int_{k_L}^{k_R}\frac{dk}{2\pi}\frac{k^2}{g^2+k^2}+\frac{g}{\pi}\int_0^{k_R}dk\frac{k \sin(2k|x|)-g\cos(2kx)}{g^2+k^2}\nn\\
& + 2 g^2 e^{2 g|x|}\left(\int_0^{k_R}+\int_0^{k_L}\right)\frac{dk}{\pi}\frac{k^2}{(g^2+k^2)^2} \\
&
\hspace*{-1.cm}=\frac{k_R}{\pi}-\int_{k_L}^{k_R}\frac{dk}{2\pi}\frac{k^2}{g^2+k^2}+\frac{g}{\pi}e^{2g|x|}{\rm Im} E_1(2(g+ik_R)|x|)\nn \\
&-\frac{2g^2}{\pi}e^{2g|x|}\left(\int_{k_R}^{\infty}+\int_{k_L}^{\infty}\right)dk\frac{k^2}{(k^2+g^2)^2} \;. \nonumber 
\end{align}
For $x<0$ the formula for $\rho_{\infty}(x<0)$ is again obtained from \eqref{densitysteadystate<0} simply by permuting $k_L$ and $k_R$.
%the same formula as in Eq. (\ref{densitysteadystate>0}) with the additional term
%\be 
%\delta \rho_{\infty}(x) = 2 g^2 e^{2 g|x|}\left(\int_0^{k_R}+\int_0^{k_L}\right)\frac{dk}{\pi}\frac{k^2}{(g^2+k^2)^2} \label{delta_rho_gneg}
%\ee 
Finally the current in the NESS is still given by formula \eqref{current_ness}, which is invariant under the change $g \to -g$,
and is thus not affected by the bound state. 

In the case $k_L=k_R=k_F$ one can compare with the equilibrium result for $g<0$ obtained in \cite{DLMS2021}. 
For the density it was found there that at equilibrium for $g<0$
\bea 
 \rho_{\rm eq}(x) &=&  \frac{k_F}{\pi} + \frac{g}{\pi} \int_0^{k_F} dk \frac{ k \sin(2 k |x|) - g \cos(2 k x) }{ k^2 + g^2} + \frac{g}{\pi} e^{2g|x|} \nn \\
& = & \frac{k_F}{\pi}+\frac{g}{\pi}e^{2g|x|}{\rm Im} E_1(2(g+ik_F)|x|) \;, \label{rho_eq_gneg}
\eea
which is compatible with \eqref{rho_eq} because the function $E_1(z)$ has a branch cut on the negative real axis. Hence, by comparing Eq. (\ref{densitysteadystate<0}) and (\ref{rho_eq_gneg}), we see that, at variance with 
the case $g>0$, the NESS for $g<0$ {\it differs} from the equilibrium ground state. This is because the mean occupation number $n_g$ of the
bound state which is unity in the ground state, and which is defined in the NESS from $\delta K_{\infty}(x,x') = n_g \phi_g(x) \phi_g(x')$ is given from \ref{Kinf_gneg}
as
\bea
&&\hspace*{-0.7cm}  n_g = \left(\int_0^{k_R}+\int_0^{k_L}\right)\frac{dk}{\pi}\frac{2 |g| k^2}{(g^2+k^2)^2} 
= \frac{1}{2} \left( F\left(\frac{k_R}{|g|}\right) + F\left(\frac{k_L}{|g|}\right) \right) \\
&& F(u) =\frac{2}{\pi} \left( {\rm arctan}(u) - \frac{u}{1+u^2}\right) \;,
\eea
is strictly smaller than unity (including in the case $k_L=k_R$). 
Hence the last term in the last line of Eq. \eqref{densitysteadystate<0} is proportional to $1-n_g$. 
{Therefore the post-quench bound state is always partially empty in the NESS, see \cite{Rossi2021} for
a related effect in a similar model based on a GGE calculation.} 

{Note that in the present model there is a single bound state. In the case of multiple bound states it has been found in related models that the NESS can present persistent oscillations in time 
\cite{Prosen2018,Gamayun1,Rossi2021}. To obtain such an effect in the present model would require to consider two delta impurities.}

%{\blue Greg and Pierre here is what numeric result suggest:
%\bea
%e^{2g|x|}Im(E_1(2(g+ik_F)|x|))&&=\int_0^{k_F}dk\frac{k\sin(2k|x|)-g\cos(2k x)}{(k^2+g^2)}+4ge^{2g|x|}\int_0^{+\infty}dk\frac{k^2}{(k^2+g^2)^2}\\
%&&=\int_0^{k_F}dk\frac{k\sin(2k|x|)-g\cos(2k x)}{(k^2+g^2)}-\pi e^{2g|x|}
%\eea
%The important part here is the $\infty$ bound in the third integral. In our case the integral was going to $k_F$ as $\int_0^{k_F}dk\frac{k^2}{(k^2+g^2)^2}$. Now
%\bea
%&&\frac{k_F}{\pi}+\frac{g}{\pi}e^{2g|x|}Im(E_1(2(g+ik_F)|x|))=\frac{k_F}{\pi}+\frac{g}{\pi}\int_0^{k_F}dk\frac{k\sin(2k|x|)-g\cos(2k x)}{(k^2+g^2)}-g e^{2g|x|}\\
%&&\frac{k_F}{\pi}+\frac{g}{\pi}e^{2g|x|}Im(E_1(2(g+ik_F)|x|))=\rho_{eq,g>0}+|\phi_g(x)|^2
%\eea
%One can conjecture that the right part of the equality correspond to $\rho_{eq}(x)$ leading to an extension of David formulation for $g<0$.

%Knowing this we have for $g<0$
%\bea
%&&\rho(x>0)=\frac{k_R}{\pi}-\int_{k_L}^{k_R}\frac{dk}{2\pi}\frac{k^2}{g^2+k^2}+\frac{g}{\pi}\int_0^{k_R}dk\frac{k\sin(2kx)-g\cos(2kx)}{k^2+g^2}+\frac{2g^2}{\pi}e^{2g|x%|}(\int_{0}^{k_R}+\int_{0}^{k_L})dk\frac{k^2}{(g^2+g^2)^2}\\
%&&=\frac{k_R}{\pi}-\int_{k_L}^{k_R}\frac{dk}{2\pi}\frac{k^2}{g^2+k^2}+\frac{g}{\pi}e^{2gx}ImE_1(2(g+ik_R)x)-\frac{2g^2}{\pi}e^{2g|x|}(\int_{k_R}^{\infty}+\int_{k_L}^{\%infty})dk\frac{k^2}{(g^2+g^2)^2}
%\eea
%}

\subsubsection{Relaxation to the NESS for $g>0$}

We have also obtained the large time decay of the kernel $K(x,x',t)$ towards its value in the NESS {for $g>0$}.
The result for $\Delta K(x,x',t)=K(x,x',t)- K_\infty(x,x')$ is given in \eqref{DeltaKt}. From
it one extract the density and the current. The decay of the density is found to be of the schematic form
\bea \label{densdecay_intro}
&& \rho(x,t) - \rho_\infty(x) = - \left(\frac{1}{k_L}+\frac{1}{k_R}\right) \frac{(1+ g|x|)^2}{\pi^2 g^4 t^3} \\
&& 
+ \frac{1}{g^4 t^{5/2}} \left( \chi_R\left(\frac{g}{k_R},k_R x\right) \cos\left(\frac{k_R^2 t}{2} - \frac{\pi}{4} \right) 
+ \chi_L\left(\frac{g}{k_L},k_L x\right) \cos\left(\frac{k_L^2 t}{2} - \frac{\pi}{4}\right) \right) + o\left(\frac{1}{t^{5/2}}\right) \nn 
\eea 
where the functions $\chi_{L/R}(z)$ can be read off from Eq. \eqref{densdecay} in the Appendix \ref{App_NESS}. Hence the
leading decay is $t^{-5/2}$ modulated by oscillations, together with a $1/t^3$ term which is non oscillating.
We also find that the current has also a leading algebraic decay as $t^{-5/2}$ modulated by oscillations.
Note that power law decays with oscillations have been obtained in other systems of noninteracting fermions 
\cite{Kulkarni,DubailViti2016}.

\subsection{Large time regime with $\xi=x/t$ fixed (ray regime)} 

$\quad$ \\

The second regime corresponds to both $x,t \to \infty$ with a fixed ratio $\xi =x/t$, i.e. along rays. 
In this case the density and the current reach finite limits, which are only functions of the scaling variable $\xi$
\be 
\rho(x,t) \to \tilde \rho(\xi)  \quad , \quad J(x,t) \to \tilde J(\xi)  \;. 
\ee 
Note that the fermion number conservation Eq. (\ref{conservation}) implies that these two functions must be related via
\be  \label{conservation2} 
\partial_\xi \tilde J(\xi) = \xi \partial_\xi \, \tilde \rho(\xi) \;.
\ee 
All the results below in that regime hold for any $g$ (positive or negative) since the bound state (that exists for $g<0$) does not contribute in that limit. 

\vspace*{0.5cm}

\begin{figure}[t]
\centering
\includegraphics[width = \linewidth]{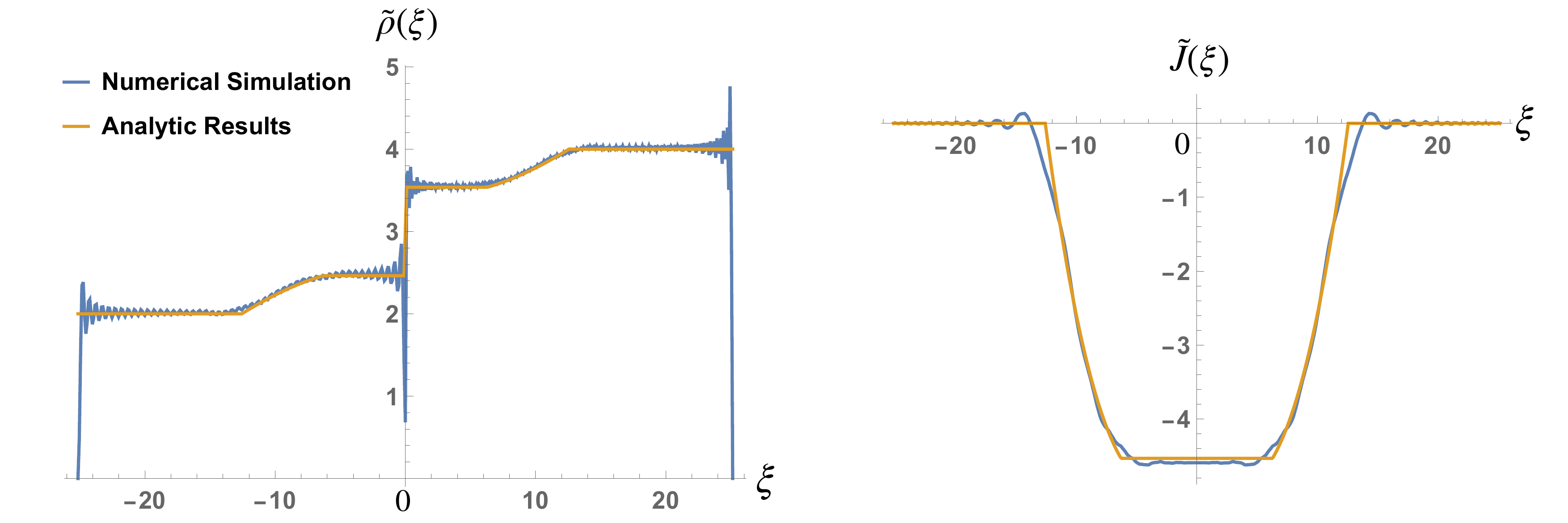}
\caption{Asymptotic density (left panel) and current (right panel) at large time in the
regime of rays $\xi = x/t$ fixed. (Orange) Plots of $\tilde \rho(\xi)$ and $\tilde J(\xi)$ as a function of $\xi=x/t$ as given in Eqs. \ref{tilderho} and \ref{tildeJ} for
$\rho_R=4$, $\rho_L=2$ and $g=10$. 
(Blue) For comparison, $\rho(x,t)$ and $J(x,t)$ are plotted versus $\xi=x/t$ for $t = \frac{\ell}{4\pi \rho_R}= t_{\ell}/2$ and $\ell=50$.
In this problem there are 2 pairs of light cones at $|\xi|=k_R=4\pi$ and $|\xi|=k_L=2\pi$. On this scale the density exhibits a jump at $\xi=0$, which
is rounded on a scale $x=O(1)$ in the NESS (see Fig.~\ref{densityprofile}) with a perfect matching as $\xi \to 0^\pm$, see Eq. (\ref{Jxinew})
(the oscillations visible here for $\xi \approx 0$ are actually part of the NESS regime). }
\label{densitycurrentprofile}
\end{figure}

\noindent {\bf Density}. We find through explicit calculation of the large time limit, that the scaling function for the density reads 
\bea \label{rhoxinew} 
\tilde \rho(\xi)=\frac{k_L+k_R}{2\pi}+{\rm sgn}(\xi)\left(\int_{k_L}^{k_R}\frac{dk}{2\pi}R(k)+\int_{k_L}^{k_R}\frac{dk}{2\pi}T(k)\Theta(|\xi|-k)\right) \;.
\eea
This gives, using the explicit formulae for $R(k)$ and $T(k)$ in (\ref{RTformula}), for $k_R>k_L$,   
\begin{eqnarray} \label{tilderho} 
\tilde \rho(\xi)=\left\{
    \begin{array}{ll}
       \rho_L & \mbox{if } \xi<-k_R \\
       \rho_L+\frac{\rho_R+\xi/\pi}{2}+\frac{g}{2\pi}(\arctan(-\frac{\xi}{g})-\arctan(\frac{\pi \rho_R}{g})) & \mbox{if } -k_R<\xi<-k_L \\
       \frac{\rho_L+\rho_R}{2}- \frac{g}{2\pi}(\arctan(\frac{\pi \rho_R}{g})-\arctan(\frac{\pi \rho_L}{g}))& \mbox{if } -k_L<\xi<0 \\
       \frac{\rho_L+\rho_R}{2}+\frac{g}{2\pi}(\arctan(\frac{\pi \rho_R}{g})-\arctan(\frac{\pi \rho_L}{g})) & \mbox{if } 0<\xi<k_L \\
       \frac{\rho_R+\xi/\pi}{2}+\frac{g}{2\pi}(\arctan(\frac{\pi \rho_R}{g})-\arctan(\frac{\xi}{g})) & \mbox{if } k_L<\xi<k_R \\
       \rho_R & \mbox{if } \xi>k_R
    \end{array}
\right. \;.
\end{eqnarray}
A plot of $\tilde \rho(\xi)$ is shown in the right panel of Fig. \ref{densitycurrentprofile}, together with an exact evaluation at finite time illustrating 
the convergence. Note that $\tilde \rho(\xi)$ is a continuous function of $\xi$ except at $\xi=0$ where it has a jump discontinuity. 
The values on each side of the jumps at $\xi=0^\pm$ are found to agree with the large distance limit of the density obtained the NESS regime, i.e. 
\be  \label{matching}
\tilde \rho(0^\pm) = \lim_{x\to \pm \infty}\rho_{\infty}(x) \;.
\ee 
This matching shows that there is no additional intermediate regime between the NESS $x=O(1)$, and the regime of rays $x= O(t)$.
\\

\vspace*{0.5cm}

\noindent {\bf Current}. For the scaling function of the current we obtain, again through explicit
calculation of the large time limit
\bea \label{Jxinew} 
\tilde J(\xi)=-\int_{k_L}^{k_R}\frac{dk}{2\pi}\; k\; T(k)\Theta(k-|\xi|) \;,
\eea
where $T(k)$ is given in (\ref{RTformula}). More explicitly, for $k_R>k_L$ it gives
\begin{equation} \label{tildeJ} 
\tilde J(\xi)=\left\{
\begin{array}{ll}
0 & \mbox{if } \xi<-k_R \\
\frac{1}{2\pi}(\frac{\xi^2}{2}-\mu_R+\frac{g^2}{2}\ln(\frac{g^2+2\mu_R}{g^2+\xi^2})) & \mbox{if }  -k_R<\xi<-k_L \\
\frac{1}{2\pi}(\mu_L-\mu_R+\frac{g^2}{2}\ln(\frac{g^2+2\mu_R}{g^2+2\mu_L})) & \mbox{if } -k_L<\xi<k_L \\
\frac{1}{2\pi}(\frac{\xi^2}{2}-\mu_R+\frac{g^2}{2}\ln(\frac{g^2+2\mu_R}{g^2+\xi^2})) & \mbox{if } k_L<\xi<k_R \\
0 & \mbox{if } \xi>k_R \\
\end{array}
\right. \;,
\end{equation}
where we recall that the Fermi energies are $\mu_{L/R}= \frac{k_{L/R}^2}{2} = \frac{\pi^2}{2} \rho_{L/R}^2$. 
A plot of $\tilde J(\xi)$ is shown in the right panel of Fig. \ref{densitycurrentprofile}, together with an exact evaluation at finite time illustrating the
convergence. The function $\tilde J(\xi)$ is a continuous function of $\xi$ everywhere. One can check that the conservation equation Eq. (\ref{conservation2}) 
is obeyed, including at the point $\xi=0$ (the delta function in $\partial_\xi \tilde \rho(\xi)$ is cancelled by the factor $\xi$ in Eq. (\ref{conservation2})). 
Note that in this model there are two pairs of light cones at $\xi=\pm k_L$ and $\xi=\pm k_R$ respectively. Outside these
two light cones ($|\xi| > \max(k_R,k_L)$) the current vanishes at large time. Inside these two light cones 
($|\xi| < \min(k_R,k_L)$) the current is constant and equal to its value in the NESS (and so is the density).

\vspace*{0.5cm}

\noindent {\bf Kernel}.
We have found by explicit calculation that in this ray regime the kernel $K(x = \xi t,x' = \xi' t,t)$ vanishes unless 
$\xi=\xi'$ or $\xi=-\xi'$. More precisely one obtains the limiting scaling forms for $y,y'=O(1)$ 
\bea \label{Kxi}
\lim_{t \to +\infty} K(\xi t + y,\pm \xi t + y') = K^{\pm}_{\xi}(y,y') \;.
\eea 
The expression for $K^{+}_{\xi}(y,y') $ is obtained as
\bea \label{Kp_xixi}
 K^{+}_{\xi}(y,y') &=&  \int_0^{k_R} \frac{dk}{\pi} \cos(k(y-y')) \Theta(\xi) + \int_0^{k_L} \frac{dk}{\pi} \cos(k(y-y')) \Theta(-\xi) 
\\
& - & {\rm sign}(\xi) \int_{k_L}^{k_R} \frac{dk}{2 \pi} T(k) 
e^{- i {\rm sign}(\xi) k (y-y') } \Theta(k - |\xi|) \nn 
\eea 
where we recall that $T(k) = k^2/(k^2+g^2)$. Note that $K^{+}_{\xi}(y,y')$ is a function of $y-y'$ only. From this expression (\ref{Kp_xixi}) in the limit of coinciding points one recovers the density $\tilde \rho(\xi)$
in Eq. \eqref{rhoxinew} (using $R(k)=1-T(k)$) and the current $J(\xi)$ in Eq. \eqref{Jxinew} using \eqref{Jdef}. It is important to note that
the kernel \eqref{Kp_xixi} matches exactly in the limit $\xi \to 0^+$ with the result 
\eqref{Ksc10} for the
kernel $K_\infty(x,x')$ of the NESS in the large distance limit $x,x'>0$. 
\\

The expression for $K^{-}_{\xi}(y,y')$ is given by 
\bea  \label{Kxi_m}
&& K^{-}_{\xi}(y,y') = 
%- \frac{\rho_{L}}{2} \frac{\sin(k_{L}(y+y') )}{k_{L}(y+y') } - \frac{\rho_{R}}{2} \frac{\sin(k_{R}(y+y') )}{k_{R}(y+y') } 
%\\
% && + 
i \,  {\rm sign}(\xi) \int_{k_L}^{k_R} \frac{dk}{2 \pi} \frac{g k}{g^2 + k^2} e^{- i {\rm sign}(\xi) k (y+y') }  \Theta(k - |\xi|) \;.
\eea
This measures the quantum correlation between opposite rays. Once again 
the kernel \eqref{Kxi_m} matches exactly in the limit $\xi \to 0^+$ with the result \eqref{Kposneg20} 
for the
kernel $K_\infty(x,x')$ of the NESS in the limit of very separated points $x>0,x'<0$.

It is important to remark that the results for the density \eqref{rhoxinew}, the current \eqref{Jxinew} 
and the kernel at coinciding rays \eqref{Kp_xixi} can be also obtained using the semi-classical method
as we will see in Section \ref{sec:wigner}. However the result for $K^{-}_{\xi}(y,y')$ for opposite rays in Eq. (\ref{Kxi_m}) cannot
be obtained from this semi-classical method. Indeed, it contains additional information about correlations
at a finite distance from the two opposite rays which correspond to fermion wave-functions which
are split between reflected and transmitted waves by the defect.

\subsection{Finite temperature generalisation}\label{sec_finiteT}

$\quad$ \\

The previous result can be generalised to an initial state with non zero temperature $T_{L/R}$ different on both sides of the impurity 
and with associated chemical potential $\mu_{L/R}$. This amounts to take as an initial kernel at $t=0$
\bea 
&&K_{L}(x,x') = \Theta(-x) \Theta(-x') \sum_{n=1}^{\infty} \frac{4}{\ell}f_L\left(\frac{2\pi n}{\ell}\right)\sin\left(\frac{2\pi n x}{\ell}\right)\sin\left(\frac{2\pi n x'}{\ell}\right) \label{KLT} \\
&&K_{R}(x,x') = \Theta(x) \Theta(x') \sum_{n'=1}^{\infty} \frac{4}{\ell}f_R\left(\frac{2\pi n'}{\ell}\right)\sin\left(\frac{2\pi n' x}{\ell}\right)\sin\left(\frac{2\pi n' x'}{\ell}\right)  \label{KRT} \;,
\eea
where 
\bea \label{Fermi}
f_{L/R}(k)=\frac{1}{\exp(\beta_{L/R}(\frac{k^2}{2}-\mu_{L/R}))+1} \;,
\eea
is the Fermi factor with $\beta_{L/R}=1/T_{L/R}$. The density of fermions in the initial state is now related to the chemical potentials via
\be 
\rho_{L/R} = \int_0^{+\infty}  \frac{dk}{\pi} f_{L/R}(k) \;.
\ee 
The calculation proceeds in the same way as for $T_{L/R}=0$ with a few additional details
which are given in Appendix \ref{app:finiteT}. We present here only the result for $g>0$. 

In the NESS regime $x=O(1)$ the full kernel is presented in 
\eqref{kernelfiniteT} and from it we find the asymptotic density at finite temperature, for $x>0$:  
\begin{align}\label{densitysteadystate>0T}
\rho_{\infty}(x>0)&&=\int_0^{\infty}\frac{dk}{\pi} \left({f_R(k)}-\frac{(f_R(k)-f_L(k))}{2}T(k)+{gf_R(k)} \frac{k \sin(2kx)-g\cos(2kx)}{g^2+k^2} \right)\;,
\end{align}
and the expression for $x<0$ can be obtained from the symmetry \eqref{symmetry} with here $\mu_L \leftrightarrow \mu_R$ and $T_L \leftrightarrow T_R$.
Note that the density retains a cusp at $x=0$ even at finite temperature. 
The current in the NESS at finite temperature is obtained as
\bea\label{currentsteadystate>0T}
J_{\infty} =-\int_0^{\infty}dk\frac{f_R(k)-f_L(k)}{2\pi}k \, T(k) \;.
\eea
Its low temperature expansion is performed in Appendix \ref{App_lowT} and we display it here in the case $\mu_L=\mu_R= \frac{k_F^2}{2}$
\be \label{Jinf_g}
J_{\infty} = \frac{\pi}{6}(T_L^2 -T_R^2) \frac{g^2}{(g^2 + k_F^2)^2} 
+ \frac{7 \pi^3}{15} \frac{g^2}{(g^2 + k_F^2)^4} (T_L^4-T_R^4) \;.
+ O(T_R^6,T_L^{6}) 
\ee 
Note that in the absence of impurity, for $g=0$, one has instead (for any $\mu_{L/R}$) 
\be \label{Jinf_g0}
J_\infty|_{g=0} = \frac{ \mu_L - \mu_R}{2 \pi}  + \frac{1}{2 \pi} (T_L e^{- \mu_L/T_L} - T_R e^{- \mu_L/T_R} ) 
\ee 
which generalizes the standard zero temperature result. Note that the $g=0$ case has been studied for the same model
in a high temperature limit $k_R^2/T_R \sim k_L^2/T_L \ll 1$ where interesting equilibration properties were found \cite{ColluraKarevski2014}. 
In our case ($g\neq 0$), one can show that the current is proportional to $T_R \sim T_L$ at high temperature. 
\\

\begin{figure}[t]
\centering
\includegraphics[width=\linewidth]{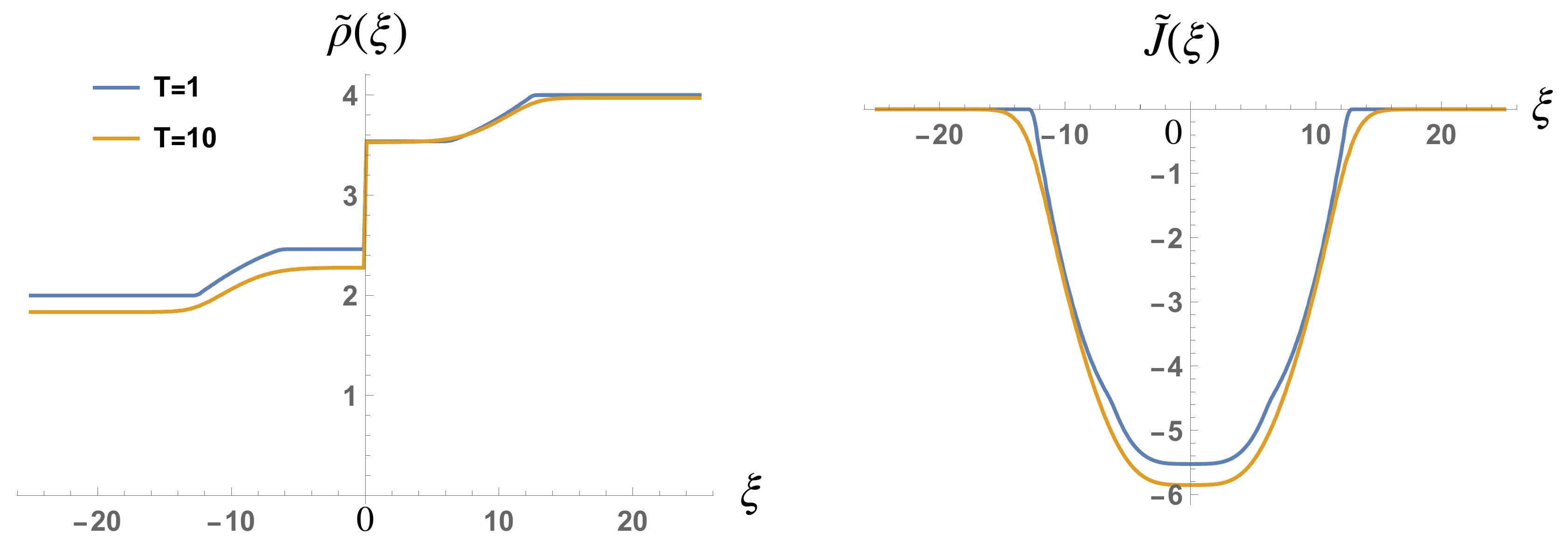}
\caption{Plot of the asymptotic density $\tilde \rho(\xi)$ and current $\tilde J(\xi)$ in the ray regime as a function of $\xi=x/t$ for non zero temperature as given by Eqs. \ref{tilderhoT} and \ref{tildeJT} for
$\rho_R=4$, $\rho_L=2$ and $g=10$. The temperatures are identical on both sides with $T_L=T_R$ equal to 1 (orange curve) and 10 (blue curve). 
We notice that the singularities at the light cones $\xi = \pm k_{R/L}$ are rounded compared to the zero temperature case shown in Fig. \ref{densitycurrentprofile}.}
\label{Fig_finiteT}
\end{figure}

In the ray regime $\xi=x/t=O(1)$ we find the density
\bea\label{tilderhoT}
\tilde \rho(\xi)=\int_0^{\infty}\frac{dk}{2 \pi} \Big( {f_R(k)+f_L(k)}+{\rm sgn}(\xi){\left(f_R(k)-f_L(k)\right)}\,(R(k)+T(k)\Theta(|\xi|-k)) \Big) \;,
%\red{+\frac{f_R(k)+f_L(k)}{4\pi}\frac{k^2-g^2}{g^2+k^2} \mbox{ if } \xi=0}
\eea
and the current 
\bea\label{tildeJT}
\tilde J(\xi)=-\int_0^{\infty}dk \frac{f_R(k)-f_L(k)}{2\pi}kT(k)\Theta(k-|\xi|)
\eea
These results are plotted in Fig. \ref{Fig_finiteT}.

\subsection{Energy current}

$\quad$ \\

Let us a recall the definition of the heat current in quantum mechanics for a single particle with a Hamiltonian $\hat H = - \frac{1}{2} \partial_x^2 + V(x)$ and described by the wavefunction $\psi(x,t)$, see e.g. \cite{Mathews1974}. The local energy $q(x,t)$ and heat current $j_q(x,t)$ are given by
\bea 
&& q(x,t)= \frac{1}{2} (\partial_x \psi(x,t)^*) (\partial_x \psi(x,t)) + \psi(x,t)^* V(x) \psi(x,t) \;, \label{def_Q} \\
&& j_q(x,t) =  -\frac{1}{2} {\rm Re} \left( 
i (\hat H \psi)(x,t)^* \partial_x \psi(x,t) \right) \;. \label{def_JQ}
\eea 
The total averaged energy is recovered from $\int dx \, q(x,t)= \langle \psi | \hat H | \psi \rangle$ while $q(x,t)$ and $j_q(x,t)$ obey the conservation equation
(from the Schr\"odinger equation) 
\be  \partial_t q(x,t) + \partial_x j_q(x,t) = 0 \;. \label{eq_conserv}
\ee 
For noninteracting fermions with the initial condition considered here (see Section \ref{sec:dyn}), each state $\psi_{L/R}^n(x,t)$ 
evolves independently, hence the total local energy $Q(x,t)$ and energy current $J_Q(x,t)$ are given by the corresponding sums 
of the one-body contributions weighted by the Fermi factors.
The time dependent energy current is thus given by
\bea \label{def_Ecurrent}
J_Q^{L/R}(x,t) =\sum_{n=1}^{\infty}f_{L/R}(k_n) 
{\rm Im} \left( 
(\hat H_g \psi_{L/R}^n)(x,t)^* \partial_x \psi_{L/R}^n(x,t) \right) \;,
\eea
where $k_n=\frac{2 \pi n}{\ell}$ and $\hat H_g$ is the single-particle Hamiltonian with a delta-impurity in Eq. (\ref{def_H}). 

The explicit calculation is performed using methods which are similar the ones used for the density and the particle current. They are summarized in Appendix \ref{sec:heat}. One finds that in the large time limit the energy current $J_Q(x,t)$ converges to an asymptotic constant value 
$J_{Q,\infty}$ which reads 
\bea \label{def_JQ}
J_{Q,\infty} =-\int_0^{\infty}\frac{dk}{2\pi}(f_{R}(k)-f_{L}(k))kE(k)T(k)  \quad , \quad E(k)= \frac{k^2}{2} \;.
\eea
In the low temperature limit and in the case $\mu_L=\mu_R= \frac{k_F^2}{2}$ the current has the following expansion (see Appendix \ref{App_lowT})
\be 
J_{Q,\infty} = \frac{\pi}{12} \frac{k_F^2(2g^2+k_F^2)}{(g^2+k_F^2)^2}  (T_L^2-T_R^2) - \frac{7 \pi^3}{30} \frac{g^4}{(g^2 + k_F^2)^4} (T_L^4-T_R^4) 
+ O(T_R^6,T_L^{6})  \label{JQ_lowT} \;.
\ee 
It is interesting to compare the first term in this low-temperature expansion with a result for the energy current obtained in \cite{BernardDoyonViti2015} 
\be  
{J^{\rm CFT}_{Q,\infty}} = c\,\frac{\pi}{12} \cos^2 \alpha   (T_L^2-T_R^2) \;,
\ee 
where $c$ is the central charge. A simple derivation of this result was given in \cite{BernardDoyonViti2015} for free Majorana fermions (corresponding to $c=1/2$). In that work $\cos^2 \alpha$ is the transmission coefficient of the defect, i.e. the analog of $T(k_F)$ here.
However the coefficient of $T_L^2-T_R^2$ that we obtain here in (\ref{JQ_lowT}) is not equal to $\frac{\pi}{12} T(k_F)$ as
would be predicted by the model of \cite{BernardDoyonViti2015} taking into account that here $c=1$. The discrepancy can be understood as follows.
From similar methods as in Appendix \ref{App_lowT} one can show that for a general dispersion relation $E(k)$ our formula \eqref{JQ_lowT} becomes
to leading order for small $T_L$ and $T_R$
\be \label{JQ_d}
{J_{Q,\infty}} \simeq \frac{\pi}{12} \frac{1}{k_F} \left( \partial_k ( E(k) T(k) ) \right)|_{k=k_F} \, (T_L^2-T_R^2) \;.
\ee 
Performing the derivative with respect to $k$ in (\ref{JQ_d}) we obtain two terms. The first one is $\propto E'(k_F) T(k_F)/k_F = T(k_F)$ and coincides with the 
result of \cite{BernardDoyonViti2015}. The additional term $\propto E(k_F) T'(k_F)/k_F$ is non zero when the transmission
coefficient depends on the momentum $k$, as it is the case here. This additional term is negligible compared to the first one only when the impurity strength is small, i.e., $g \ll k_F$.
\\

Note that in the absence of the impurity, i.e. for $g=0$, the low-temperature expansion of $J_{Q, \infty}$ reads for arbitrary $\mu_{L/R}$
\be \label{JQ_geq0}
J_{Q,\infty}|_{g=0} = \frac{ \mu_L^2 - \mu_R^2}{4 \pi} + \frac{\pi}{12}(T_L^2-T_R^2) - \frac{1}{2 \pi} (T_L^2 e^{- \mu_L/T_L} - T_R^2 e^{- \mu_R/T_R} )  + O(e^{-2\mu_L/T_L}, e^{-2\mu_R/T_R})
\ee 
The expansion formula for $g \neq 0$ and $\mu_L \neq \mu_R$ are given in the Appendix \ref{App_lowT}. 

\section{Exact calculation at finite time and finite size} \label{sec:exact}

In this section we derive an exact formula for the kernel at any time $t$ in a system with of size $\ell$. This will be the starting point for
the asymptotic analysis for $\ell \to +\infty$ and subsequently for the computation of the large time limit performed in the following sections.

\subsection{Eigenbasis of $\hat H_g$} 

$\quad$ \\

We first consider here $g>0$, and a finite interval $x \in [-\ell/2,\ell/2]$ and specify further $\hat H_g$ by imposing the vanishing of the wavefunctions at $x =\pm \ell/2$. 
The eigenfunctions of $\hat H_g$ are either even or odd in $x$, respectively labelled by a subscript '$+$' or '$-$'. The odd eigenfunctions do not feel the
delta impurity (since they vanish at the location of the impurity) hence they read 
\be \label{def_lm}
\phi_{-,q}(x) =\sqrt{\frac{2}{\ell}}\sin( q x) \quad , \quad q \in \Lambda_- = \left\{ \frac{2\pi n}{\ell} , n \in \mathbb{N}^*  \right\} \;,
\ee
where we denote $\Lambda_-$ the lattice of possible values for the wavevector $q$. 

\begin{figure}[t]
    \centering
    \includegraphics[width = 0.8\linewidth]{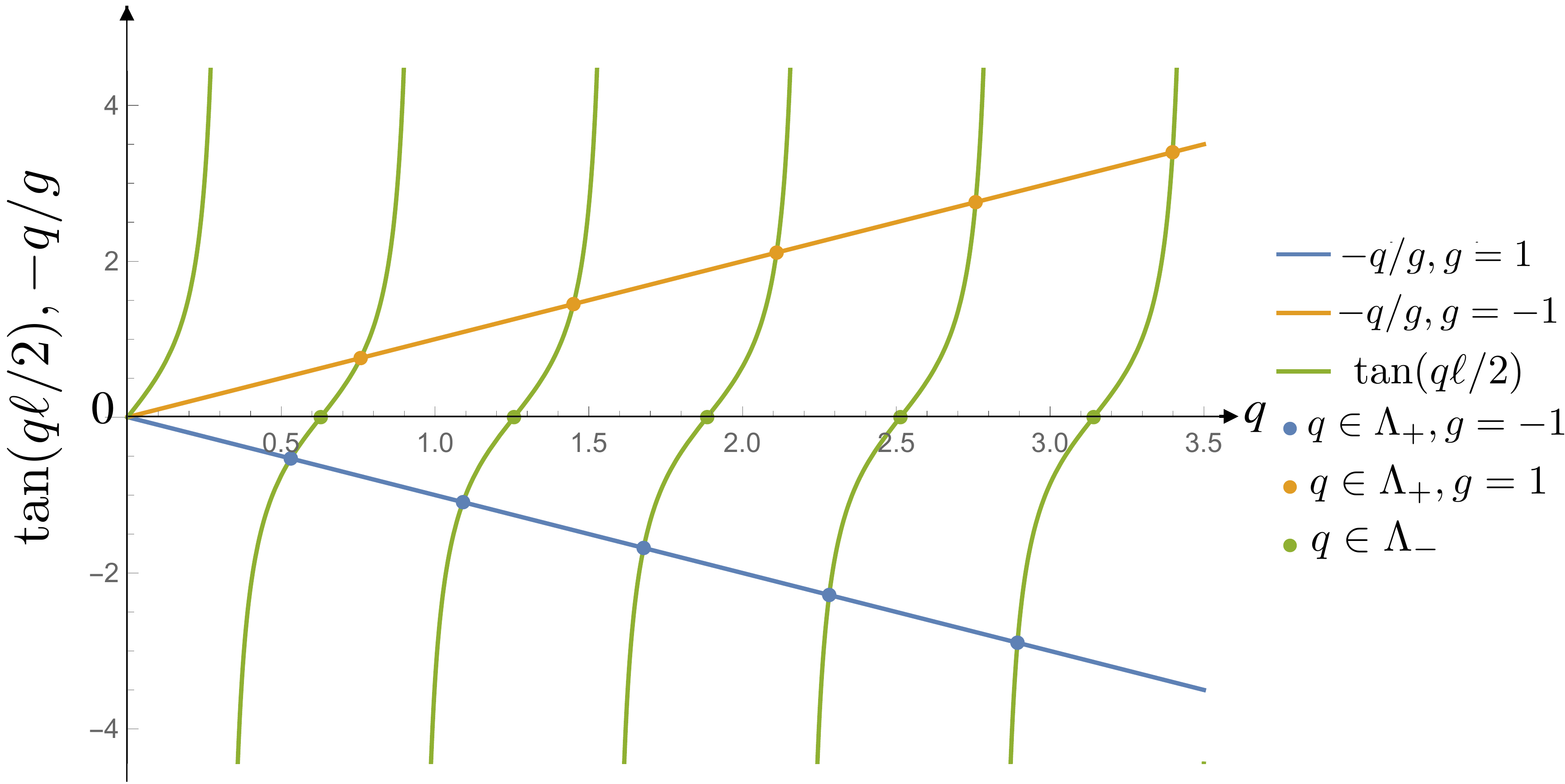}
    \caption{{Graphical representation of the quantization condition in Eq. (\ref{quantification0}). It is plotted here for $g=\pm 1$ and $\ell=10$.
    The intersection points $-\frac{q}{g}=\tan(\frac{q \ell}{2})$ generate the lattice $q \in \Lambda_+$ (\ref{def_lp}). The lattice $\Lambda_-$ corresponds to the roots of the equation  $0=\tan(\frac{q \ell}{2})=\sin(\frac{q \ell}{2})$ (see Eq. (\ref{def_lm})). Therefore the two lattices $\Lambda_+$ and $\Lambda_-$ are intertwined. For $g<0$ the situation is almost the same but now $-\frac{q}{g}$ has a positive slope. Note that for $g<0$, there is an additional bound state which cannot be shown on this figure.}  
    }
    \label{quantization}
\end{figure}

The even eigenfunctions are also plane vaves, and denoted by $\phi_{+,q}(x)$, but with a different quantization condition on $q$. They read 
\begin{align}
\phi_{+,q}(x)&= \frac{1}{\sqrt{({g^2+q^2)\frac{\ell}{2}+g}}} (q \cos(qx)+ g \sin(q|x|)) \label{def_pm} \;,
%&=\frac{1}{\sqrt{\frac{\ell}{2}+\frac{g}{g^2+q^2}}}\cos(qx+{\rm sgn}(x)\Theta_q) 
\end{align}
%where the phase shift angle $\theta_q \in [- \pi,\pi]$ is defined via $q - i g = \sqrt{q^2 + g^2} e^{i \theta_q}$ 
with the quantification condition (see Fig. (\ref{quantization}))
\be \label{quantification0}
q \cos\left(\frac{q \ell}{2}\right)+g\sin\left(\frac{q \ell}{2}\right)=0 \quad \Leftrightarrow e^{-i q \ell}=-\frac{q-ig}{q+ig} \;,
\ee
i.e., $q \ell= -2 {\rm atan} ({q}/{g}) + m \pi$,
which defines the lattice of possible wavevectors $q \in \Lambda_{+}$
\begin{align}
\Lambda_{+}&=\left\{q,q \cos(\frac{q \ell}{2})+g\sin(\frac{q \ell}{2})=0 \cap q > 0\right \} \;. \label{def_lp}
\end{align}
Note that $q$ and $-q$ correspond to the same state, hence the condition $q>0$. Equivalently the states can be labeled by the strictly positive integers $m \in \mathbb{N}^*$ (see Fig. (\ref{quantization})). 

Finally, both the odd and even eigenstates $\phi_{\pm,q}(x)$ are associated to the eigenenergy 
\be \label{energy}
E(q) = \frac{q^2}{2} \;.
\ee 

%\begin{figure}[h]
%\begin{center}
%	\includegraphics[width=8cm]{2stickyBMs.pdf}
%	\includegraphics[width=8cm]{5stickyBMs.pdf}
%\end{center}
%\caption{Left: Simulation of $2$ sticky Brownian motions $x_1(t)$ and $x_2(t)$ starting from $x=0$ at $t=0$ up to time $t=1$ (taken from \cite{barraquand2019large}). %Right: Simulation of $5$ sticky Brownian motions, note that for most of the time, at least two trajectories are stuck together. Both simulations were performed using a %discretization of $n$-tuples of sticky Brownian motions (here $n=2$ or $5$) as $n$ random walks in the same space-time iid random environment (see more details in %Section \ref{sec:convergencediscretetocontinuous} about this discretization).}
%\label{fig:stickyBM}
%\end{figure}

\subsection{Time dependent kernel}

$\quad$ \\

As discussed above, see Eq. (\ref{def_Kernel_t}), the time dependent kernel splits into two parts 
\be  \label{sumK} 
K(x,x';t) =  K_L(x,x';t)+K_R(x,x';t)
\ee
where each component evolves independently 
\be 
K_L(x,x';t) = \sum_{n=1}^{N_L} \psi_n^{L*}(x,t) \psi_n^L(x',t)  \quad , \quad K_R(x,x';t) = \sum_{n=1}^{N_R} \psi_n^{R*}(x,t) \psi_n^R(x',t) \;. 
\ee
Since the $\psi_n^{L/R}$'s evolve according to the Schr\"odinger equation with Hamiltonian $\hat H_g$ in Eq. (\ref{def_H}) these components can 
be rewritten using the real time Green's function 
\be \label{green}
G(x,y,t) =\langle {x}|e^{-i \hat H_g t}|y \rangle =\sum_{\sigma=\pm ,q \in \Lambda_{\sigma}}\phi_{\sigma,q}(x)\phi_{\sigma,q}^*(y)e^{-i E(q)t} \;,
\ee
where the eigenfunctions $\phi_{\sigma,q}(x)$ of $\hat H_g$ are given in
\eqref{def_lm} and \eqref{def_pm}. This leads to 
\bea 
&& K_L(x,x';t) = \sum_{n=1}^{N_L} \psi_n^{L*}(x,t) \psi_n^L(x',t)  = \int_{- \ell/2}^{0} dy dy'G^*(x,y,t)G(x',y',t)K_{L}(y,y') \;, \label{KL_start}\\
&& K_R(x,x';t) = \sum_{n=1}^{N_R} \psi_n^{R*}(x,t) \psi_n^R(x',t)  = \int_{0}^{\ell/2} dy dy'G^*(x,y,t)G(x',y',t)K_{R}(y,y') \;. \label{KR_start}
\eea 
The time evolution of the total kernel is thus obtained as the sum as in \eqref{sumK}. 
%\begin{align}
%K(x,x',t) %& %=\int_{- \ell/2}^{ \ell/2} dy dy'G^*(x,y,t)G(x',y',t)K_0(y,y')\nn\\
%&=\int_{0}^{ \ell/2} dy dy'G^*(x,y,t)G(x',y',t)K_{R}(y,y') +\int_{- \ell/2}^{0} dy dy'G^*(x,y,t)G(x',y',t)K_{L}(y,y')\nn\\
%&=K_R(x,x',t)+K_L(x,x',t)
%\end{align}

Let us first study $K_R(x,x';t)$ in Eq. (\ref{KR_start}). Inserting the decomposition \eqref{green} of the Green's function together with the explicit expression of $K_R(y,y')$ in Eq. (\ref{KR}) we obtain 
\begin{align}\label{KR_bis}
K_{R}(x,x',t)&=\int_{0}^{ \ell/2} dy dy'\sum_{\sigma_a = \pm,k_a\in \Lambda_{\sigma_a}}\phi_{\sigma_a,k_a}^*(x)\phi_{\sigma_a,k_a}(y)e^{iE(k_a)t}\nn\\
&\sum_{\sigma_b=\pm ,k_b\in\Lambda_{\sigma_b}}\phi_{\sigma_b,k_b}(x')\phi_{\sigma_b,k_b}^*(y')e^{-iE(k_b)t}\sum_{k\in\Lambda_{-},k\leq k_R}\frac{4}{ \ell}\sin(ky)\sin(ky')\nn\\
&=\sum_{\sigma_a =\pm,k_a\in \Lambda_{\sigma_a}}\sum_{\sigma_b=\pm,k_b\in \Lambda_{\sigma_b}}\sum_{k\in\Lambda_{-},k\leq k_R}\phi^*_{\sigma_a,k_a}(x)\phi_{\sigma_b,k_b}(x')e^{i(E(k_a)-E(k_b))t}\nn\\
&\int_{0}^{ \ell/2} dy \sqrt{\frac{4}{ \ell}}\phi_{\sigma_a,k_a}(y)\sin(ky)\int_{0}^{ \ell/2} dy'\sqrt{\frac{4}{ \ell}}\phi^*_{\sigma_b,k_b}(y')\sin(ky') \;,
%&=\sum_{\sigma_a = \pm,k_a\in \Lambda_{\sigma_a}}\sum_{\sigma_b=\pm,k_b\in \Lambda_{\sigma_b}}\sum_{k\in\Lambda_{-},k\leq k_R}\phi_{\sigma_a,k_a}(x)\phi_{\sigma_b,k_b}^*(x')e^{i(E(k_a)-E(k_b))t}\nn\\
%&\times \left(\frac{1}{\sqrt{2}}\delta_{\sigma_a,-}\delta_{k,k_{a}}+\delta_{\sigma_a,+}\frac{2^{3/2}}{ \ell}\frac{kk_{a}}{(k^2-k_{a}^2)\sqrt{g^2+k_{a}^2+\frac{2g}{ \ell}}}\right)\nn\\
%&\times \left(\frac{1}{\sqrt{2}}\delta_{\sigma_b,-}\delta_{k,k_{b}}+\delta_{\sigma_b,+}\frac{2^{3/2}}{ \ell}\frac{kk_{b}}{(k^2-k_{b}^2)\sqrt{g^2+k_{b}^2+\frac{2g}{ \ell}}}\right) \;.
\end{align}
where, in the third and fourth lines, we have just reorganized the discrete sums and the integrals over $y$ and $y'$. The same manipulations can
be performed for $K_L(x,x',t)$. The overlap integrals over $y$ and $y'$ can be performed explicitly, which gives 
%which can be performed explicitly, which gives
%\bea \label{KR}
%&& K_{R}(x,x',t) = \sum_{\sigma_a = \pm,k_a\in \Lambda_{\sigma_a}}\sum_{\sigma_b=\pm,k_b\in \Lambda_{\sigma_b}}\sum_{k\in\Lambda_{-},k\leq %k_R}\phi_{\sigma_a,k_a}(x)\phi_{\sigma_b,k_b}^*(x')e^{i(E(k_a)-E(k_b))t}\nn\\
%&&\times \left(\frac{1}{\sqrt{2}}\delta_{\sigma_a,-}\delta_{k,k_{a}}+\delta_{\sigma_a,+}\frac{2^{3/2}}{ \ell}\frac{kk_{a}}{(k^2-k_{a}^2)\sqrt{g^2+k_{a}^2+\frac{2g}{ %\ell}}}\right) \nn \\
%&&\times \left(\frac{1}{\sqrt{2}}\delta_{\sigma_b,-}\delta_{k,k_{b}}+\delta_{\sigma_b,+}\frac{2^{3/2}}{ \ell}\frac{kk_{b}}{(k^2-k_{b}^2)\sqrt{g^2+k_{b}^2+\frac{2g}{ %\ell}}}\right) \;.
%\eea 
%
\begin{align}\label{KernelLR}
K_{R/L}(x,x',t)&=\sum_{\sigma_a = \pm,k_a\in \Lambda_{\sigma_a}}\sum_{\sigma_b=\pm, k_b\in \Lambda_{\sigma_b}}\sum_{k\in\Lambda_{-}, k\leq {k_{R/L}}}\phi^*_{\sigma_a,k_a}(x)\phi_{\sigma_b,k_b}(x')e^{i(E(k_a)-E(k_b))t}\nn\\
&\times \left(\frac{1}{\sqrt{2}}\delta_{\sigma_a,-}\delta_{k,k_{a}}\pm\frac{2^{3/2}}{ \ell}\frac{kk_{a}}{(k^2-k_{a}^2)\sqrt{g^2+k_{a}^2+\frac{2g}{ \ell}}}\delta_{\sigma_a,+}\right)\nn\\
&\times \left(\frac{1}{\sqrt{2}}\delta_{\sigma_b,-}\delta_{k,k_{b}}\pm\frac{2^{3/2}}{ \ell}\frac{kk_{b}}{(k^2-k_{b}^2)\sqrt{g^2+k_{b}^2+\frac{2g}{ \ell}}}\delta_{\sigma_b,+}\right) \;,
\end{align}
where, in the last two factors in brackets, the $+$ sign refers to $K_R$ and the $-$ to $K_L$. To compute the full kernel we add the two halves and develop the product to get
\bea 
&&\hspace*{-2cm}K(x,x',t)=\underbrace{\left(\sum_{n=1}^{N_R}+\sum_{n=1}^{N_L}\right)\frac{1}{2}\frac{2}{\ell}\sin\left(\frac{2\pi n x }{\ell}\right)\sin\left(\frac{2\pi n x' }{\ell}\right)}_{A=A_R+A_L} \label{exact_K}\\
&&\hspace*{-2cm}+\left(\sum_{k\in\Lambda_{-},k\leq k_{R}}+\sum_{k\in\Lambda_{-},k\leq k_{L}}\right)\sum_{k_{a}\in\Lambda_{+},k_{b}\in\Lambda_{+}}2\left(\frac{2}{\ell}\right)^3\frac{k_{a}\cos(k_{a}x)+g\sin(k_{a}|x|)}{g^2+k_{a}^2+\frac{2g}{\ell}}\nn\\
&&\hspace*{-2cm}\times \underbrace{\frac{k_{b}\cos(k_{b}x')+g\sin(k_{b}|x'|)}{g^2+k_{b}^2+\frac{2g}{\ell}}\frac{k^2k_{a} k_{b}}{(k_{a}^2-k^2)(k_{b}^2-k^2)}e^{i(E(k_{a})-E(k_{b}))t}}_{B=B_R+B_L}\nn\\
&&\hspace*{-2cm}\underbrace{+\sum_{k\in\Lambda_{-},k_{L}<k\leq k_{R}}\sum_{k_{b}\in\Lambda_{+}}(\frac{2}{\ell})^2\sin(kx)\frac{k_{b}\cos(k_{b}x')+g\sin(k_{b}|x'|)}{g^2+k_{b}^2+\frac{2g}{\ell}}\frac{k k_{b}}{k^2-k_{b}^2}e^{i(E(k)-E(k_{b}))t}}_{C}\nn\\
&&\underbrace{\hspace*{-2cm}+\sum_{k\in\Lambda_{-},k_{L}<k\leq k_{R}}\sum_{k_{a}\in\Lambda_{+}}(\frac{2}{\ell})^2\sin(kx')\frac{k_{a}\cos(k_{a}x)+g\sin(k_{a}|x|)}{g^2+k_{a}^2+\frac{2g}{\ell}} \frac{k k_{a}}{k^2-k_{a}^2}e^{i(E(k_{a})-E(k))t}}_{D} \;, \nn 
\eea
 which is the sum of four terms denoted $A=A_R+A_L,B=B_R+B_L,C,D$ as indicated in the equation. These terms satisfy the symmetries: 
$A(x,x')$ is real and symmetric, $B(x,x')=B^*(x',x)$ and $D(x,x')=C^*(x',x)$. 

\vspace*{0.5cm}

{\bf Remarks:} \\

\vspace*{0.2cm}
\noindent (i) {\it Symmetries:} Since the initial kernel is unchanged under the simultaneous transformation $(k_L,k_R,x) \to (k_R,k_L,-x)$ and because of 
the invariance of $\hat H_g$ under parity transformation $x \to -x$, it follows that $K(x,x',t)$ is also unchanged under $(k_L,k_R,x) \to (k_R,k_L,-x)$.
The density $\rho(x,t)$ has thus the same invariance and the current satisfies $J_{k_L,k_R}(x,t) = - J_{k_R,k_L}(-x,t)$. This property is the 
finite time analog of the relations valid for the NESS stated in Eq. \ref{symmetry}.

\vspace*{0.3cm}

\noindent (ii) {\it Contribution to the current:} Since the term $A$ is real, it does not contribute to $J(x,t)$, i.e., $J_A(x,t)=0$ (see Eq. (\ref{Jdef})).
It is easy to see that the term $B$ gives only an odd contribution to the current, i.e. $J_B(x,t)=- J_B(-x,t)$. Hence the term $B$ cannot contribute
to $J_\infty$, the current in the NESS, which is uniform. It does however contribute to $\rho_\infty(x)$. The current in the NESS is thus
only determined by $C+D$ which gives an even contribution at all time $t$, i.e., $J_{C+D}(x,t)=J_{C+D}(-x,t)$.

\section{Thermodynamic limit $\ell \to +\infty$} \label{sec:thermo}

Our first goal is to obtain a formula for the kernel for the infinite system, i.e., $\lim_{\ell \to +\infty} K(x,x',t)$ 
for fixed $x,x',t$, which, for notational convenience, we will also denote by $K(x,x',t)$. Here we consider the case of a repulsive impurity $g>0$ -- the analysis of an attractive impurity $g<0$ is performed in Appendix \ref{App:gneg}. Let us recall that we take the limit $\ell\to\infty$ while fixing the left and right initial densities $\frac{2 N_{L/R}}{\ell}=\rho_{L/R}= \frac{k_{L/R}}{\pi}$. 
This is possible thanks to a contour integration trick
that we explain below. From the exact expression for the kernel in (\ref{exact_K}), one can write
\be \label{K_ABCD}
K(x,x',t) = A_L(x,x') + A_R(x,x') + C(x,x',t) + C^*(x',x,t) + B_L(x,x',t) + B_R(x,x',t) \;,
\ee 
where we have used the relation $D(x,x',t)=C^*(x',x,t)$. We now give the expressions of the limits when $\ell=+\infty$ of the terms $A_{L/R},B_{L/R}$ and $C$ separately. \\

\vspace*{0.5cm}

\noindent{\bf The term ${A(x,x')}$}: From \eqref{exact_K} we see that it is time independent and, in the large $\ell$ limit, it is given by the reflected sine-kernel (see e.g. \cite{us_hardwall_epl, us_hardwall_jstat})
\bea \label{ref_sine}
&& A_{L/R}(x,x')= \lim_{\ell\to\infty}\frac{1}{\ell}\sum_{n=1}^{N_{L/R}}\sin\left(\frac{2\pi n x }{\ell}\right)\sin\left(\frac{2\pi n x' }{\ell}\right) =\int_0^{k_{L/R}}\frac{dk}{2 \pi} \sin(kx)\sin(kx')\nn\\
&&=\frac{\rho_{L/R}}{4}\left(\frac{\sin(k_{L/R}(x-x'))}{k_{L/R}(x-x')} -\frac{\sin(k_{L/R}(x+x'))}{k_{L/R}(x+x')}\right) \;.
\eea

%\bea
%\lim_{\ell\to\infty}A_{L/R}(x,x')=\frac{\rho_{L/R}}{4}\left(\frac{\sin(k_{L/R}(x-x'))}{k_{L/R}(x-x')} -\frac{\sin(k_{L/R}(x+x'))}{k_{L/R}(x+x')}\right)
%\eea

\vspace*{0.5cm}

\noindent{\bf The term ${C(x,x',t)}$}: From \eqref{exact_K} the term $C$ is given by a double sum 
\be \label{C_expl}
C=C(x,x',t) = \frac{4}{\ell^2} \sum_{k_b \in \Lambda_+} \sum_{k \in \Lambda_-, k =k_L^+}^{k_R} \frac{h_{x,x',t}(k,k_b)}{k-k_b} 
\ee 
with $k_L= 2 \pi N_L/\ell$, $k_L^+= 2 \pi (N_L+1)/\ell$ and $k_R= 2 \pi N_R/\ell$ and we recall that the lattices $\Lambda_-$ and $\Lambda_+$ are defined in Eqs. (\ref{def_lm}) and (\ref{def_lp}) respectively. We have also defined the function 
\footnote{Since we are eventually interested in the large $\ell$ limit we omitted in \eqref{ff} the unimportant extra factor $2g/\ell $ in the denominator in
\eqref{exact_K}, which
should be restored to obtain finite $\ell$ formula.} 
\be \label{ff} 
h_{x,x',t}(k,k_b) = \sin(k x) \frac{ k_b \cos(k_b x') + g \sin(k_b |x'|)}{g^2 + k_b^2} \frac{k k_b}{k+k_b} e^{\frac{i}{2} (k^2 - k_b^2) t} \;.
\ee 
\begin{figure}[t]
\centering
\includegraphics[width = 0.4\linewidth]{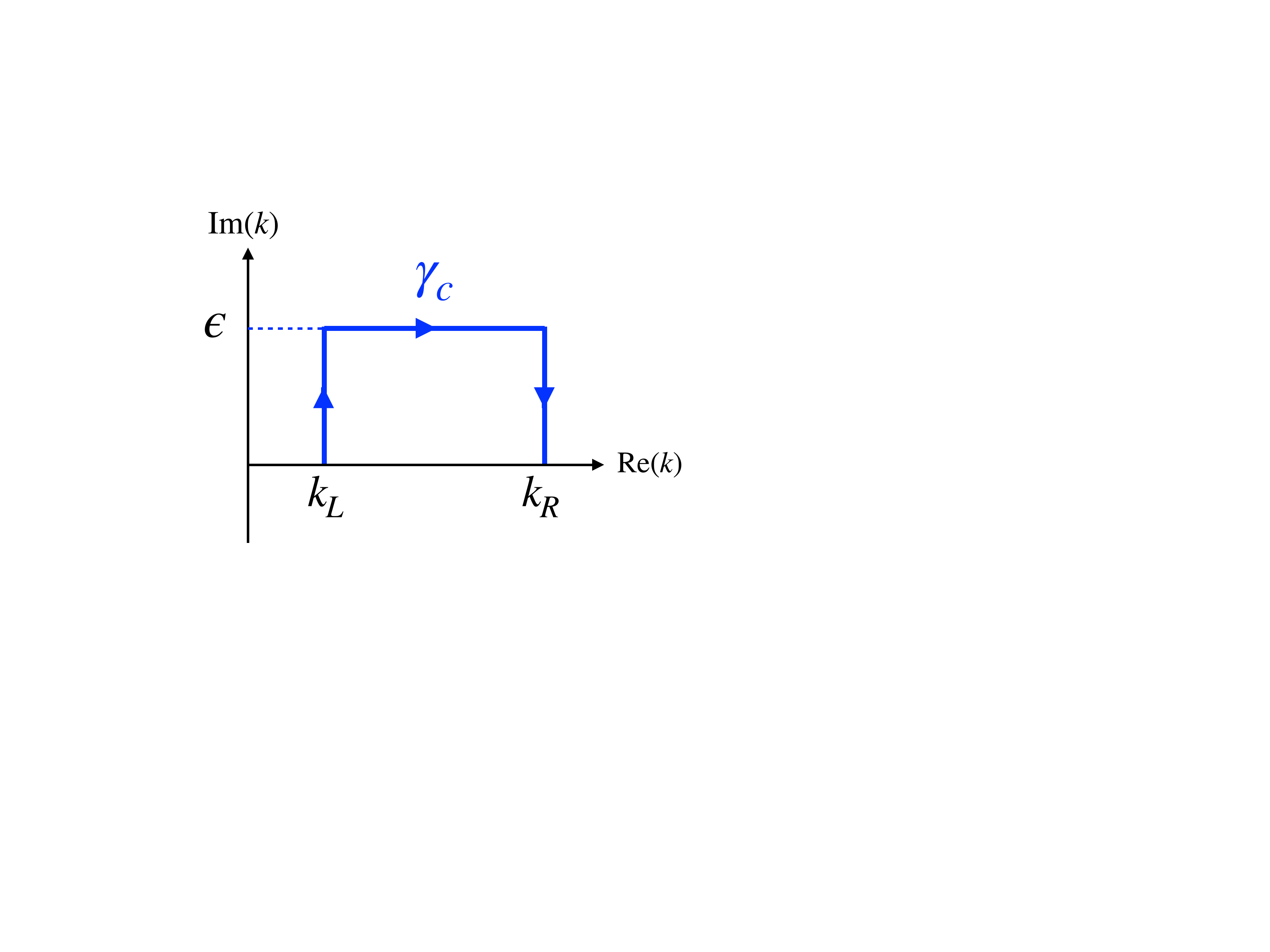}
\caption{Illustration of the contour $\gamma_c$ over which the integral over $k$ in the first line of (\ref{res33}) is performed.}\label{Fig_contour1}
\end{figure}
Taking the limit $\ell\to\infty$ of \eqref{C_expl} to obtain a double integral is very delicate due to the presence 
in the sum of a pole $\frac{1}{k-k_b}$ and the fact that the two lattices $\Lambda_-$ and $\Lambda_+$ are intertwined (see Fig.  \ref{quantization}). 
Let us first state the result and give its derivation below. One finds
\bea \label{res33} 
&& \lim_{\ell\to\infty}  C(x,x',t)  = \int_0^{+\infty} \frac{dk_b}{\pi} \bigg[ 
\left( \int_{k_L}^{k_L+i \epsilon} \frac{dk}{\pi} + \int_{k_L+i \epsilon}^{k_R+i \epsilon}\frac{dk}{\pi} - \int_{k_R}^{k_R+i \epsilon}\frac{dk}{\pi}
\right) \frac{h_{x,x',t}(k,k_b)}{k-k_b}  \bigg] \nonumber \\
&& \hspace*{2.5cm}  + \int_{k_L}^{k_R} \frac{dk_b}{\pi} \left( i + \frac{g}{k_b} \right) h_{x,x',t}(k_b,k_b) \;.
\eea 
The first line of \eqref{res33} is an integral over a contour that we denote $\gamma_c$ consisting in straight lines in the complex $k$-plane, forming a half-rectangle as represented in Fig. \ref{Fig_contour1}. Its value does not depend on the parameter $\epsilon>0$ since 
$h_{x,x',t}(k,k_b)$ as a function of $k$ does not have poles in the strip $[k_L,k_R] + i \mathbb{R}^+$. 
In addition to the derivation given below we have carefully checked numerically this formula (\ref{res33}) for a variety of function $h$
which share the same properties.

\begin{figure}[t]
\centering
\includegraphics[width = 0.7\linewidth]{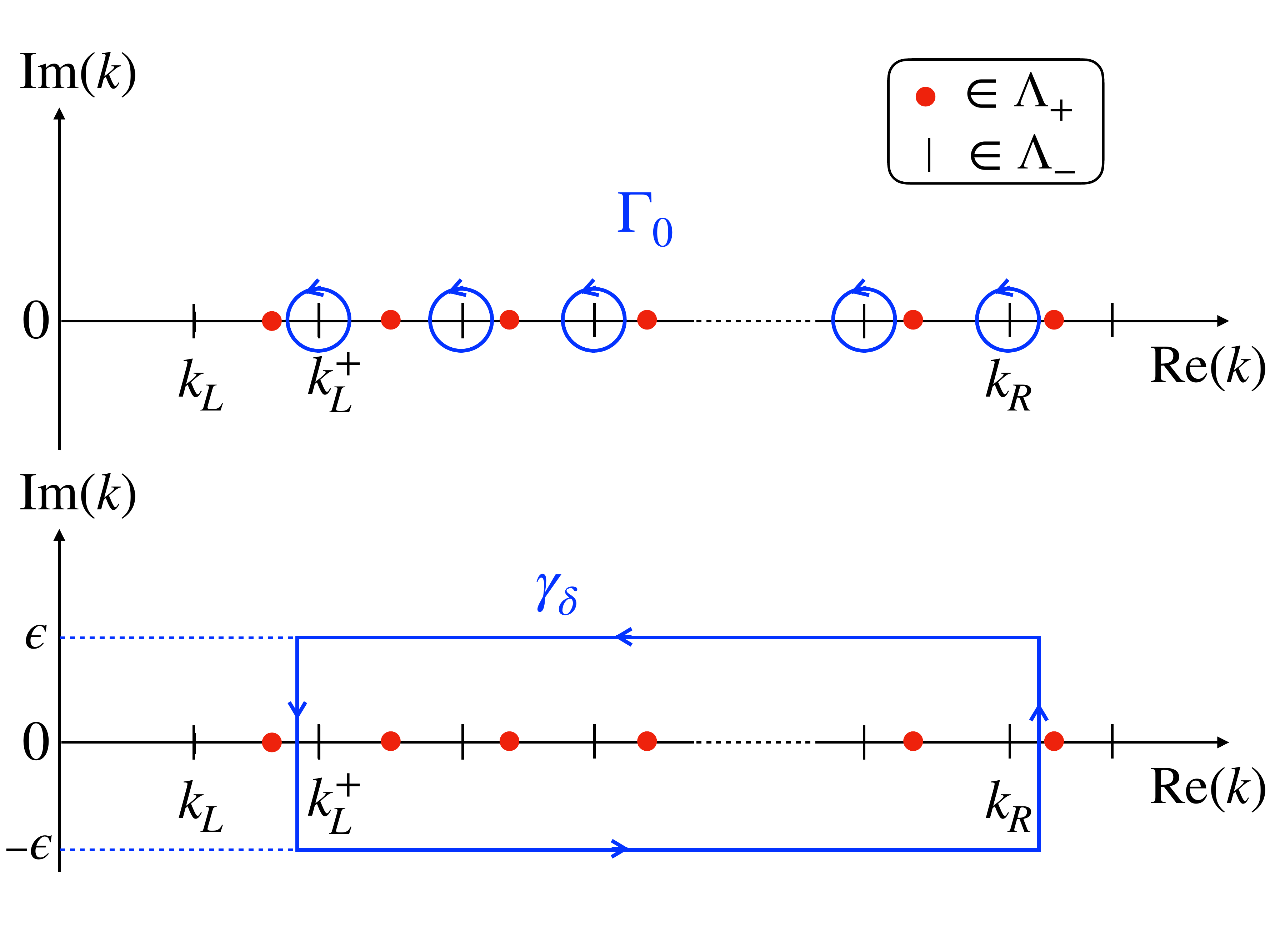}
\caption{Illustration of the contours $\Gamma_0$ (upper panel) and $\gamma_\delta$ (lower panel) used in Eqs. (\ref{C_contour_g0}) and (\ref{Cexact}) respectively.}\label{Fig_contour_exact}
\end{figure}

Let us now give the derivation of this result in Eq.~(\ref{res33}). The idea is inspired from \cite{Prosen2018}, although some details are different here. The trick is to replace the discrete sum over $k$ in \eqref{C_expl} by a contour integral as follows
\be \label{C_contour_g0} 
C(x,x',t) = \frac{4}{\ell^2} \sum_{k_b \in \Lambda_+} \int_{\Gamma_0} \frac{dk}{2 \pi} \frac{\ell}{e^{i \ell k} -1 } \frac{h_{x,x',t}(k,k_b)}{k-k_b} \;,
\ee 
where the contour $\Gamma_0$ is a union of very small circles centered around the points $k=\frac{2 \pi n}{\ell}$ with $N_L+1 \leq n \leq N_R$
and oriented counterclockwise (see Fig. \ref{Fig_contour_exact}). The circles should be small enough so that the contour does not enclose any point $k=k_b \in \Lambda_+$. 
We now deform the contour $\Gamma_0$ into the closed counterclockwise contour $\gamma_\delta$ which is the rectangle with the four corners $k_L^+- \frac{2 \pi \delta}{\ell} -i \epsilon$, $k_L^+- \frac{2 \pi \delta}{\ell}+i \epsilon$,
$k_R + \frac{2 \pi \delta}{\ell} + i \epsilon$, $k_R+ \frac{2 \pi \delta}{\ell} - i \epsilon$, represented in Fig. \ref{Fig_contour_exact}.
The parameter $0<\delta<1$ is chosen small enough so that the contour does not contain any point $k_b$ of $\Lambda_+$ located to
the left of $k_L^+$ and to the right of $k_R$. During this deformation one encounters only the poles at $k= k_b \in \Lambda_+ \cap ]k_L^+,k_R[$.
Taking into account the residues associated to these poles one obtains
\be \label{Cexact} 
C(x,x',t) = \frac{4}{\ell^2} \bigg( \sum_{k_b \in \Lambda_+}  \oint_{\gamma_\delta} \frac{dk}{2 \pi} \frac{\ell}{e^{i \ell k} -1 } \frac{h_{x,x',t}(k,k_b)}{k-k_b}  
- 2 \pi i \sum_{k_b \in \Lambda_+ \cap ]k_L^+,k_R[} \frac{\ell}{2\pi} \frac{1}{e^{i \ell k_b }-1} h_{x,x',t}(k_b,k_b) \bigg) \;.
\ee 

Until now this is an exact rewriting of $C(x,x',t)$ in Eq. (\ref{C_expl}) valid for any $\ell$. For any $k_b \in \Lambda_+$, using the second relation in
\eqref{quantification0}, one can evaluate the factor
\be\label{Theta}
\frac{1}{e^{i \ell k_{b}}-1} 
%=-\frac{1}{2}-\frac{i}{2}\cot(\pi k_{b}/\delta k)\nn\\
%&=-\frac{1}{2}-\frac{i}{2}\cot(\frac{k_{b}\ell}{2})\nn\\
=-\frac{1}{2}+\frac{i}{2}\frac{g}{k_{b}} \;.
\ee
We now take the large $\ell$ limit on Eq. \eqref{Cexact}. The factor $\frac{1}{e^{i \ell k} -1 } \to 0$ for ${\rm Im} \, k<0$
and $\frac{1}{e^{i \ell k} -1 } \to -1$ for ${\rm Im} \,k>0$. Hence the $k$ integral in \eqref{Cexact} over the counterclockwise closed contour $\gamma_\delta$
becomes the $k$ integral over the clockwise half-rectangle in \eqref{res33}
\footnote{Proving the convergence for the vertical parts of the contour $\gamma_c$ is non-trivial, which is why we did careful numerical checks
of \eqref{res33}. This problem does not occur in \cite{Prosen2018,CEF12}.}
In addition the sum over $k_b$ becomes an integral in
the large $\ell$ limit. This leads to the result \eqref{res33}.

\vspace*{0.5cm}

\noindent{\bf The term ${B(x,x',t)}$}: We now give the large $\ell$ limit of $B_L(x,x')$ and $B_R(x,x')$ defined in \eqref{exact_K}. 
Each term can be decomposed as the sum of two terms 
\be \label{Bsum} 
B_{R/L}(x,x',t) = B_{R/L}^{\text{off-diag}}(x,x',t) + B_{R/L}^{\text{diag}}(x,x')
\ee 
where the first term comes from the terms $k_a \neq k_b$ in the triple sum in \eqref{exact_K},
while the second one comes from the terms $k_a=k_b$ and does not depend on time. 
As detailed in the Appendix \ref{App_largel}, we obtain the first term in \eqref{Bsum} as
\bea \label{BR_offdiag_large}
&& B_R^{\text{off-diag}}(x,x',t)=  \int_0^{\infty}\int_0^{\infty}\frac{dk_a}{\pi}\frac{dk_b}{\pi} F_{x,x'}(k_a,k_b) e^{\frac{1}{2} i(k_{a}^2 -k_{b}^2)t} \\
&&\times \left[\frac{1}{2 \pi}\left(\frac{k_b \log(\frac{k_b+k_R}{|k_b-k_R|})-k_a \log(\frac{k_a+k_R}{|k_a-k_R|})
}{k_a^2-k_b^2})+  \frac{g}{2(k_a^2-k_b^2)}\left(\Theta(k_R-k_a) - \Theta(k_R-k_b)\right)\right) \right]  \;, \nn 
\eea
where we have defined the function
\be   \label{def_Fkakb}
F_{x,x'}(k_a,k_b)= 2 \frac{k_a\cos(k_ax)+ g\sin(k_a|x|)}{g^2+k_a^2} \frac{k_b\cos(k_bx')+ g\sin(k_b|x'|)}{g^2+k_b^2}k_a k_b \;.
\ee 
The second term in \eqref{Bsum} is obtained in the large $\ell$ limit as
\bea  \label{BR_diag_large}
B_R^{\rm diag}(x,x') = \int_0^{k_R} \frac{dk_a}{2 \pi} F_{x,x'}(k_a,k_a) \frac{(g^2 + k_a^2)}{2 k_a^2} \;.
\eea 
The terms $B_L^{\text{off-diag}}(x,x',t)$ and $B_L^{\rm diag}(x,x')$ are simply obtained from
$B_R^{\text{off-diag}}(x,x',t)$ and $B_R^{\rm diag}(x,x')$ by substituting $k_R \to k_L$ in Eqs. (\ref{BR_offdiag_large}) and (\ref{BR_diag_large}).

\section{Large time limit: stationary state at fixed position $x$} \label{sec:largetime}

In the previous section we have obtained the expression of the kernel in the thermodynamic limit $\lim_{\ell \to +\infty} K(x,x',t)$
in \eqref{K_ABCD} together with \eqref{Bsum}, as a sum of several terms. Now we can take the limit $t \to +\infty$ of each term in 
this expression for fixed positions $x$ and $x'$ to obtain
\be \label{K_infty}
K_\infty(x,x') = \lim_{t \to +\infty} \lim_{\ell \to +\infty}  K(x,x',t) \;.
\ee 
In \eqref{K_ABCD} and \eqref{Bsum} the terms $A_{L/R}(x,x')$ and $B_{R/L}^{\rm diag}(x,x')$ are independent of time.
As shown in Appendix \ref{App_NESS}, the terms $B_{L/R}^{\rm off-diag}(x,x',t)$ decay to zero at large time as power laws in time.
Finally $C(x,x',t)$ goes to a finite limit $C_\infty(x,x')$ where only the last term in \eqref{res33} (coming from the residues) survives at large time,
as discussed in Appendix \ref{App_NESS} where the time decay is studied. From the last term in \eqref{res33} we obtain
\begin{align} \label{resCinfty} 
C_\infty(x,x')=\frac{1}{2}\int_{k_L}^{k_R}\frac{dk}{\pi}(g+ik)\sin(kx)\frac{k\cos(kx')+ g\sin(k |x'|)}{g^2+k^2} \;.
\end{align}

As a result the kernel at infinite time, i.e., in the NESS, is obtained as the sum
\be  \label{recap_Kinf}
K_\infty(x,x') = A_R(x,x') + A_L(x,x') + B_R^{\rm diag}(x,x') + B_L^{\rm diag}(x,x') + E(x,x')  \;,
\ee 
where the terms $A_{L/R}(x,x')$ are given in \eqref{ref_sine}, 
\be 
E(x,x') = C_\infty(x,x') + C_\infty(x',x)^* \;,
\ee 
where $C_\infty(x,x')$ is given in \eqref{resCinfty}, 
and we recall 
\begin{align}
B^{\rm diag}_{R/L}(x,x')&= \int_{0}^{k_{R/L}} \frac{dk}{2 \pi} \frac{k\cos(kx)+ g\sin(k|x|)}{g^2+k^2 } (k\cos(kx')+ g\sin(k |x'|)) \;. 
%2\int\frac{dk}{\pi^2}k^2\frac{k\cos(kx)+g\sin(k|x|)}{g^2+k^2}\frac{k\cos(kx')+g\sin(k|x'|)}{g^2+k^2}\nn\\
%&\left[\frac{1}{\pi}\frac{k k_R+(k_R^2-k^2) arctanh(\frac{k_R}{k})}{2k(k^2-k_R^2)}+-2i(-\frac{1}{2}+\frac{i}{2}\frac{g}{k})\frac{1}{4k}\Theta(k_R-k)\right]
\end{align}

\vspace*{0.5cm}

\noindent{\bf Stationary density.} The stationary density can be obtained from the kernel using the relation $\rho_\infty(x) = K_\infty(x,x)$. 
It can be written as the sum $\rho_\infty(x) =  \rho_A(x) + \rho_B(x) + \rho_E(x)$ of the contributions
of the terms in \eqref{recap_Kinf} with
\begin{align}\label{n-separated}
\rho_A(x)&=\frac{\rho_R}{4}\left(1-\frac{\sin(2 k_R x)}{2 k_R x}\right)+\frac{\rho_L}{4}\left(1-\frac{\sin(2 k_L x)}{2 k_L x}\right)\\
\rho_B(x) %&=(\int_0^{k_R}+\int_0^{k_L})\frac{dk}{4\pi}\frac{(g^2+k^2)+(k^2-g^2)\cos(2 k x)+2 gk \sin(2k|x|)}{g^2+k^2}\nn\\
&=\frac{\rho_R}{4}+\frac{\rho_L}{4}+\left(\int_0^{k_R}+\int_0^{k_L}\right)\frac{dk}{4\pi}\frac{(k^2-g^2)\cos(2 k x)+2 gk \sin(2k|x|)}{g^2+k^2}\\
\rho_E(x)&=g\int_{k_L}^{k_R}\frac{dk}{\pi}\sin(kx)\frac{k \cos(kx)+ g\sin(k|x|)}{g^2+k^2}\nn\\
&=\int_{k_L}^{k_R}\frac{dk}{2\pi}\frac{g}{g^2+k^2}(k\sin(2 k x)+{\rm sgn}(x)g(1-\cos(2 k x))) \;.
\end{align}
%The final result is
%\bea 
%&&\rho_\infty(x)=\frac{\rho_R}{2}(1-\frac{\sin(2 k_R x)}{4 k_R x})+\frac{\rho_L}{2}(1-\frac{\sin(2 k_L x)}{4 k_L x})\nn\\
%&&+(\int_0^{k_R}+\int_0^{k_L})\frac{dk}{4\pi}\frac{(k^2-g^2)\cos(2 k x)+2 gk\sin(2k|x|)}{g^2+k^2}\nn\\
%&&+\int_{k_L}^{k_R}\frac{dk}{2\pi}\frac{g}{g^2+k^2}(k\sin(2 k x)+g \, {\rm sgn}(x)(1-\cos(2 k x))) 
%\eea
%Note that the last term is odd in $x$. 
Let us rewrite $\rho_A(x)$ using the identity
\be 
\frac{\rho_R}{4} \frac{\sin(2 k_R x)}{2 k_R x} = \frac{\sin(2 k_R x)}{8 \pi x}
= \int_0^{k_R} \frac{dk}{4 \pi}  \cos(2 k x)  \;.
\ee 
Summing the different contributions to $\rho_\infty(x)$, one obtains
\bea 
&&\rho_\infty(x)=\frac{k_R+ k_L}{2 \pi} + {\rm sgn}(x) \int_{k_L}^{k_R}\frac{dk}{2\pi}\frac{g^2}{g^2+k^2} \\
&&+\left(\int_0^{k_R}+\int_0^{k_L}\right)\frac{dk}{4\pi}\frac{(- 2 g^2)\cos(2 k x)+2 g k\sin(2k|x|)}{g^2+k^2}\nn\\
&&+\int_{k_L}^{k_R}\frac{dk}{2\pi}\frac{g}{g^2+k^2}(k\sin(2 k x)- g \, {\rm sgn}(x)\cos(2 k x)) \;. 
\eea 
In that expression the constant parts can be rewritten for $x>0$ and $x<0$ respectively as
\be 
\frac{k_R+k_L}{2 \pi} + \int_{k_L}^{k_R} \frac{dk}{2 \pi} \frac{g^2}{g^2 + k^2} = \frac{k_R}{\pi} - \int_{k_L}^{k_R} \frac{dk}{2 \pi} \frac{k^2}{g^2 + k^2} \quad, \quad (x>0)
\ee 
\be 
\frac{k_R+k_L}{2 \pi} - \int_{k_L}^{k_R} \frac{dk}{2 \pi} \frac{g^2}{g^2 + k^2} = \frac{k_L}{\pi} - \int_{k_R}^{k_L} \frac{dk}{2 \pi} \frac{k^2}{g^2 + k^2} \quad, \quad (x<0) \;,
\ee 
as in Eq. (\ref{densitysteadystate>0}). Next, regrouping the terms proportional to $\cos(2 k x)$ and $\sin (2 k x)$ in the integrals one can
check that it is identical to the expression given in Eq.~(\ref{densitysteadystate>0}).

\noindent{\bf Stationary current.} For the current, using \eqref{Jdef} one obtains in the large time limit
\begin{align}
J_A(x)&=0\\
J_B(x)&=0\\
J_E(x)&= - \int_{k_L}^{k_R} \frac{dk}{2 \pi} \frac{k^3}{g^2 + k^2} = \frac{1}{4\pi}(k_L^2-k_R^2 +\frac{g^2}{2}\ln(\frac{g^2+k_R^2}{g^2+k_L^2}))
\end{align}
where the first two contributions vanish since $A$ and $B$ are both real, and the third simplifies. The total result for the current in the NESS regime can
thus be written as 
\begin{equation}
J_{\infty}(x)= J_\infty =  \frac{1}{4\pi}\left(k_L^2-k_R^2 +\frac{g^2}{2}\ln\left(\frac{g^2+k_R^2}{g^2+k_L^2}\right)\right)
\end{equation}
Using $\mu_{R/L}= \frac{1}{2} k_{R/L}^2$ one can rewrite this result as in \eqref{current_ness} in terms of the Fermi energies on both sides.
\\

\noindent{\bf Stationary kernel.} Putting all terms together in \eqref{recap_Kinf} 
and using standard trigonometric relations we find for $x, x'>0$
the result for $K_\infty(x>0,x'>0)$ given in Eq. \eqref{Kpospos_intro}. 
%\bea  
%&&  K_\infty(x>0,x'>0) = \int_{0}^{k_R}\frac{dk}{\pi} \cos \left(k(x-x')\right) - 
%\int_{k_L}^{k_R}\frac{dk}{2 \pi} \frac{k^2}{k^2 + g^2} \cos (k(x-x')) \\
%&+ & \int_{0}^{k_R}\frac{dk}{\pi} \frac{ g k \sin \left(k(x+x')\right) - g^2 \cos\left(k(x+x')\right) }{k^2 + g^2} + i \int_{k_L}^{k_R} \frac{dk}{2 \pi} %\frac{k^2}{g^2+k^2}\, \sin{(k(x-x'))} \nonumber 
%\eea 
Note that the first term in 
\eqref{Kpospos_intro} is the sine-kernel associated to the right side of the system, namely ${\sin(k_R(x-x'))}/{(\pi (x-x'))}$.
One can rewrite this result as (we recall that here $g>0$) 
\bea \label{Kernel_NESS}
K_\infty(x>0, x'>0)&=&\int_0^{k_R}\frac{dk}{\pi}\cos(k(x-x'))-\int_{k_L}^{k_R}\frac{dk}{2\pi}\frac{k^2}{g^2+k^2}e^{-ik(x-x')} \nn \\
%(-\cos(k(x-x'))+i\sin(k(x-x')))\nn\\
&+&\frac{g}{\pi}e^{g (|x|+|x'|)}{\rm Im} \, E_1((g+ik_R)(|x|+|x'|)) \;.
\eea 

Similarly, for $x>0$ and $x'<0$ one finds the result for $K_\infty(x>0,x'<0)$ given in Eq. \eqref{Kposneg_intro}. 
%\bea \label{Kposneg} 
%&& K_\infty(x>0,x'<0) = ( \int_{0}^{k_R} + \int_{0}^{k_L}  ) \frac{dk}{2 \pi} \frac{k^2}{k^2 + g^2} \cos (k(x-x')) 
%\\
%&& + \int_{k_L}^{k_R} \frac{dk}{2 \pi} \frac{k g}{k^2 + g^2} \sin (k(x+x')) + 
%( \int_{0}^{k_R} + \int_{0}^{k_L}  ) \frac{dk}{2 \pi} \frac{k g}{k^2 + g^2} \sin (k(x-x')) \nonumber 
%\\
%&& + i \int_{k_L}^{k_R} \frac{dk}{2 \pi} \frac{k}{k^2 + g^2} (k \sin k(x-x') - g \cos k(x-x') + g \cos k(x+x')) \nonumber 
%\eea
Once again, it is equivalent to the following expression
\bea  
\hspace*{-0.5cm} &&K_\infty(x>0,x'<0)= \frac{g}{2\pi}e^{g (|x|+|x'|)}
({\rm Im} E_1((g+ik_R)(|x|+|x'|)) + {\rm Im} E_1((g+ik_L)(|x|+|x'|))) \nn \\
\hspace*{-0.5cm} && +\left(\int_0^{k_R}+\int_0^{k_L}\right)\frac{dk}{2\pi}\cos(k(x-x')) \nn \\
&& + \int_{k_L}^{k_R}\frac{dk}{2\pi}\frac{k}{g^2+k^2}(g\sin k (x+x')+i(k\sin k (x-x')-2 g\sin(k x)\sin(k x'))) \;. \label{Kernel_NESS2} 
\eea
\\

{\it Large distance limits}. Interestingly this stationary kernel has several non-trivial limits far from the defect. Indeed consider first the expression \eqref{Kpospos_intro}
for
$x,x' \to +\infty$ with $x-x'$
fixed. Let us use the following property. For any smooth function with bounded $f''(k)$, as $u \to +\infty$ one has
\bea  \label{IPP}
&& \int_0^{k_R} dk f(k) \sin(k u) = \frac{f(0)-f(k_R) \cos k_R u}{u} + o\left(\frac{1}{u}\right) \\
&& \int_0^{k_R} dk f(k) \cos(k u) = \frac{f(k_R) \sin k_R u}{u} + o\left(\frac{1}{u}\right)  \;.
\eea 
This can be shown by performing two successive integrations by parts with respect to $k$. 
This shows that the last term in \eqref{Kpospos_intro}
\bea 
&& \int_{0}^{k_R}\frac{dk}{\pi} \frac{ g k \sin \left(k(x+x')\right) - g^2 \cos\left(k(x+x')\right) }{k^2 + g^2} \\
&& \simeq \frac{-1}{\pi (x+x')} \frac{g}{k_R^2 + g^2} (k_R \cos(k_R (x+x')) + g \sin(k_R (x+x'))) \;, \nn
\eea
decays to zero at large distance. The asymptotic behavior is thus given by \eqref{Ksc10}.

Another interesting limit is $x \to +\infty$, $x' \to - \infty$ with $x+x'=O(1)$ fixed. 
In that case, by a similar calculation as above, the terms depending on $x-x'$ in
\eqref{Kposneg_intro}
decay to zero leading to the asymptotics given in \eqref{Kposneg20}.
%\bea \label{Kposneg2} 
%K_\infty(x,x') \to \int_{k_L}^{k_R} \frac{dk}{2 \pi} \frac{k g}{k^2 + g^2} ( \sin (k(x+x')) + i \cos k(x+x') ) 
%\eea

\section{Large time limit with $\xi=x/t, \xi'=x'/t$ fixed} \label{sec:light}

In this section we study the large time limit of the kernel in the regimes of rays, i.e. $x,x',t \to \infty$
with $\xi=x/t$ and $\xi'=x'/t$ fixed and $O(1)$. We start from the expression \eqref{K_ABCD} for the kernel $K(x,x',t)$ 
in the $\ell=+\infty$ limit.
It is a sum of terms of type $A,C,B$ which are given in Eq. \eqref{ref_sine} for $A$,  
in Eq. \eqref{res33} for $C$ and in Eqs. \eqref{Bsum}, 
\eqref{BR_offdiag_large} and
\eqref{BR_diag_large} for $B$. In these expressions, we will set $x=\xi t + y$ and $x'=\xi t + y'$ and study the large $t$ limit at fixed $\xi,\xi' = O(1)$
and fixed $y,y'=O(1)$. As we show below the terms $A_{L/R}$ and $B_{L/R}^{\rm diag}$ are simple to analyze.
The study of the term $C$ requires a modification of the contour integral trick used for the NESS, as pointed 
out in \cite{Prosen2018}. We now examine
these terms independently. \\
\noindent {\bf The term ${A(x,x')}$}. From \eqref{ref_sine} this term decays to zero unless $\xi=\pm \xi'$. In the first case one finds
\bea \label{ref_sine2}
&& A_{L/R}(\xi t + y,\xi t + y') \simeq \frac{\rho_{L/R}}{4} \frac{\sin(k_{L/R}(y-y') )}{k_{L/R}(y - y') }  \;,
\eea
while in the second case one finds 
\bea \label{ref_sine2}
&& A_{L/R}(\xi t + y,- \xi t + y') \simeq - \frac{\rho_{L/R}}{4} \frac{\sin(k_{L/R}(y+y') )}{k_{L/R}(y + y') }  \;.
\eea

\vspace*{0.5cm}

\begin{figure}[t]
\centering
\includegraphics[width=0.9\linewidth]{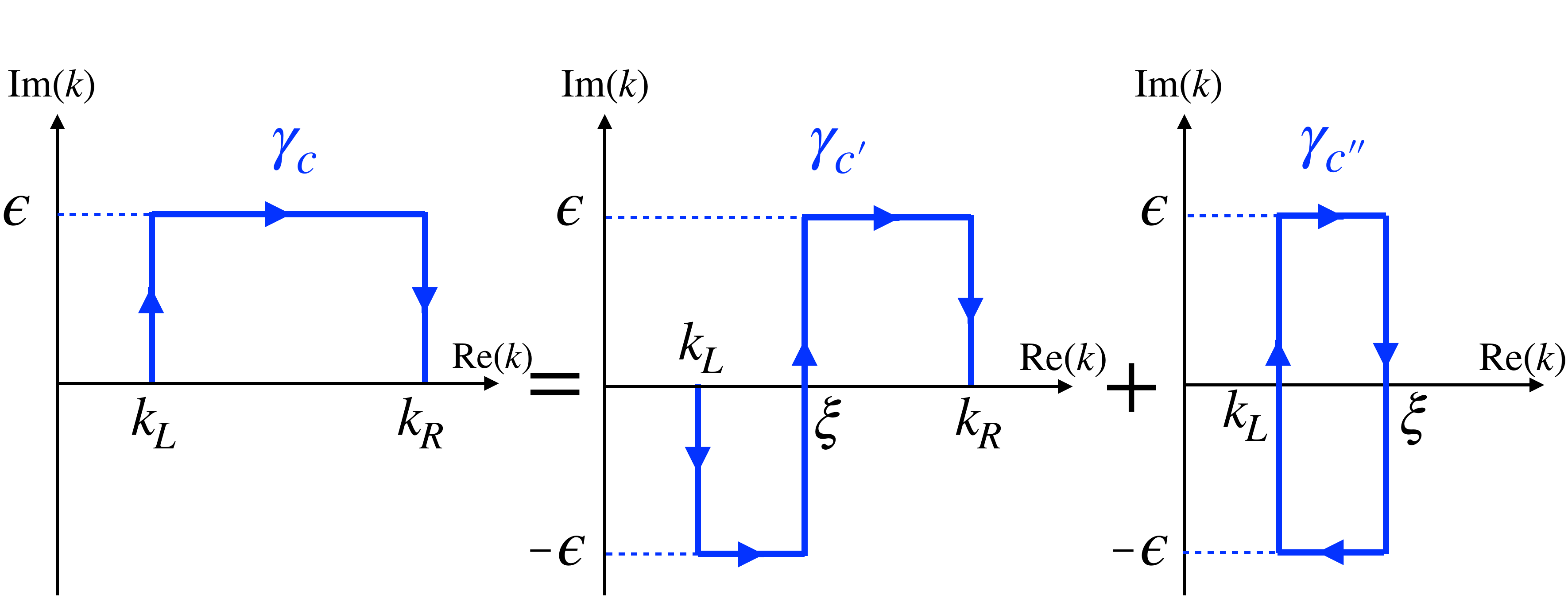}
\caption{Illustration of the contour $\gamma_c$, $\gamma_{c'}$ and $\gamma_{c''}$.
To be precise the contour $\gamma_{c'}$ includes the vertical line at $\xi+ \frac{y}{t}$ (instead of $\xi$ as indicated in the figure) but this difference is irrelevant in
the large $t$ limit.}\label{Fig_gammac}
\end{figure}

\noindent {\bf The term ${C(x,x',t)}$}. Inserting $x=\xi t + y$ and $x'=\xi t + y'$ in \eqref{res33} we will treat separately the contour integral
in the first line and the residue part in the second line of that equation. In the contour integral over $k$ in \eqref{res33} one must be careful with the factor
which comes from the factor containing $\sin(k x)$ in the function $h_{x,x',t}(k,k_b)$ in \eqref{ff}, and which reads schematically
\be \label{sinus} 
\frac{e^{i k (\xi t + y)} - e^{- i k (\xi t + y)}  }{2 i( k-k_b) } e^{\frac{i}{2} k^2 t} \;.
\ee 
Previously, for $\xi=0$, this term was decaying to zero at large $t$ since ${\rm Im} (k) >0$ along the contour. 
For $\xi \neq 0$ let us rewrite \eqref{sinus} setting $k = \bar k + i q$, as 
\be \label{sinus2} 
\frac{e^{\frac{i}{2}(\bar k^2-q^2)t}}{2 i( k-k_b) } \left( e^{- q ((\xi + \bar k) t + y) } e^{i \bar k (\xi t + y)} -  e^{q ((\xi - \bar k) t + y)}
e^{- i \bar k (\xi t + y)}    \right) \;.
\ee 
Let us recall that on the contour in \eqref{res33} $q={\rm Im}(k) >0$ and $\bar k \in [k_L,k_R]$. Consider first $\xi>0$. 
In that case the first term in \eqref{sinus2} can be discarded at large time, and the second term diverges at large time
if $\bar k = {\rm Re} k < \xi $. Since $\bar k \geq k_L$, if $\xi< k_L$ we can still discard the contribution
of the contour integral at large time. If $\xi > k_L$ 
we need to deform the contour to be able to handle the large time 
limit of \eqref{res33}. Consider first the case $k_L < \xi < k_R$. Denoting $\gamma_c$ the original contour, one writes
$\int_{\gamma_c} = \int_{\gamma_{c'}}+ \oint_{\gamma_{c''}}$ where the contours are represented in Fig. \ref{Fig_gammac}. 
One has
\be \label{sumcontour} 
\int_{\gamma_{c'}} dk = - \int_{k_L- i \epsilon}^{k_L} dk +
\int_{k_L- i \epsilon}^{\xi+ \frac{y}{t} - i \epsilon} dk +
\int_{\xi+ \frac{y}{t} - i \epsilon}^{\xi+ \frac{y}{t} + i \epsilon} dk
+
\int^{k_R+ i \epsilon}_{\xi+ \frac{y}{t} + i \epsilon} dk 
- \int^{k_R + i \epsilon}_{k_R} dk
\ee 
and, as we argue below, the contribution of $\gamma_{c'}$ vanishes at large time. On the other hand, 
since $\gamma_{c''}$ is a closed contour which surrounds the interval $[k_L,\xi+ \frac{y}{t}]$, its total contribution
to $C(\xi t+y , \xi' t+ y',t)$ 
is given by a sum of residues at $k=k_b$ and reads (it involves only the second term in \eqref{sinus} since the
first term can be discarded for any $\xi>0$) 
\be \label{118} 
\int_0^{+\infty} \frac{dk_b}{\pi}  \oint_{\gamma_{c''}}\frac{dk}{\pi} \frac{h_{\xi t+ y,\xi' t+y,t}(k,k_b)}{k-k_b}   =
- 2 i \int_{k_L}^{\xi+ \frac{y}{t}} \frac{dk_b}{\pi} \frac{-1}{2 i} e^{- i k_b (\xi t + y)} \tilde h_{\xi' t + y',t}(k_b,k_b) \;,
\ee 
where 
\be \label{ffnew} 
\tilde h_{x',t}(k,k_b) = \frac{ k_b \cos(k_b x') + g \sin(k_b |x'|)}{g^2 + k_b^2} \frac{k k_b}{k+k_b} e^{\frac{i}{2} (k^2 - k_b^2) t} 
\ee 
It thus adds to the residue term on the second line of \eqref{res33}, noting however 
that in \eqref{res33} $h_{x,x',t}(k,k_b)= \sin(k x) \tilde h_{x',t}(k,k_b)$.

Let us now discuss the contribution of the contour $\gamma_{c'}$ in Fig. \eqref{Fig_gammac} denoting $k=\bar k + i q$.
The general idea is that this new contour has been chosen such that its contribution decays to zero at large time. 
For the first and last term in \eqref{sumcontour}, it is clear that the large time limit vanishes since $q (\xi - \bar k)<0$
in both cases on the new contour. It is convenient to take again the limit $\epsilon \to +\infty$ which makes
the contributions of the horizontal parts (second and fourth terms in \eqref{sumcontour}) vanish. Let us analyze the contribution to
$C$ of the remaining integral (the third integral in the right hand side of Eq. (\ref{sumcontour}))
along the vertical axis at $\bar k=\xi+ \frac{y}{t}$,
setting $y=y'=0$ the conclusion remains unchanged for the general case. Writing $q=\xi + i q$, it reads (retaining only the second term in Eq. (\ref{sinus2})) 
%\be 
%\int_0^{+\infty} \frac{dk_b}{\pi} \int_{\xi  - i \infty}^{\xi+   i \infty} \frac{dk}{\pi} 
%\frac{-1}{2 i} e^{- i k \xi t} 
%\frac{ k_b \cos(k_b \xi' t) + g \sin(k_b |\xi'| t)}{g^2 + k_b^2} \frac{k k_b}{k^2-k_b^2} e^{\frac{i}{2} (k^2 - k_b^2) t} 
%\ee 
\be \label{122}
\frac{-1}{2} \int_0^{+\infty} \frac{dk_b}{\pi} \int_{ - \infty}^{+ \infty} \frac{dq}{\pi} 
\frac{ k_b \cos(k_b \xi' t) + g \sin(k_b |\xi'| t)}{g^2 + k_b^2} \frac{(\xi+ i q) k_b}{(\xi + i q)^2-k_b^2} e^{- \frac{i}{2} (\xi^2+q^2+k_b^2) t}  \;.
\ee 
One can argue that this integral vanishes at large time. By expressing the sine and cosine terms in (\ref{122}) as the sum of exponentials, the resulting exponential terms have the form
$e^{ - \frac{i}{2} (q^2 + (k_b  \pm \xi')^2 + \xi^2 - (\xi')^2 ) t }$
which have a saddle point at $q=0$, $k_b = \pm \xi'$. By expanding around these saddle points $k_b=\pm \xi'+p$
one finds that the integral decays algebraically at large $t$.
%there are two cases: either $\xi' \neq \pm \xi$ and under rescaling the decay is like $1/t$, or $\xi' = \pm \xi$ in which case it can be brought to a form proportional %to
%the reduced integral 
%\be 
%\int dp dq \frac{e^{- \frac{i}{2} (q^2 + p^2) t} }{i q + p}
%\ee 
%which decays as $1/\sqrt{t}$.

For $\xi<0$ it is the second term in \eqref{sinus2} which vanishes at large time. For the first term in \eqref{sinus2} one can perform the same
manipulations as above with $\xi \to |\xi|$ and the right hand side in \eqref{118} is changed to  
\be \label{118bis} 
 {\rm sign}(\xi) \int_{k_L}^{|\xi|} \frac{dk_b}{\pi} e^{-  {\rm sign}(\xi) i k_b (\xi t + y)} \tilde f_{\xi' t + y',t}(k_b,k_b) \;,
\ee 
where we have used that the shift by $y/t$ in the bounds of the integral is irrelevant in the following.

Discarding all the terms which vanish at large time, we are left with the sum of \eqref{sinus2} and the contribution from the second line in \eqref{res33} which leads to 
\bea 
&& C(\xi t + y, \xi' t+ y') \simeq \bigg(
 {\rm sign} (\xi)  \int_{k_L}^{\max(\min(|\xi|,k_R),k_L)} \frac{dk_b}{\pi} e^{ - {\rm sign}(\xi) i k_b (\xi t + y)}  \\
&& +
\int_{k_L}^{k_R} \frac{dk_b}{\pi} ( i + \frac{g}{k_b}) \sin(k_b (\xi t + y)) 
\bigg)
 \frac{k_b}{2}  \frac{ k_b \cos(k_b (\xi' t +y')) + g \sin(k_b |\xi' t +y'|)}{g^2 + k_b^2} \;. \nn
\eea 
It vanishes at large time unless $\xi'=\xi$ or $\xi'=-\xi$. For $\xi'=\xi$ we find using standard trigonometric identities
\bea 
&& C(\xi t + y, \xi t+ y') \simeq {\rm sign} (\xi) \int_{k_L}^{\max(\min(|\xi|,k_R),k_L)} \frac{dk_b}{4 \pi} \frac{k_b(k_b- i g)}{g^2 + k_b^2} 
e^{- i {\rm sign}(\xi) k_b (y-y') }   \\
&& + \int_{k_L}^{k_R} \frac{dk_b}{\pi}  
 \frac{g + i k_b}{4 (g^2 + k_b^2)} (k_b \sin( k_b(y-y')) + ({\rm sign} (\xi)) g \cos( k_b(y-y') ) \;,
  \nonumber 
\eea 
from which one computes the combination $C(\xi t + y, \xi t+ y') + C(\xi t + y', \xi t+ y)^*$ which is needed
for the kernel (see Eq. (\ref{K_ABCD})).
%One finds 
%\bea 
%&& C(\xi t + y, \xi t+ y') + C(\xi t + y', \xi t+ y)^*  
%\simeq {\rm sign} \xi \int_{k_L}^{\max(\min(|\xi|,k_R),k_L)} \frac{dk_b}{2 \pi} \frac{k_b^2}{g^2 + k_b^2} 
%e^{- i {\rm sign}(\xi) k_b (y-y') }  \nn  \\
%&& +  \int_{k_L}^{k_R} \frac{dk_b}{2 \pi}  
% \frac{i k_b^2 \sin( k_b(y-y')) + ({\rm sign} \xi) g^2 \cos( k_b(y-y')) }{g^2 + k_b^2} 
%\eea 
%\\
%\bea
%\frac{{\rm sign} \xi}{2 i} \int_{k_L}^{\max(\min(|\xi|,k_R),k_L)} \frac{dk_b}{\pi} 
%\left( i + \frac{g}{k_b} \right)  \frac{k_b}{4(g^2 + k_b^2)} (k_b+i g) e^{ ({\rm sign}(\xi) i k_b (y+y')} \nn
%\eea 
%\bea   
%\frac{{\rm sign} \xi}{2 i} \int_{k_L}^{\max(\min(|\xi|,k_R),k_L)} \frac{dk_b}{\pi} 
%\left( i + \frac{g}{k_b} \right) 
% \frac{k_b}{4 (g^2 + k_b^2)} 
% ( (k_b + i g) e^{i {\rm sign}(\xi)) k_b (y-y') } \nn
%\eea 
For $\xi'=-\xi$ we find similarly
\bea 
&& C(\xi t + y, - \xi t+ y') \simeq {\rm sign} \xi \int_{k_L}^{\max(\min(|\xi|,k_R),k_L)} 
\frac{dk_b}{4 \pi} \frac{k_b(k_b- i g)}{g^2 + k_b^2} e^{- i {\rm sign}(\xi) k_b (y+y') }\\
&& +  \int_{k_L}^{k_R} \frac{dk_b}{4 \pi} 
 \frac{i k_b+ g}{g^2 + k_b^2} (k_b \sin( k_b(y+y')) + g ({\rm sign} \xi ) \cos( k_b(y+y') ) 
  \nonumber 
\eea 
from which one computes the combination $C(\xi t + y, - \xi t+ y') + C(-\xi t + y', \xi t+ y)^*$ which is needed
for the kernel (see Eq. (\ref{K_ABCD})).

%Now we have 
%\bea 
%&& C(\xi t + y, - \xi t+ y') + C(-\xi t + y', \xi t+ y)^* \simeq \\
%&& 
%- i {\rm sign} \xi \int_{k_L}^{\max(\min(|\xi|,k_R),k_L)} 
%\frac{dk_b}{2 \pi} \frac{k_b g}{g^2 + k_b^2} e^{- i {\rm sign}(\xi) k_b (y+y') } \nn
%\\
%&& + \int_{k_L}^{k_R} \frac{dk_b}{2 \pi} \frac{g k_b}{g^2 + k_b^2} 
%  (\sin( k_b(y+y')) + i  ({\rm sign} \xi ) \cos( k_b(y+y') ) \nn
%\eea 
%\\

\vspace*{0.5cm}

\noindent {\bf The term ${B(x,x',t)}$}. Let us start with $B_R^{\rm diag}(x,x',t)$. From \eqref{BR_diag_large} one has 
\bea  \label{BR_diag_largebis}
B_R^{\rm diag}(\xi t + y,\xi' t + y') = \int_0^{k_R} \frac{dk_a}{2 \pi} \frac{\tilde F_{\xi t + y,\xi' t + y'}(k_a)}{g^2+k_a^2}
\eea 
where 
\be  
\tilde F_{x,x',t}(k_a)= (k_a\cos(k_ax)+ g\sin(k_a|x|) ) ( k_a\cos(k_a x')+ g\sin(k_a |x'|) ) 
\ee 
Using trigonometric identities, one finds that it vanishes at large time except for $\xi'=\pm \xi$. One finds for $\xi'=\xi$
\bea 
B_R^{\rm diag}(\xi t + y,\xi t + y') = \int_0^{k_R} \frac{dk_a}{4 \pi} \cos k_a(y-y') 
\eea 
and for $\xi'=- \xi$
\bea 
B_R^{\rm diag}(\xi t + y,- \xi t + y') = \int_0^{k_R} \frac{dk_a}{4 \pi} \cos k_a(y+y') \;.
\eea 
\\

Finally note that the term $B_{L/R}^{\rm off-diag}(x,x',t)$, which was shown to vanish in the NESS
regime (see previous Section \ref{sec:largetime}), is expected to also vanish in the present ray regime 
(although we will not study here its decay in detail). 
\\

\noindent {\bf Final result}. Adding all the terms computed above we obtain the final result for the 
kernel $K^{\pm}_{\xi}(y,y')= \lim_{t \to +\infty} K(\xi t + y,\pm \xi t + y')$. For $K^+_\xi$ we obtain, recalling the identity $\frac{\rho_R}{4} \frac{\sin(k_R x)}{k_R x} = \int_0^{k_R} \frac{dk}{4 \pi}  \cos(k x)$,
\bea 
&& K^{+}_{\xi}(y,y') = \frac{\rho_{L}}{2} \frac{\sin(k_{L}(y-y') )}{k_{L}(y - y') } + \frac{\rho_{R}}{2} \frac{\sin(k_{R}(y-y') )}{k_{R}(y - y') } 
\\
&& + {\rm sign} (\xi) \int_{k_L}^{\max(\min(|\xi|,k_R),k_L)} \frac{dk_b}{2 \pi} \frac{k_b^2}{g^2 + k_b^2} 
e^{- i {\rm sign}(\xi) k_b (y-y') } \nn \\
&& +  \int_{k_L}^{k_R} \frac{dk_b}{2 \pi}  
 \frac{i k_b^2 \sin( k_b(y-y')) + ({\rm sign} \xi) g^2 \cos( k_b(y-y')) }{g^2 + k_b^2} \;. \nn
\eea 
This can be reorganized, leading to the result given in the text in \eqref{Kp_xixi}. From
this formula we obtain the density {$\tilde \rho(\xi) = K^{+}_{\xi}(y,y)$}, which is independent
of $y$, and given in the text in \eqref{rhoxinew}. Similarly one obtains the current {$\tilde J(\xi)={\rm Im} K^{+}_{\xi,01}(y,y)$}
leading to the expression \eqref{Jxinew}. 
%and it is also equal to 
%\bea 
%&& K^{+}_{\xi}(y,y') =  \int_0^{k_R} \frac{dk}{\pi} \cos(k(y-y')) \Theta(\xi>0) + \int_0^{k_L} \frac{dk}{\pi} \cos(k(y-y')) \Theta(\xi<0) 
%\\
%&& - {\rm sign} \xi \int_{\xi^*}^{k_R} \frac{dk_b}{2 \pi} \frac{k_b^2}{g^2 + k_b^2} 
%e^{- i {\rm sign}(\xi) k_b (y-y') } 
%\eea 
%\\

%One finds for the density $\rho^{\pm}_\xi = K^{\pm}_{\xi}(y,y)$
%\bea 
%&& \rho^{+}_\xi = \frac{1}{2} (\rho_{L}  + \rho_{R} ) + {\rm sign} \xi 
%\left( \int_{k_L}^{\max(\min(|\xi|,k_R),k_L)} \frac{dk_b}{2 \pi} \frac{k_b^2}{g^2 + k_b^2} +  \int_{k_L}^{k_R} \frac{dk_b}{2 \pi}  
% \frac{g^2}{g^2 + k_b^2} \right) 
%\eea 
%which is equivalent to \eqref{rhoxinew} up to a small change {\red please recheck} where one defines 
%\be 
%\xi^* = \max(\min(|\xi|,k_R),k_L)
%\ee 
%that is
%\bea \label{rhoxinew2} 
%\rho(\xi)=\frac{k_L+k_R}{2\pi}+{\rm sgn}(\xi)\left(\int_{k_L}^{k_R}\frac{dk}{2\pi}R(k)+\int_{k_L}^{k_R}\frac{dk}{2\pi}T(k)\Theta(\xi^*-k)\right)
%\eea
%\\

%Consider now the current $J^{\pm}_\xi={\rm Im} K^{\pm}_{\xi,01}(y,y)$. One has
%\be 
%J^{+}_\xi= - \int_{\xi^*}^{k_R} \frac{dk_b}{2 \pi} \frac{k_b^3}{g^2 + k_b^2}
%\ee 
%which coincides with \eqref{Jxinew} {\red up to the change $|\xi| \to \xi^*$}. 
%\\

For $K^-_\xi$, i.e. $\xi'=-\xi$, we obtain 
\bea 
&& K^{-}_{\xi}(y,y') =
%= - \frac{\rho_{L}}{2} \frac{\sin(k_{L}(y+y') )}{k_{L}(y+y') } - \frac{\rho_{R}}{2} \frac{\sin(k_{R}(y+y') )}{k_{R}(y+y') } 
%\\
% && 
- i \, {\rm sign} (\xi) \int_{k_L}^{\max(\min(|\xi|,k_R),k_L)} 
\frac{dk_b}{2 \pi} \frac{k_b g}{g^2 + k_b^2} e^{- i {\rm sign}(\xi) k_b (y+y') } \nn
\\
&& + \int_{k_L}^{k_R} \frac{dk_b}{2 \pi} \frac{g k_b}{g^2 + k_b^2} 
  (\sin( k_b(y+y')) + i  ({\rm sign} \xi ) \cos( k_b(y+y') )
\eea
These two terms can be combined and one obtains the result \eqref{Kxi_m}. 
%\bea 
%&& K^{-}_{\xi}(y,y') = 
%- \frac{\rho_{L}}{2} \frac{\sin(k_{L}(y+y') )}{k_{L}(y+y') } - \frac{\rho_{R}}{2} \frac{\sin(k_{R}(y+y') )}{k_{R}(y+y') } 
%\\
% && + 
%i  ({\rm sign} \xi ) \int_{\max(\min(|\xi|,k_R),k_L)}^{k_R} \frac{dk_b}{2 \pi} \frac{g k_b}{g^2 + k_b^2} e^{- i {\rm sign}(\xi) k_b (y+y') }  
%\eea
%\\

%\begin{figure}
%\centering
%\includegraphics[scale=.7]{schema-2.pdf}
%\caption{The integral contour is deformed in a way such that the integral goes to zero in the large time limit. If the contour cross a pole, an additional term has to %be added.}
%\end{figure}

\section{Wigner function and semi-classical approach} \label{sec:wigner}

It turns out that the results for the density and current obtained here by an exact computation in the regime of large time and $\xi=x/t$ fixed, agree with the prediction of a semi-classical approach. A similar agreement was observed for the discrete model studied in \cite{Prosen2018}. Here in addition we show that the agreement extends to the full kernel $K(x,x',t)$ at large time with $x=\xi t + y$, $x'= \xi' t + y'$ obtained here in Section \ref{sec:light} with $\xi'=\xi$ and $y,y'=O(1)$ from an exact computation. However, as we discuss below, this semi-classical approach does not allow to predict the non trivial correlations
obtained here for $\xi'=-\xi$. Let us present here this approach.

The many-body Wigner function $W(x,k,t)$ was defined and studied in e.g. \cite{DLMS2018} in the case of noninteracting fermions (see also \cite{Kulkarni}). It was shown to 
be related to the kernel via (we recall that we set here $\hbar=1$)
\be \label{KtoW} 
W(x,k,t) = \int_{-\infty}^{\infty} \frac{dy}{2 \pi} e^{i k y} K(x+ \frac{y}{2}, x- \frac{y}{2},t) \;,
\ee 
and we recall that the time-dependent mean fermion density is simply
\be \label{densW} 
\rho(x,t)= \int_{-\infty}^{\infty} dk \, W(x,k,t) \;.
\ee
The Wigner function for fermions in an external potential obeys an exact time evolution equation, see e.g. \cite{DLSM2019}. 
For smooth potentials one can define a semi-classical limit 
($\hbar \to 0$, large $N$ and fixed $N\hbar$ see e.g. \cite{Kulkarni}) and this equation simplifies into the Liouville equation
$\partial_t W = - k \partial_x W + V'(x) \partial_k W$. For a delta function potential this limit cannot be defined in that way, but one
can define instead a semi-classical form for the Wigner function by introducing transmission and reflection coefficients, as 
was done in \cite{Prosen2018}. Let us now recall this approach.

\begin{figure}
%\begin{subfigure}[b]{0.55\textwidth}
\centering
\includegraphics[width=0.9\linewidth]{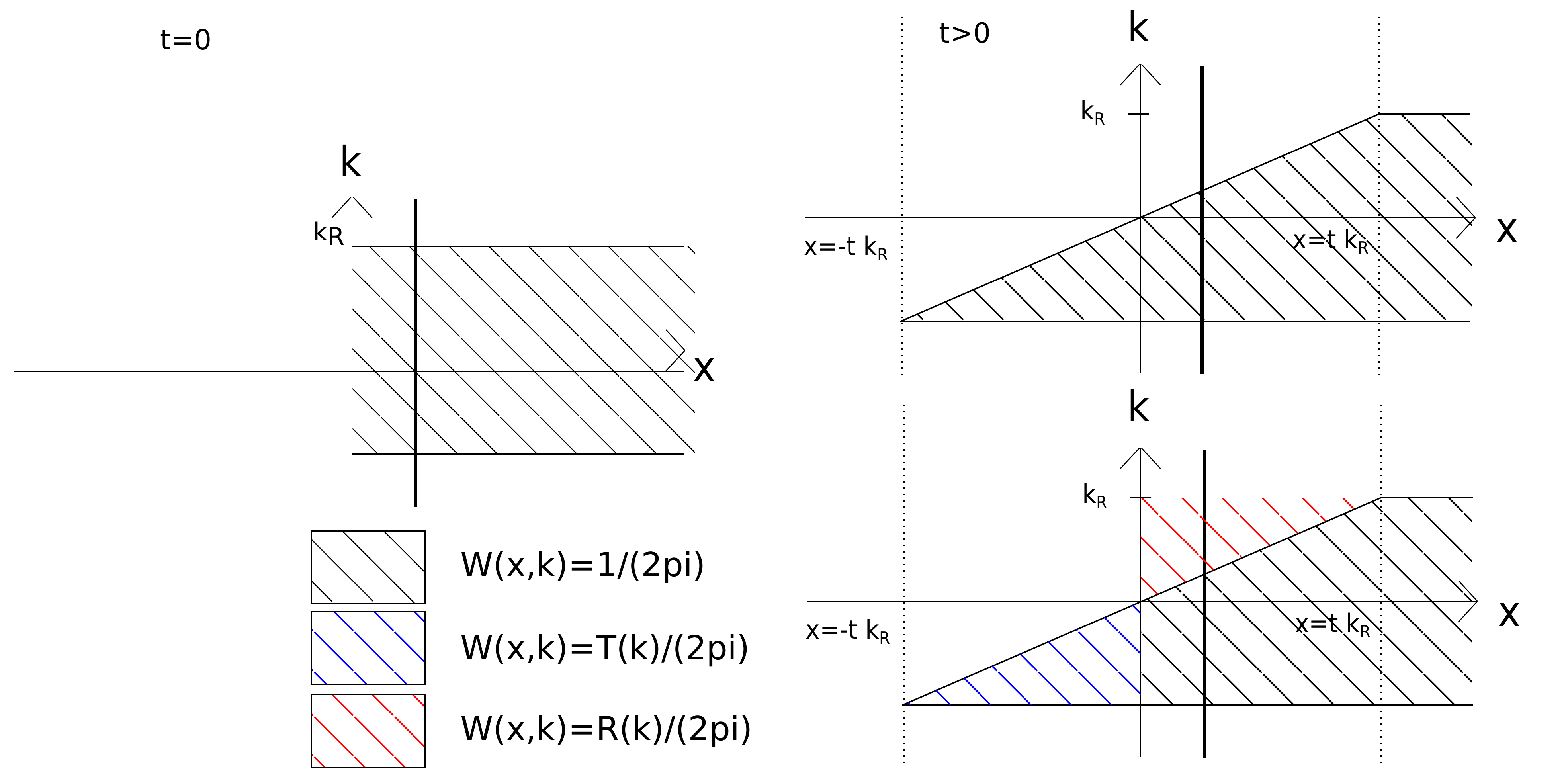}
\caption{{
Semi-classical evolution of the Wigner function. On the left panel we have indicated the initial condition for which the Wigner function is equal to $\frac{1}{2\pi}$ in the black hatched area and $0$ otherwise. For simplicity we choose $k_L=0$ and $k_R > 0$. On the right top panel the Wigner function is propagated at time $t$ without reflection at $x=0$ (this corresponds to $g=0$ or $T(k)=1$). On the bottom right panel we show the evolution with $g\neq 0$: the blue hatched area corresponds to transmitted part, while the red hatched area corresponds to the reflected part. Therefore the Wigner function is affected by the corresponding factors $R(k)$ and $T(k)$ (see Eqs. (\ref{Wigner_sc1}) and (\ref{Wigner_sc2})).
Note that to compute the density at a given point $x$ one has to integrate the Wigner function over the vertical line passing through $x$, as indicated in the figure.
%for example if $x<tk_R$ one has to integrate on $[-k_R,\frac{x}{t}]$.
}} \label{fig_semi_c1}
\end{figure}
Consider first the easier case with no defect. The fermions evolve freely with speed $v(k)$, and in the present model $v(k)=k$.
The Liouville equation applies to this free evolution with $V'(x)=0$ and the Wigner function is simply transported along the
classical trajectories. It is thus obtained from the trajectories run backward in time as
\be 
W(x,k,t) = W_0(x-v(k) t,k) 
\ee 
where $W_0(x,k)$ is the Wigner function at time $t=0$. As described in Section \ref{sec:initial}, here 
we consider an initial condition which is the product of two independent
fermionic ground states, separated by an infinite hard wall at $x=0$. On each side of the wall one considers the 
ground state at Fermi energy $\mu_{L}= \frac{k_L^2}{2}$ for $x<0$, respectively $\mu_R= \frac{k_R^2}{2}$ for $x>0$, of noninteracting
fermions in the absence of an external potential, with vanishing wave function at $x=0$. The corresponding
kernel is given in Eqs. \eqref{K0}-\eqref{KR}. As a consequence
\be 
W_0(x,k) = W_0^R(x,k) \Theta(x>0) + W_0^L(x,k) \Theta(x<0) 
\ee 
where the functions $W_0^{L/R}(x,k)$ are obtained from $K_{L/R}(x,x')$ in \eqref{KL},\eqref{KR} by the same transformation as in \eqref{KtoW}. 
The exact calculation of these functions is performed in Appendix C.1 of \cite{WignerBenjamin2021}. Away from a small layer of width
$x = O(1/k_{L/R})$ near the wall, it takes the form predicted by the local density approximation (LDA)
\be \label{W_LDA}
W_0(x,k) \simeq  \frac{1}{2 \pi}  \Theta(k_R - |k| ) \Theta(x>0) + \frac{1}{2 \pi}  \Theta(k_L - |k| ) \Theta(x<0) 
\ee 
which is discontinuous at $x=0$, as represented in Fig. \ref{fig_semi_c1} (left panel). We will use this form (\ref{W_LDA}) from now on. 

%. Then in order to compute the density $\rho(x,t)$ one has to integrate over all position $x'$ and momentum $k$ while imposing the condition $x-x'=v(k)t$ this leads to %\bea
%\rho(x,t)&&=\int_{-\infty}^{\infty}\frac{dk}{2\pi}\int_{-\infty}^{\infty}dx'\hat \rho_0(x',k)\delta(x-x'-v(k)t)\nn\\
%&&=\int_{-\infty}^{\infty}\frac{dk}{2\pi}\hat \rho_0(x-v(k)t,k)\nn\\
%&&=\int_{-\infty}^{\infty}\frac{dk}{2\pi}\hat \rho_{0L}(k)\theta(v(k)-\xi)+\hat \rho_{0R}(k)\theta(\xi-v(k))
%\eea
%where $\hat \rho_{0}(k)$ is the initial momentum density at $t=0$ which is then decomposed into $\hat \rho_{0L/R}(k)$ for $x<0$ and $x>0$ respectively and $\xi=x/t$.

\begin{figure}[t]
\centering
\includegraphics[width=.7\linewidth]{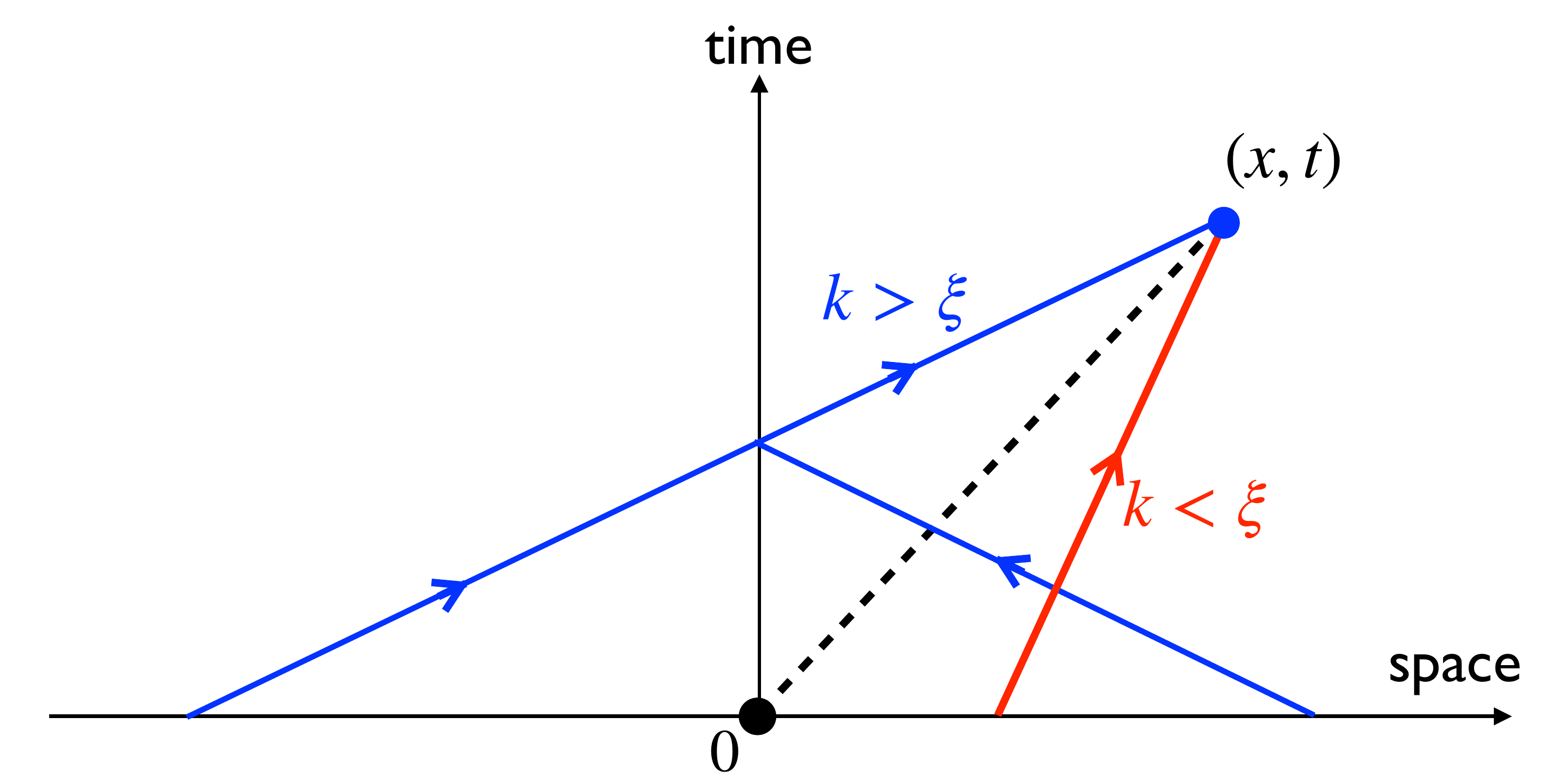}
\caption{Illustration of some trajectories in the semi-classical method. The defect is at position $x=0$ (black dot). 
Consider a fermion which at time $t$ is at position $x = \xi t$ (blue dot)
with momentum $k$. There are several possibilities for its initial position: (i) $k<\xi$ and it was to the right 
of the defect at time zero, (ii) $k>\xi$ and either it was to the left of the defect and has been transmitted (with probability $T(k)$)
or it was to the right of the defect and has been reflected (with probability $R(k)=1-T(k)$). The dashed line corresponds to $k=\xi$. Here we assume $T(k)=T(-k)$.}
\label{semi-classic-scheme}
\end{figure}

Now we add the defect at $x=0$. Consider a fermion at position $(x,t)$ with $x>0$ and with momentum $k$. If the condition
$\xi=\frac{x}{t} > v(k)$ holds, then the fermion has traveled to position $x$ without crossing the defect, as shown by the red solid line in Fig. \ref{semi-classic-scheme}.
On the other hand, if the condition $\xi<v(k)$ holds, there are two possibilities (indicated by the blue solid lines in Fig. \ref{semi-classic-scheme}): (i) either the fermion was at $t=0$ on the half-line $x<0$ to the
left of the defect, and it has been transmitted, or (ii) it was at $t=0$ on the half-line $x>0$ to the right of the defect, and it has
been reflected, see Fig. \ref{semi-classic-scheme}. The picture is similar for $x<0$. Setting $v(k)=k$ this leads to the prediction, for $x>0$
\be \label{Wigner_sc1}
W(x,k,t) = \frac{1}{2 \pi} ((T(k) \Theta(k_L - |k| ) + R(k) \Theta(k_R - |k| ) ) \Theta(k> \frac{x}{t}) +  \Theta(k_R - |k| ) \Theta(k< \frac{x}{t})  ) 
\ee 
and for $x<0$
\be \label{Wigner_sc2}
W(x,k,t) = \frac{1}{2 \pi} ((T(k) \Theta(k_R - |k| ) + R(k) \Theta(k_L - |k| ) ) \Theta(k< \frac{x}{t}) +  \Theta(k_L - |k| ) \Theta(k> \frac{x}{t})  ) 
\ee 
%\be \label{semi_cl1}
%\rho(x>0,t) = \int dk \bigg[ (T(k) \hat \rho_{0L}(k) + R(k) \hat \rho_{0R}(k) ) \theta(k> \xi) + \hat \rho_{0R}(k) \theta(k<\xi) \bigg]
%\ee 
where $T(k)$ and $R(k)=1 - T(k)$ are the transmission and reflection coefficients of the defect. We have assumed here for 
simplicity that $T(k)=T(-k)$.
Although this applies to any defect which is symmetric under $x \to -x$, for the present delta function potential one has
\be
R(k) = \frac{g^2}{g^2 + k^2} \quad , \quad T(k) = \frac{k^2}{g^2 + k^2} \;.
\ee 
Note that the Wigner function is invariant under the symmetry $x \to -x$ and $k_L \to k_R$.

It is useful to rewrite the Wigner function using $R(k)=1-T(k)$ in the following form, with $\xi=x/t$, for $\xi>0$
\be \label{Wnew1} 
W(x,k,t) = \frac{1}{2 \pi} \Theta(k_R - |k| ) - \frac{T(k)}{2 \pi} \Theta(k_L \leq k \leq k_R) \Theta(k>\xi) \;,
\ee 
and for $\xi<0$
\be \label{Wnew2}
W(x,k,t) = \frac{1}{2 \pi} \Theta(k_L - |k| ) + \frac{T(k)}{2 \pi} 
\Theta(- k_R \leq k \leq - k_L) \Theta(k< \xi) \;.
\ee 
The density is then obtained from \eqref{densW}. Using \eqref{Wnew1} and \eqref{Wnew2} we obtain,
with $\xi=\frac{x}{t}$ fixed
\bea 
&& \rho(x>0,t) = \rho_R  - \int_{k_L}^{k_R} \frac{dk}{2 \pi} T(k) \Theta(k- \xi) \\
&& \rho(x<0,t) = \rho_L  + \int_{k_L}^{k_R} \frac{dk}{2 \pi} T(k) \Theta(k- |\xi|) \;,
\eea 
where in the second equation we have changed $k \to -k$ and used $T(k)=T(-k)$. One can check
upon rearranging and using $R(k)=1-T(k)$ that this prediction coincides with the result of the exact calculation obtained 
in \eqref{rhoxinew}.
\\

%where we have defined
%\be 
%\xi^* = \max( \min(|\xi|,k_R) , k_L) = \begin{cases} k_L \quad , \quad |\xi| \leq k_L \\
%\xi \quad , \quad k_L \leq |\xi| \leq k_R \\
%k_R \quad , \quad |\xi| \geq k_R
%\end{cases} 
%\ee 

%This leads to 
%\be 
%\rho(x>0,t) = \int_{-k_R}^{k_R} \frac{dk}{2 \pi} = \rho_R \quad , \quad \xi > k_R > k_L > 0 \nn\\
%\ee 
%\be 
%\rho(x>0,t) = \int_{-k_R}^{\xi} \frac{dk}{2 \pi} + \int_{\xi}^{k_R} R(k) \frac{dk}{2 \pi} \quad , \quad  k_R > \xi > k_L > 0\nn \\
%\ee 
%\be 
%\rho(x>0,t) = \int_{-k_R}^{\xi} \frac{dk}{2 \pi} + \int_{\xi}^{k_R} \frac{dk}{2 \pi} R(k) 
%+ \int_{\xi}^{k_L} \frac{dk}{2 \pi} T(k) \quad , \quad  k_R > k_L > \xi > 0 \\
%\ee 
%\red{maybe we should erase the blue paragraph below. For the eq above(42) $\rho$ written in a weird way not showing that $\rho$ is independent of $\xi$ was it %intentional for pedagogic reasons or smthing?}
%\blue{In a more compact way we have
%\be
%\rho(x>0,t) =\frac{k_L+k_R}{2\pi}+\int_{k_L}^{k_R}\frac{dk}{2\pi}R(k)+\int_{k_L}^{k_R}\frac{dk}{2\pi}T(k)\Theta(\xi-k)
%\ee
%And we get back equation Eq. (\ref{tilderho}})
%\\

Let us now extend this prediction to the full kernel using the semi-classical Wigner function. 
By inverse Fourier transform of \eqref{KtoW} one has
\be \label{WtoK} 
K(x,x',t) = \int dk e^{- i k (x-x')} W\left(\frac{x+x'}{2},k,t\right)  \;.
\ee 
Let us consider the large time limit in the ray regime with $x= \xi t + y$ and $x'= \xi t + y'$, where $y,y'=O(1)$.
In that case one can approximate \eqref{WtoK} as
\be 
K(\xi t + y,\xi t + y',t) \simeq \int dk \, e^{- i k (y-y')} W(\xi t,k,t) \;.
\ee 
Using again \eqref{Wnew1}, for $t \to +\infty$ and $\xi>0$ one obtains
\bea  \label{Ksc1} 
&& K(\xi t + y,\xi t + y',t)  \simeq \int_0^{k_R} \frac{dk}{\pi} \cos(k(y-y')) - \int_{k_L}^{k_R} \frac{dk}{2 \pi} T(k) e^{- i k (y-y')} 
\Theta(k-\xi) 
\eea 
and using \eqref{Wnew2} for $\xi<0$ 
\bea  \label{Ksc2} 
&& K(\xi t + y,\xi t + y',t) \simeq \int_0^{k_L} \frac{dk}{\pi} \cos(k(y-y')) + \int_{k_L}^{k_R} \frac{dk}{2 \pi} T(k) e^{ i k (y-y')} \Theta(k-|\xi|) \;,
\eea 
where we have used $k \to -k$ and the symmetry $T(k)=T(-k)$. Remarkably, this semi-classical prediction 
coincides with the exact limiting kernel in \eqref{Kp_xixi} obtained by explicit calculation for the delta function potential model 
in the ray regime for any $\xi \neq 0$,
including $\xi=0^+$ and $\xi=0^-$. It is natural to expect that this semi-classical method will predict the correct behavior 
for more general potential, although this remains to explored. 
%\bea  \label{Ksc} 
% && K(\xi t + y,\xi t + y',t) \\
% && \simeq \int_{\min(\xi,k_L)}^{k_L} \frac{dk}{2 \pi} T(k) e^{- i k (x-x')}  + 
%\int_{\min(\xi,k_R)}^{k_R} \frac{dk}{2 \pi} (1-T(k)) e^{- i k (x-x')} + 
%\int_{-k_R}^{\min(\xi,k_R)} \frac{dk}{2 \pi} e^{- i k (x-x')} \nonumber \\
% && = \begin{cases} 
%\int_0^{k_R} \frac{dk}{\pi} \cos(k(x-x')) - \int_{k_L}^{k_R} \frac{dk}{2 \pi} T(k) e^{- i k (x-x')} \quad , \quad 0<\xi<k_L<k_R \\
%\int_0^{k_R} \frac{dk}{\pi} \cos(k(x-x')) - \int_{\xi}^{k_R} \frac{dk}{2 \pi} T(k) e^{- i k (x-x')} \quad , \quad k_L<\xi<k_R  \\
% \int_0^{k_R} \frac{dk}{\pi} \cos(k(x-x')) \quad , \quad k_L<k_R< \xi 
%\end{cases} \nonumber 
%\eea 
However, as discussed in the previous section, the semi-classical approach, in the present form, does not predict the kernel in the NESS regime (except in the limit $x, x' \to + \infty$ with $x-x' = O(1)$) nor the kernel $K_\xi^-(y,y')$ which is obtained for $\xi'= - \xi$.

\section{Conclusion} \label{sec:conclu}

In summary, we have studied a quantum quench of noninteracting fermions in one dimension in the presence
of a delta impurity at the origin whose strength is changed at $t=0$ from $g=+\infty$ to a finite value $g$.
The initial state consists in two Fermi gases separately at equilibrium on the two half-axis, with different bulk densities 
and temperatures $T_{L/R}$. We have obtained exact results in the thermodynamic limit (i.e., for an infinite size system) for 
the correlation kernel, density and current (particle and energy) for any time $t$ after the quench.
This allowed us to study analytically the large time
limit of these quantities and their relaxation properties. We found that there are two distinct regimes:

(i) For fixed position
$x$ and large time, the system relaxes to a non equilibrium steady state (NESS) characterized by a uniform particle transport current
$J_\infty$ and energy current $J^Q_\infty$. We have obtained explicit expressions for the kernel, density and $J_\infty,J^Q_\infty$ in this regime 
as a function of $g$. 
In particular, we showed that the density in the NESS exhibits a non-trivial spatial dependence near the defect, with a jump in the first derivative at $x=0$, oscillations which extend far from the defect (which are non equilibrium analogs of Friedel oscillations)
and different plateau values at $x=\pm \infty$. We obtain the temperature dependence of
the currents $J_\infty$ and $J^Q_\infty$. In particular, in the low temperature limit, we find $J^Q_\infty \sim T_L^2-T_R^2$, with a prefactor that matches with the conformal field theory universal prediction in the small $g$ limit.

(ii) The ray regime $x \sim \xi t$, at fixed $\xi$, where the density and current reach asymptotic finite
values which depend only on $\xi$. At zero temperature, their dependence on $\xi$ shows slope discontinuities along the
two pairs of light cones at $\xi = \pm k_{L/R}$. We also obtain the spatial dependence of the asymptotic kernel, denoted 
$K^+_\xi$, in the vicinity of the ray $\xi$. We also obtain $K^-_\xi$, which describes
the correlations between two points with two opposite rays $\xi$ and $-\xi$.

It is important to note that these two regimes match smoothly, as we have shown, which indicates
the absence of possible intermediate regime. This means that in the ray regime, as $\xi \to 0^\pm$ one recovers
the asymptotic densities $\rho_\infty(\pm \infty)$ of the NESS, as well as the current $J_\infty$. Finally, in the case (i) we have studied in detail the convergence in time $t$ towards the NESS. 
We have found that at zero temperature the kernel and the density decay as $t^{-5/2}$ modulated by oscillations,
superposed to a $t^{-3}$ non-oscillating part, towards their values in the NESS. 

The present study is the analog for the non-equilibrium dynamics of our recent work in~\cite{DLMS2021} where we
studied the ground state of noninteracting fermions in the presence of a delta impurity. The form obtained 
here for the kernel and density in the NESS can be compared with the one obtained at equilibrium in \cite{DLMS2021}. 
Although they have some common features (such as the cusp in the density and Friedel-like oscillations) their overall form is different. 
In the limit $k_R=k_L$ (at zero temperature) they become identical, however only in the case 
$g>0$. This can be qualitatively understood since for $g>0$ the excitations which are present in the
initial state can propagate to infinity, while for $g<0$ there is a bound state which cannot carry
propagating excitations. 

Our work has close relations to the work of Ljubotina, Sotiriadis and Prosen \cite{Prosen2018},
which has inspired the analytical methods used here. However, in addition to being defined on a lattice,
the initial state considered there is completely uncorrelated, while in our case there are
the correlations of a Fermi gas with two Fermi momenta $k_R$ and $k_L$. This leads to
a different structure in the ray regime with two distinct pairs of light cones. 
As pointed out in Ref. \cite{Prosen2018}, the results for the density and the current in the ray regime can be obtained from a semi-classical
method. However the density in the NESS (at finite $x$) cannot be obtained from this method. 
We have shown that the same properties hold in the present continuum model. In addition we
have computed the kernels $K_\xi^\pm$: while $K_\xi^+$ can indeed be predicted
by the semi-classical method, the kernel $K_\xi^-$ which measures correlations in
opposite rays cannot be predicted by this approach, and requires exact methods. 

The present study unveils a number of correlation kernels, which have a non trivial form at the Fermi scale $k_{L/R}$,
different from the universal sine kernel (which is recovered here only outside the light cones). At variance
with standard kernels of RMT, which also describe trapped fermions at equilibrium, the present
ones found here carry currents, hence they are not purely real. It would be interesting to 
find analog in the context of RMT. 

Finally, we obtained here the kernel in the NESS, but it would be interesting to obtain also the full
density matrix and compute other observables such as the entanglement entropy. {This is left for future studies}.

\subsection*{Acknowledgements} We thank D. S. Dean for suggesting this calculation and helpful comments.
We thank D. Bernard, A. Krajenbrink, S. N. Majumdar and L. Zadnik for 
useful discussions. We acknowledge support from ANR grant ANR-17-CE30-0027-01 RaMaTraF.

%\section{Appendix}

\begin{appendix}

%\renewcommand\thesection{Appendix \Alph{section}.}
%\renewcommand\thesubsection{\Alph{section}.\arabic{subsection}}
%\renewcommand\theremark{\Alph{section}.\arabic{remark}}

%\section{More general continuous model}
%\label{sec:interpolation}

\setcounter{equation}{0}
\setcounter{figure}{0}
\renewcommand{\theequation}{A\arabic{equation}}
\renewcommand{\thefigure}{A\arabic{figure}}

\section{Large $\ell$ limit, term $B$} \label{App_largel}

In this appendix we give some more details on the evaluation of the large $\ell$ limit of the
term $B(x,x',t)= B_R(x,x',t) + B_L(x,x',t)$ defined in the text in \eqref{exact_K}.
As discussed in the text [see Eq. (\ref{Bsum})] each $B_{L/R}$ is split into two parts, an off-diagonal one with $k_a \neq k_b$
and a diagonal one with $k_a=k_b$. Below, we discuss them separately.

\subsection{The off-diagonal part} 

$\quad$\\

Let us focus on the off-diagonal term $B_R^{\text{off-diag}}$ from \eqref{exact_K} (the term $B_L^{\text{off-diag}}$
is obtained similarly by changing $k_R \to k_L$)
\bea \label{BR_offdiag_1}
&& B_R^{\text{off-diag}}(x,x',t)= \left(\frac{2}{\ell}\right)^3 \sum_{k\in\Lambda_{-},k\leq k_R}\sum_{\underset{k_{a}\neq k_{b}}{k_{a}\in\Lambda_{+},k_{b}\in\Lambda_{+}}}2\frac{k_{a}\cos(k_{a}x)+g\sin(k_{a}|x|)}{g^2+k_{a}^2}\nn\\
&& \times \frac{k_{b}\cos(k_{b}x')+g\sin(k_{b}|x'|)}{g^2+k_{b}^2}\frac{k^2k_{a} k_{b}}{(k_{a}^2-k^2)(k_{b}^2-k^2)}e^{i(E(k_{a})-E(k_{b}))t} \;,
\eea
where the discrete sum has been restricted to $k_a \neq k_b$ (and the term $2 g/\ell$ which is subdominant at large $\ell$
has been removed). As in the text we denote
\be  \label{def_F}
F_{xx'}(k_a,k_b)= 2 \frac{k_a\cos(k_ax)+g\sin(k_a|x|)}{g^2+k_a^2} \frac{k_b\cos(k_bx')+g\sin(k_b|x'|)}{g^2+k_b^2}k_a k_b \;.
\ee 
The large $\ell$ analysis is very similar to the one presented in the text for the term $C$ in Section \ref{sec:thermo}. Namely we replace the discrete sum over $k$ 
in \eqref{BR_offdiag_1} by a contour integral using the same trick as in \eqref{Cexact}. The contour $\gamma_\delta$ in Fig. \ref{Fig_contour_exact} is now a rectangle with $k_L$ replaced by $0$, which in the large $\ell$ limit becomes again a half-rectangle, similar to $\gamma_c$ in Fig. \ref{Fig_contour1} (with $k_L$ set to zero). In addition to this contour integral there are residues which we need to take into account. 
Since $k_a \neq k_b$ one sees that there are only simple poles at either $k=k_a$ or $k=k_b$ and their residues must be summed up.
Hence we arrive at the following formula in the large $\ell$ limit
\bea \label{BBB} 
 && B_R^{\text{off-diag}}(x,x',t)= \int_0^{\infty}\int_0^{\infty}\frac{dk_a}{\pi}\frac{dk_b}{\pi} F_{xx'}(k_a,k_b) e^{i(E(k_{a})-E(k_{b}))t}\nn\\
&& 
+ \left( \int_{0}^{i\epsilon} \frac{dk}{\pi} + \int_{+i\epsilon}^{k_R+i\epsilon}\frac{dk}{\pi} 
- \int_{k_R}^{k_R+i \epsilon} \frac{dk}{\pi} \right) \frac{k^2}{(k_a^2-k^2)(k_b^2-k^2)} \\
&& -2i \bigg[ \Theta(k_R-k_a) (-\frac{1}{2}+\frac{i}{2}\frac{g}{k_a}){\rm Res}_{k=k_a}
+ \Theta(k_R-k_b) (-\frac{1}{2}+\frac{i}{2}\frac{g}{k_b}){\rm Res}_{k=k_b} \bigg] 
\frac{k^2}{(k_a^2-k^2)(k_b^2-k^2)} \nn
\eea

Let us now compute each term separately in \eqref{BBB}. The last line (the residue part) is found to be equal to 
\be  \label{reskakb} 
\frac{1}{2(k_a^2-k_b^2)}(\Theta(k_R-k_a) (i k_a + g) - \Theta(k_R-k_b) (i k_b + g)) \;.
\ee 
On the other hand the contour integral can be calculated explicitly, namely 
\bea \label{int_tanh}
&& \left( \int_{0}^{i\epsilon} \frac{dk}{\pi} + \int_{+i\epsilon}^{k_R+i\epsilon}\frac{dk}{\pi} 
- \int_{k_R}^{k_R+i \epsilon} \frac{dk}{\pi} \right) \frac{k^2}{(k_a^2-k^2)(k_b^2-k^2)}
\\
&& = \frac{1}{\pi(k_a^2 - k_b^2)} 
\bigg[ \frac{k_b}{2} \left( \log( \frac{k_b+k}{|k_b-k|} ) + i \pi \Theta(k-k_b) \right) 
- \frac{k_a}{2} \left( \log( \frac{k_a+k}{|k_a-k|} ) + i \pi \Theta(k-k_a) \right)  \bigg]_{0}^{k_R} \;, \nn
\eea 
which holds for any $\epsilon>0$. When adding to \eqref{reskakb} we see that the imaginary part cancels and we are left with
the result given in the text in \eqref{BR_offdiag_large}. 

\subsection{The diagonal part} 

$\quad$\\

We now turn to the diagonal part, i.e., the term with $k_a=k_b$ in \eqref{exact_K}, focusing on $B_R^{\rm diag}$ 
(the term $B_L^{\text{diag}}$
is obtained similarly by changing $k_R \to k_L$). It reads
\begin{align} \label{BRR} 
B_R^{\rm diag}(x,x')&=\left(\frac{2}{\ell}\right)^3 \sum_{k\in\Lambda_{-},k\leq k_R}\sum_{k_{a}\in\Lambda_{+}}
F_{xx'}(k_a,k_a) \frac{k^2}{(k_{a}^2-k^2)^2} \;,
\end{align}
where $F_{xx'}(k_a,k_b)$ is given in \eqref{def_F}. To take the large $\ell$ limit, we again use
the contour integral trick used above. We convert the discrete sum over $k$ into a contour integral, denoted by $B_R^{\rm diag,1}(x,x')$, plus a residue term, denoted by $B_R^{\rm diag,2}(x,x')$.
However, instead of having two simple poles as in the off-diagonal terms, 
we now have a double pole at $k=k_a$, which changes the result for the residue.

Let us first consider the contour integral term. It has the form, in the large $\ell$ limit
\be
B_R^{\rm diag,1}(x,x') \simeq \left(\frac{2}{\ell}\right)^3 \ell^2 \int_{0}^{+\infty} \frac{dk_a}{2 \pi} 
F_{xx'}(k_a,k_a) I(k_a,k_R) 
\ee 
where the expression of the integral $I(k_a,k_R)$ is obtained from (\ref{int_tanh}) by setting $k_b=k_a$. We get
\bea
&& I(k_a,k_R) =
\left( \int_{0}^{i\epsilon} \frac{dk}{\pi} + \int_{+i\epsilon}^{k_R+i\epsilon}\frac{dk}{\pi} 
- \int_{k_R}^{k_R+i \epsilon} \frac{dk}{\pi} \right) \frac{k^2}{(k_{a}^2-k^2)^2} \\
&& = \frac{k_R}{2 \pi(k_a^2-k_R^2)} - \frac{1}{4 \pi k_a } \log \frac{k_a + k_R}{|k_a-k_R|} - \frac{i}{4 k_a} \Theta(k_R-k_a)  \;. \nn
\eea 
The remaining integral over $k_a$ is well defined (in the principal value sense around $k_R$). Hence 
this term clearly vanishes in the large $\ell$ limit since it is of order $O(1/\ell)$. This is because 
the initial {\it triple} sum in $B_R^{\rm diag,1}(x,x')$ has become a {\it double} integral, and hence an extra factor $1/\ell$.

Let us now consider the residue term. It can be read off from \eqref{BRR} as
\be  \label{BR_v}
B_R^{\rm diag,2}(x,x')  = \left(\frac{2}{\ell}\right)^2 \int_0^{+\infty} \frac{dk_a}{\pi} F_{xx'}(k_a,k_a)
\int_{\Gamma_1} \frac{d k}{2 \pi} \frac{\ell}{e^{i k \ell}-1} \frac{v(k,k_a)}{(k-k_a)^2}
\quad , \quad v(k,k_a) = \frac{k^2}{(k+k_{a})^2} \;,
\ee 
where $\Gamma_1$ is a contour in the complex $k$-plane consisting of the union of small circles oriented clockwise centered around $k=k_a$ with $0 < k_a \leq k_R$.
We now use the residue formula for a double pole with a counterclockwise contour around $z=0$, namely
\be 
\oint \frac{dz}{2 i \pi z^2} h(z) = h'(0) \;,
\ee 
which, applied to the integral over $k$ in Eq. (\ref{BR_v}), leads to 
\bea \label{double}
 \int_{\Gamma_1} \frac{d k}{2 \pi} \frac{\ell}{e^{i k \ell}-1} \frac{v(k,k_a)}{(k-k_a)^2}
&=& - i \frac{d}{dk} \left( v(k,k_a) \frac{\ell}{e^{i k \ell}-1} \right) \Bigg|_{k=k_a} \nn \\
&=& \frac{k_a^2 + g^2}{16 k_a^2} \ell^2 + \frac{g + i k_a}{8 k_a^2} \ell \;,
\eea 
where, in the last line, we have computed the derivative with respect to $k$, set $k=k_a$ and used the quantification
condition \eqref{quantification0} $e^{i \ell k_a} = - \frac{k_a+ i g}{k_a- i g}$. In the large $\ell$
limit, the first term $\propto \ell^2$ in (\ref{double}) dominates and plugging it into \eqref{BR_v}, we finally obtain the residue part 
as
\be 
B_R^{\rm diag,2}(x,x')  \approx  \int_0^{k_R} \frac{dk_a}{\pi} F_{xx'}(k_a,k_a) \frac{k_a^2 + g^2}{4 k_a^2} \;.
\ee 
Since we found that $B_R^{\rm diag,1}(x,x')$ vanishes, we finally obtain the result for the limit $\ell \to +\infty$ given in the
text for $B_R^{\rm diag}(x,x')$ in \eqref{BR_diag_large}. 

\setcounter{equation}{0}
\setcounter{figure}{0}
\renewcommand{\theequation}{B\arabic{equation}}
\renewcommand{\thefigure}{B\arabic{figure}}

\section{Convergence to the NESS at large time ($x,x'=O(1)$)}\label{App_NESS}

In this appendix we extract the leading decay in time of the kernel to its stationary value in the NESS.
We start from the result for the kernel at infinite $\ell$ given as a sum of terms in \eqref{K_ABCD},
where the $B_{L/R}$ terms are themselves decomposed in \eqref{Bsum}.
In this sum only the terms $C(x,x',t)$ and $B_{L/R}^{\rm off-diag}(x,x',t)$ depend on time.
Below we study them separately.

\subsection{Time decay of the term $C$} 

$\quad$ \\

Let us define $\Delta C(x,x',t)=C(x,x',t) - C_\infty(x,x')$.
We start from the expression \eqref{res33} for $C(x,x',t)$. Only the contour integral in the first line depends on time.
As discussed above it is independent of $\epsilon>0$ so we can choose here $\epsilon=+\infty$, 
in which case the contribution of the second integral in
the first line of \eqref{res33} vanishes (since it contains a factor $e^{- \epsilon p t}$ where $k = i \epsilon + p$ with $p \in [k_L,k_R]$ on this contour). Hence we obtain
\bea 
\Delta C(x,x',t)= \int_0^{+\infty} \frac{dk_b}{\pi} \bigg[ 
\left( \int_{k_L}^{k_L+i \infty} \frac{dk}{\pi}  - \int_{k_R}^{k_R+i \infty}\frac{dk}{\pi}
\right) \frac{h_{x,x',t}(k,k_b)}{k-k_b}  \bigg]  
\eea 
where $h_{x,x',t}(k,k_b)$ is defined in \eqref{ff}. We now notice that it is convenient to take the time derivative
of this expression so that the two integrals decouple, leading to
\bea\label{Cdecay}
&& \partial_t \Delta C(x,x',t) = \frac{i}{2} G_b(t) ( G_L(t) - G_R(t))  \\
&& G_b(t) = \int_0^{+\infty} \frac{dk_b}{\pi} k_b
 \frac{ k_b \cos(k_b x') + g \sin(k_b |x'|)}{g^2 + k_b^2}  e^{- \frac{i}{2} k_b^2 t}  \label{Gb} \\
&& G_{L/R}(t) = i \int_0^{+\infty} \frac{dp}{\pi} 
(k_{L/R}+ i p) \sin((k_{L/R}+ i p) x)  e^{\frac{i}{2} (k_{L/R} + i p)^2 t} 
\eea 
At large $t$, the integral in $G_b(t)$ is dominated by the
vicinity of $k_b=0$, while the integral in $G_L(t)$ is dominated by the vicinity of $p=0$. One finds, assuming $g>0$
and $x$ fixed
\be \label{asymptG} 
G_b(t) \simeq \frac{1}{g^2 \sqrt{2 \pi}} (1 + g |x'|) 
\frac{1}{(i t)^{3/2}} \quad , \quad G_{L/R}(t) = \frac{i}{\pi t} \sin(k_{L/R} x) e^{ \frac{i}{2} k_{L/R}^2 t} \;.
\ee 
Putting all together and integrating (up to subdominant terms) we obtain the estimate
\be \label{DeltaC} 
\Delta C(x,x',t)\simeq \frac{e^{- i \frac{\pi}{4}}}{ (\pi)^{3/2} g^2 \sqrt{2}} (1 + g |x'|) \frac{1}{t^{5/2}} \left( \frac{1}{k_L^2} e^{\frac{i}{2} k_L^2 t} {\sin(k_L x)}
- \frac{1}{k_R^2} e^{\frac{i}{2} k_R^2 t} {\sin(k_R x)} \right) \;,
\ee 
which we have also checked numerically. 
\\

{\bf Remark}. One can also start from the {\it discrete} double sum expression \eqref{C_expl} at finite $\ell$ for $C(x,x',t)$ 
and take a time derivative, which leads to a product of two decoupled discrete sums
\bea 
&& \partial_t C(x,x',t) =  \frac{i}{2} H_b(t) H(t) \\
&& H_b(t) = \frac{2}{\ell} \sum_{k_b \in \Lambda_+} k_b
\frac{ k_b \cos(k_b x) + g \sin(k_b |x|)}{g^2 + k_b^2} e^{- \frac{i}{2} k_b^2 t}  \\
&& H(t) = \frac{2}{\ell} \sum_{k \in \Lambda_-, k =k_L^+}^{k_R} 
k \sin(k x) e^{\frac{i}{2} k^2 t} \;.
\eea 
For each term the large $\ell$ limit is straightforward, hence here there is no need for the
contour integral trick. The formula for $H_b(t)$ gives immediately the formula for $G_b(t)$ in Eq. (\ref{Gb}) in the large $\ell$ limit.
The formula for $H(t)$ becomes, in the large $\ell$ limit
\bea \label{Htt} 
H(t) = \int_{k_L}^{k_R} \frac{dk}{\pi} k \sin(k x) e^{\frac{i}{2} k^2 t}  \;,
\eea 
which can be shown to be equal to $G_L(t)-G_R(t)$ upon changing the contour of integration. 
Note that using the integration by part method described in \eqref{IPP} one shows
that for any smooth function with bounded $h''(k)$ 
\be \label{identity2} 
\int_{0}^{k_R} dk \,k \, h(k) e^{\frac{i}{2} k^2 t} = \frac{i}{t} (h(0) - h(k_R) e^{\frac{i}{2} k_R^2 t}) + o\left(\frac{1}{t}\right) \;, 
\ee 
from which we can obtain the asymptotics of $H(t)$ from \eqref{Htt}, recovering
the result in \eqref{asymptG}.
\\

\subsection{Time decay of the term $B$}  \label{sec:Blargetime}

$\quad$\\

We start from the expression for $B_R^{\text{off-diag}}(x,x',t)$ in \eqref{BR_offdiag_large}
together with the definition of $F_{x,x'}(k_a,k_b)$ in
\eqref{def_Fkakb}. The same analysis can be made for
$B_L^{\text{off-diag}}(x,x',t)$. As we show below, the large time limit is dominated by two contributions: one where $k_a,k_b$ in the
double integral in \eqref{BR_offdiag_large} are both close to zero, and one where one of these 
momenta is close to $k_R$. Before presenting the detailed computation, let us show how the
first contribution can be obtained by a simple argument. Upon rescaling $k_a \to k_a/\sqrt{t}$
and $k_b \to k_b/\sqrt{t}$, this gives straightforwardly by estimating the behavior of the
integrand near $k_a=k_b=0$, 
\bea
B_R^{\text{off-diag}}(x,x',t)&\simeq&  - \frac{2}{t^3 g^4 \pi k_R} (1+ g|x|) (1+ g|x'|) \int_0^{\infty}\int_0^{\infty}\frac{dk_a}{\pi}\frac{dk_b}{\pi} k_a^2 k_b^2 
  e^{\frac{1}{2} i(k_{a}^2 -k_{b}^2)} \nn \\
&=& - \frac{1}{\pi^2 g^4 k_R} (1+ g|x|) (1+ g|x'|) \frac{1}{t^3} \;. \label{previous} 
\eea
In this calculation the terms in \eqref{BR_offdiag_large} containing the $\Theta$ functions do not contribute since they vanish when both $k_a,k_b<k_R$.
However the full calculation, to which we now turn, shows that this part does produce an additional contribution. 

Let us now again consider the time derivative of $B_R^{\text{off-diag}}(x,x')$ which allows to decouple
the two integrals over $k_a$ and $k_b$. It can be written as a sum of two parts. The first part is
\bea
&& \partial_t B_R^{\text{off-diag},1}(x,x')=i ( {\sf G}_x(t)^* {\sf H}_{x'}(t) - {\sf G}_{x'}(t) {\sf H}_{x}(t)^* )  \;,
\eea
where we have defined
\bea 
&& {\sf G}_x(t) = \int_0^{+\infty} \frac{dk_a}{\pi} k_a
 \frac{ k_a \cos(k_a x) + g \sin(k_a |x|)}{g^2 + k_a^2}  e^{- \frac{i}{2} k_a^2 t} \\
&& {\sf H}_x(t) = \int_0^{+\infty} \frac{dk_a}{\pi} k_a
 \frac{ k_a \cos(k_a x) + g \sin(k_a |x|)}{g^2 + k_a^2}  \frac{1}{2 \pi} k_a \log\left(\frac{k_a+k_R}{|k_a-k_R|}\right) e^{- \frac{i}{2} k_a^2 t} \;.
\eea  
It is easy to analyze the large time behavior of these functions, and one obtains
\be
{\sf G}_x(t)  \simeq \frac{1}{g^2 \sqrt{2 \pi}} (1 + g |x|) 
\frac{1}{(i t)^{3/2}} \quad , \quad {\sf H}_x(t) \simeq \frac{3}{\pi k_R \sqrt{2 \pi} g^2} (1 + g |x|)  \frac{1}{(i t)^{5/2}} \;.
\ee
This leads to
\bea
&& \partial_t B_R^{\text{off-diag},1}(x,x')= \frac{3}{ g^4 k_R \pi^2} (1+ g|x|) (1+ g|x'|) \frac{1}{t^4} \;,
\eea 
which agrees with the previous result \eqref{previous}. The second part reads 
\be
 \partial_t B_R^{\text{off-diag},2}(x,x') % =i {\blue \frac{g}{2}} ( ({\sf G}^{<}_x)^*(t) {\sf G}_{x'}(t) - {\sf G}^{<}_{x'}(t) {\sf G}_{x}^*(t) ) \\
% = - i {\blue \frac{g}{2}} ( ({\sf G}^{>}_x)^*(t) {\sf G}_{x'}(t) - {\sf G}^{>}_{x'}(t) {\sf G}_{x}^*(t) ) 
= i  \frac{g}{2} \left( {\sf G}_{x}(t)^* {\sf G}^{>}_{x'}(t)  - ({\sf G}^{>}_x)(t)^* {\sf G}_{x'}(t) \right) 
\ee
where we have defined
\bea 
&& {\sf G}^{>}_x(t) = \int_{k_R}^{+\infty} \frac{dk_a}{\pi} k_a
 \frac{ k_a \cos(k_a x) + g \sin(k_a |x|)}{g^2 + k_a^2}  e^{- \frac{i}{2} k_a^2 t} \;.
\eea
Using linear combinations of \eqref{identity2} (and the fact that $h(k) \to 0$ as $k \to +\infty$, we obtain
at large time 
\be
{\sf G}^{>}_x(t) \simeq 
 \frac{ k_R \cos(k_R x) + g \sin(k_R |x|)}{g^2 + k_R^2} \frac{e^{- \frac{i}{2} k_R^2 t} }{i \pi t} \;.
\ee
This leads to 
\bea 
&& \partial_t B_R^{\text{off-diag},2}(x,x') \simeq  \frac{1}{g (2 \pi)^{3/2}  t^{5/2}} 
\bigg( (1 + g |x'|) \frac{ k_R \cos(k_R x) + g \sin(k_R |x|)}{g^2 + k_R^2} \frac{1}{i^{3/2}} e^{\frac{i}{2} k_R^2 t} \nn \\
&& 
+ (1 + g |x|) \frac{ k_R \cos(k_R x') + g \sin(k_R |x'|)}{g^2 + k_R^2} \frac{1}{(-i)^{3/2}} e^{-\frac{i}{2} k_R^2 t} \bigg) \;. \nn
\eea
%\bea
%&& \partial_t B_R^{\text{off-diag},2}(x,x') \simeq  \frac{1}{g^2 \sqrt{2 \pi}} 
%\bigg[ - i (1 + g |x'|) 
%k_R
% \frac{ k_R \cos(k_R x) + g \sin(k_R |x|)}{g^2 + k_R^2} 
% \frac{e^{- \frac{i}{2} k_R^2 t} }{- i k_R t} \frac{1}{(i t)^{3/2}}  \nn 
% \\
% && + i (1 + g |x|) 
%k_R
% \frac{ k_R \cos(k_R x') + g \sin(k_R |x'|)}{g^2 + k_R^2} 
% \frac{e^{\frac{i}{2} k_R^2 t} }{i k_R t} \frac{1}{(-i t)^{3/2}}
% \bigg] 
%\eea 
%This leads to
%\bea
% && B_R^{\text{off-diag},2}(x,x') \simeq  \frac{1}{g^2 \sqrt{2 \pi}} 
%\bigg[  i (1 + g |x'|) 
%k_R
% \frac{ k_R \cos(k_R x) + g \sin(k_R |x|)}{g^2 + k_R^2} 
% \frac{e^{- \frac{i}{2} k_R^2 t} }{(\frac{1}{2} k_R^2) (k_R t)} \frac{1}{(i t)^{3/2}}  \nn 
% \\
%  && - i (1 + g |x|) 
%k_R
% \frac{ k_R \cos(k_R x') + g \sin(k_R |x'|)}{g^2 + k_R^2} 
% \frac{e^{\frac{i}{2} k_R^2 t} }{(\frac{1}{2} k_R^2) (k_R t)} \frac{1}{(-i t)^{3/2}}
% \bigg] 
%\eea 
%\\
Integrating with respect to time, and putting together the two terms, we finally obtain 
the large time behavior
\bea \label{decayBR} 
 && B_R^{\text{off-diag}}(x,x')\simeq  - \frac{1}{\pi^2 g^4 k_R} (1+ g|x|) (1+ g|x'|) \frac{1}{t^3} \\
% && + \frac{2}{g^2 \sqrt{2 \pi} k_R^2 t^{5/2}} 
%\bigg[  e^{- i \frac{\pi}{4}}  (1 + g |x'|) 
% \frac{ k_R \cos(k_R x) + g \sin(k_R |x|)}{g^2 + k_R^2} 
%e^{- \frac{i}{2} k_R^2 t}   \nn 
% \\
%  && + e^{i \frac{\pi}{4}} (1 + g |x|) 
% \frac{ k_R \cos(k_R x') + g \sin(k_R |x'|)}{g^2 + k_R^2} 
%e^{\frac{i}{2} k_R^2 t} 
% \bigg] 
&&  -  \frac{2}{g (2 \pi)^{3/2}  k_R^2 t^{5/2}} 
\bigg( (1 + g |x'|) \frac{ k_R \cos(k_R x) + g \sin(k_R |x|)}{g^2 + k_R^2}  e^{- i \frac{\pi}{4}} e^{\frac{i}{2} k_R^2 t} \nn \\
&& 
+ (1 + g |x|) \frac{ k_R \cos(k_R x') + g \sin(k_R |x'|)}{g^2 + k_R^2} e^{ i \frac{\pi}{4}}   e^{-\frac{i}{2} k_R^2 t} \bigg) \;, \nn
\eea
which is the sum of $1/t^3$ term and an oscillating $1/t^{5/2}$ term.

\subsection{Summary} 

$\quad$ \\

One can put together the terms computed above and obtain the convergence of the kernel towards
its asymptotic value in the NESS. Denoting $\Delta K(x,x',t)=K(x,x',t)-K_\infty(x,x')$ we 
obtain
\be \label{DeltaKt} 
\Delta K(x,x',t) = \Delta C(x,x',t) + \Delta C(x',x,t)^* + B_R^{\text{off-diag}}(x,x') +
B_L^{\text{off-diag}}(x,x')
\ee 
where $\Delta C$ is given in \eqref{DeltaC} and $B_R^{\text{off-diag}}(x,x')$ is given in \eqref{decayBR} (with the 
same formula for $B_L^{\text{off-diag}}(x,x')$ with $k_R$ replaced by $k_L$. We display here the large time behavior of the density 
\bea \label{densdecay}
&&\hspace*{-0.5cm} \rho(x,t) - \rho_\infty(x) = - \frac{1}{\pi^2 g^4} (1+ g|x|)^2 \frac{1}{t^3} (\frac{1}{k_L}+\frac{1}{k_R}) \\
&&\hspace*{-0.5cm} -  \frac{4 (1 + g |x|)}{g (2 \pi)^{3/2} t^{5/2}}  
 \Big( \frac{ k_R \cos(k_R x) + g \sin(k_R |x|)}{k_R^2(g^2 + k_R^2)} \cos(\frac{k_R^2 t}{2} - \frac{\pi}{4} ) 
\nn \\
&&\hspace*{-0.5cm}+ \frac{ k_L \cos(k_L x) + g \sin(k_L |x|)}{k_L^2(g^2 + k_L^2)} \cos(\frac{k_L^2 t}{2} - \frac{\pi}{4} ) \Big) \nn \\
&& \hspace*{-0.5cm} + 
\frac{4 (1 + g |x|)}{g^2 (2 \pi)^{3/2} t^{5/2}} \left( \frac{\sin(k_L x)}{k_L^2} \cos(\frac{k_L^2 t}{2} - \frac{\pi}{4} ) 
- \frac{\sin(k_R x)}{k_R^2} \cos(\frac{k_R^2 t}{2} - \frac{\pi}{4} )   \right) \;. \nn
\eea 
We note that these formulae are valid for large time at fixed $x$ and for $g>0$. We expect that this expansion
breaks down when $x$ becomes large simultaneously with $t$. Note also that this formula does not apply to the
case $g=0$. Indeed, consider e.g. the $1/t^3$ in Eq. (\ref{previous}): this decay was obtained by rescaling $k_a \to k_a/\sqrt{t}$ and $k_b \to k_b/\sqrt{t}$, and approximating the denominator $k_a^2 + g^2$ by $g^2$. Obviously, this estimate and the $1/t^3$ decay holds only if 
$t \gg 1/g^2$. 

Finally note that the decay of the time dependent current $J(x,t)-J_\infty$ can be obtained from the kernel \eqref{DeltaKt} using \eqref{Jdef}.
We find that it decays as $t^{-5/2}$ modulated by oscillating terms. It can also be obtained from the conservation equation
\eqref{conservation} and \eqref{densdecay}.

 \setcounter{equation}{0}
\setcounter{figure}{0}
\renewcommand{\theequation}{C\arabic{equation}}
\renewcommand{\thefigure}{C\arabic{figure}}

\section{Details for $g<0$} \label{App:gneg}

If $g$ is negative there is an additional eigenstate of $\hat H_g$, namely $\phi_g(x)\underset{\ell\to\infty}{\simeq}\sqrt{-g}e^{g|x|}$ of energy $E=-\frac{g^2}{2}$.
The kernel then has an additional term which we call $\delta K$.

Before we start our computation we recall the overlap of the eigenstate $\phi_{1,k'}$ and $\phi_{-1,k'}$ with the initial state
\begin{align} \label{overlaps}
R_{k',k}^{+}&:=\int_0^{\ell/2}dy\sqrt{\frac{4}{\ell}}\sin(ky)\phi_{+,k'}(y)=\frac{2^{3/2}}{ \ell}\frac{kk'}{(k^2-k'^2)\sqrt{g^2+k'^2+\frac{2g}{ \ell}}} \;, \nn \\
L_{k',k}^{+}&:=\int_{-\ell/2}^{0}dy\sqrt{\frac{4}{\ell}}\sin(ky)\phi_{+,k'}(y)=-R_{k',k}^{+} \;, \nn \\
R_{k',k}^{-}&:=\int_0^{\ell/2}dy\sqrt{\frac{4}{\ell}}\sin(ky)\phi_{-,k'}(y)=\frac{1}{\sqrt{2}}\delta_{k,k'} \;, \nn \\
L_{k',k}^{-}&:=\int_{-\ell/2}^{0}dy\sqrt{\frac{4}{\ell}}\sin(ky)\phi_{-,k'}(y)=R_{k',k}^{-} \;,
\end{align}
and we also introduce the new overlaps of $\phi_g$ with the left and right initial states
\begin{align}
R_{g,k}&:=\int_0^{\ell/2}dy\sqrt{\frac{4}{\ell}}\sin(ky)\phi_g(y)\underset{\ell\to\infty}{\simeq}\sqrt{-g}\sqrt{\frac{4}{\ell}}\frac{k}{g^2+k^2} \;, \\
L_{g,k}&:=\int_{-\ell/2}^{0}dy\sqrt{\frac{4}{\ell}}\sin(ky)\phi_g(y)=-R_{g,k} \;.
\end{align}
Let us first consider $K_R(x,x',t)$ which for $g>0$ is given by \eqref{KernelLR} and reads 
\begin{align}
K_R(x,x',t)&=\sum_{\sigma_a = \pm,k_a\in \Lambda_{\sigma_a}}\sum_{\sigma_b=\pm,k_b\in \Lambda_{\sigma_b}}\sum_{k\in\Lambda_{-},k\leq k_R}\phi^*_{\sigma_a,k_a}(x)\phi_{\sigma_b,k_b}(x')e^{i(E(k_a)-E(k_b))t}\nn\\
&\times \left(\delta_{\sigma_a,-}R_{k_a,k}^{-}+\delta_{\sigma_a,+}R_{k_a,k}^{+}\right) \left(\delta_{\sigma_b,-}R_{k_b,k}^{-}+\delta_{\sigma_b,+}R_{k_b,k}^{+}\right) \;.
\end{align}
For simplicity let us introduce the schematic notation
\bea\label{SimplifiedKR}
K_R=\sum_{k_a}\sum_{k_b} \, \cdots \;.
\eea
Now for $g<0$, we must add one additional term to the sum over $k_a$ corresponding to the bound state. 
This means that we substitute $\phi_{\sigma_a,k_a}\to \phi_g$, $E(k_a)\to E(g)=-\frac{g^2}{2}$ and the overlap 
\bea
\left(\delta_{\sigma_a,-}R_{k_a,k}^{-}+\delta_{\sigma_a,+}R_{k_a,k}^{+}\right)\to R_{g,k} \;. 
\eea
In our schematic notation \eqref{SimplifiedKR} this becomes $\sum_{k_a}\to(\sum_{k_a}+g)$. The same procedure is applied to the sum over $k_b$. The product of sums have to be expanded leading to three additional terms 
\begin{align}
K_R^{g<0}&=(\sum_{k_a}+g)(\sum_{k_b}+g) =\sum_{k_a}\sum_{k_b}+(g,\sum_{k_b})_R+(\sum_{k_a},g)_R+(g,g)_R =K_R+\delta K_R \;,
\end{align}
with $\delta K_R=(g,\sum_{k_b})_R+(\sum_{k_a},g)_R+(g,g)_R$ in terms of our schematic notations. Each term is given explicitly below. Similarly, one can obtain $\delta K_L$. The full expression of $\delta K$ is then obtained by summing the right and left contributions, namely $\delta K = \delta K_R + \delta K_L$. Among those three terms, the only one remaining in the stationary state is $(g,g)$ which is time independent and computed as follows.
\begin{align}
(g,g)(x,x')&=(g,g)_R(x,x')+(g,g)_L(x,x')\\
&=\sum_{k_n<k_R}\phi^*_g(x)\phi_g(x')R_{g,k_n}^2+\sum_{k_n<k_L}\phi^*_g(x)\phi_g(x')L_{g,k_n}^2\\
&=-g\,e^{g(|x|+|x'|)}\,\left(\sum_{k_n<k_R}+\sum_{k_n<k_L}\right)R_{g,k_n}^2\\
&=2g^2e^{g(|x|+|x'|)}(\sum_{k_n<k_R}+\sum_{k_n<k_L})\frac{2}{\ell}\frac{k_n^2}{(g^2+k_n^2)^2}\\
&\underset{\ell\to\infty}{\simeq}2g^2e^{g(|x|+|x'|)}(\int_0^{k_R}+\int_0^{k_L})\frac{dk}{\pi}\frac{k^2}{(g^2+k^2)^2} \;.
\end{align}
This term corresponds precisely to $\delta K_{\infty}$ given in Eq. (\ref{Kinf_gneg}) in the text. Now we show that the other terms do not contribute to the large time limit. 
For the other term, we have (here implicitly, in all the discrete sums, $k \in \Lambda_-$ and $k_b \in \Lambda_+$ ) 
\begin{align}
(g,\sum_{k_b})(x,x')&=(g,\sum_{k_b})_R(x,x')+(g,\sum_{k_b})_L(x,x')\\
&=\sum_{k<k_R,k_b}\phi^*_g(x)\phi_{\sigma_b,k_b}(x')e^{-i\frac{g^2+k_b^2}{2}t}R_{g,k}(\delta_{\sigma_b,-}R_{k_b,k}^{-}+\delta_{\sigma_b,+}R_{k_b,k}^{+})\\
&+\sum_{k<k_L,k_b}\phi^*_g(x)\phi_{\sigma_b,k_b}(x')e^{-i\frac{g^2+k_b^2}{2}t}L_{g,k}(\delta_{\sigma_b,-}L_{k_b,k}^{-}+\delta_{\sigma_b,+}L_{k_b,k}^{+})\nn \\
%&=(\sum_{k_n<k_R,k_b}+\sum_{k_n<k_L,k_b})\phi_g(x)\phi_{+,k_b}^*(x')e^{-i\frac{g^2+k_b^2}{2}t}R_{g,k_n}R_{k_b,k_n}^{+}\\
%&+\sum_{k_L<k_n<k_R,k_b}\phi_g(x)\phi_{-,k_b}^*(x')e^{-i\frac{g^2+k_b^2}{2}t}R_{g,k_n}R_{k_b,k_n}^{-}\\
&= (g,\sum_{k_b})(x,x')_1 + (g,\sum_{k_b})(x,x')_2 \;,
\end{align}
in terms of the quantities
\bea
&&\hspace*{-1.3cm}(g,\sum_{k_b})(x,x')_1 = (\sum_{k<k_R,k_b}+\sum_{k<k_L,k_b})\phi^*_g(x)\phi_{+,k_b}(x')e^{-i\frac{g^2+k_b^2}{2}t}\frac{R_{g,k}}{\ell}\frac{2^{3/2} k\, k_b}{(k^2-k_b^2)\sqrt{g^2+k_b^2+\frac{2g}{\ell}}} \label{piece1}\\
&&\hspace*{-1.3cm}(g,\sum_{k_b})(x,x')_2 = \frac{1}{\sqrt{2}}\sum_{k_L<k<k_R}\phi^*_g(x)\phi_{-,k}(x')e^{-i\frac{g^2+k^2}{2}t}R_{g,k} \;, \label{piece2}
\eea
where we have used the explicit expressions of the overlaps $R_{k_b,k_n}^{\pm}$ and $L_{k_b,k_n}^{\pm}$ in Eq. (\ref{overlaps}).

It is easy to take the large $\ell$ limit of the second piece (\ref{piece2}), which gives
\be 
(g,\sum_{k_b})(x,x')_2 =
-ge^{g|x|}\int_{k_L}^{k_R}\frac{dk}{\pi}\sin(k x')\frac{k}{g^2+k^2}e^{-i\frac{g^2+k^2}{2}t} \;.
\ee
At large times, this term decays to zero as $1/t$ modulated by oscillations at $k_R$ and $k_L$. To take the large $\ell$
limit of the first piece (\ref{piece1}) requires again to introduce a contour integral because of the pole at $k=k_b$.
The contours $\gamma_R$ and $\gamma_L$ are defined such that $\int_{\gamma_{R/L}}=\int_{0}^{i\epsilon}+\int_{i\epsilon}^{i\epsilon +k_{R/L}}+\int_{i\epsilon +k_{R/L}}^{k_{R/L}}$. Let us introduce the notations
\bea
&&\hspace*{0.5cm} i_{k}(x)=\frac{k\cos(kx)+g\sin(k|x|)}{\sqrt{g^2+k^2}}\\
&&\hspace*{-0.5cm} I(k_b,x')=2[\int_{\gamma_R}\frac{dk}{\pi}+\int_{\gamma_L}\frac{dk}{\pi} \\
&& +(\frac{g}{k_b}+i)(\Theta(k_b<k_R)+\Theta(k_b<k_L))\, {\rm Res}_{k=k_b}]\frac{i_{k_b}(x')}{\sqrt{k_b^2+g^2}}\frac{k}{g^2+k^2}\frac{k k_b}{k^2-k_b^2} \nn 
\eea
where $I(k_b,x')$ is a continuous function except for $k_b=k_{L/R}$ where is has a logarithmic divergence.
Using the usual arguments we obtain
\be 
(g,\sum_{k_b})(x,x')_1 =-ge^{g|x|}e^{-i\frac{g^2}{2}t} \int_0^{\infty}\frac{dk_b}{\pi}e^{-i\frac{k_b^2}{2}t}I(k_b,x') \;,
\ee
and one can argue that the large time limit of this term vanishes. 
Finally the last term can be computed using $(\sum_{k_a},g)(x,x')=(g,\sum_{k_b})^*(x',x)$ with a similar conclusion.

%Now, using the trick that allows us to compute continuous limit of sum with poles we get
%\begin{align}
%&\underset{\ell\to\infty}{\simeq}-2ge^{g|x|}\int_0^{\infty}\frac{dk_b}{\pi}[\int_{\eta_R}\frac{dk}{\pi}+\int_{\eta_L}\frac{dk}{\pi}+(\frac{g}{k_b}+i)(\chi_{k_b<k_R}+\chi_{k_b<k_L})Res_{k=k_b}]\\
%&\times\frac{i_{k_b}(x')}{\sqrt{k_b^2+g^2}}e^{-i\frac{g^2+k_b^2}{2}t}\frac{k}{g^2+k^2}\frac{k k_b}{k^2-k_b^2}-ge^{g|x|}\int_{k_L}^{k_R}\frac{dk}{\pi}\sin(k x')\frac{k}{g^2+k^2}e^{-i\frac{g^2+k^2}{2}t} \\
%&=-ge^{g|x|}e^{-i\frac{g^2}{2}t}\left( \int_0^{\infty}\frac{dk_b}{\pi}e^{-i\frac{k_b^2}{2}t}I(k_b,x')+\int_{k_L}^{k_R}\frac{dk}{\pi}\sin(k x')\frac{k}{g^2+k^2}e^{-i\frac{k^2}{2}t}\right)\\
%&\underset{t\to\infty}{\simeq}0
%\end{align}

%&=\sum_{k_n,k_a}\phi(x')f_{k_a}(x)e^{i\frac{g^2+k_a^2}{2}t}(R_{k_n}\int_0^{\ell/2}dy'\sqrt{\frac{4}{\ell}}f_{k_a}(y')\sin(k_n y')+L_{k_n}\int_{-\ell/2}^0dy'\sqrt{\frac{4}{\ell}}f_{k_a}(y')\sin(k_n y'))\\
%&=\sum_{k_n,k_a}\phi(x')f_{k_a}(x)e^{i\frac{g^2+k_b^2}{2}t}2R_{k_n}\frac{2^{3/2}}{\ell}\frac{k_n k_a}{(k_n^2-k_a^2)\sqrt{g^2+k_a^2+\frac{2g}{\ell}}}\\
%&\underset{\ell\to\infty}{\simeq}-4ge^{g|x'|}\int_0^{\infty}\frac{dk_a}{\pi}[\int_{i\epsilon}^{\infty+i\epsilon}\frac{dk}{\pi}+(\frac{g}{k_a}+i)Res_{k=k_a}]\frac{f_{k_a}(x)}{\sqrt{k_a^2+g^2}}e^{-i\frac{g^2+k_a^2}{2}t}\frac{k}{g^2+k^2}\frac{k k_a}{k^2-k_a^2}\\
%&=-4ge^{g|x'|}e^{-i\frac{g^2}{2}t}\int_0^{\infty}\frac{dk_a}{\pi}e^{-i\frac{k_a^2}{2}t}F(k_a,x)\\
%&\underset{t\to\infty}{\simeq}0
%\end{align}

\setcounter{equation}{0}
\setcounter{figure}{0}
\renewcommand{\theequation}{D\arabic{equation}}
\renewcommand{\thefigure}{D\arabic{figure}}

\section{Details for finite temperature} \label{app:finiteT}

In this appendix we give some details of the finite temperature calculation. One starts 
from the initial kernel defined in Section \ref{sec_finiteT}. The formula \eqref{KernelLR} 
for $K_{R/L}(x,x',t)$ becomes
\begin{align}
K_{R/L}(x,x',t)&=\sum_{\sigma_a = \pm,k_a\in \Lambda_{\sigma_a}}\sum_{\sigma_b=\pm,k_b\in \Lambda_{\sigma_b}}\sum_{k\in\Lambda_{-}}f_{L/R}(k)\phi^*_{\sigma_a,k_a}(x)\phi_{\sigma_b,k_b}(x')e^{i(E(k_a)-E(k_b))t}\nn\\
&\times \left(\frac{1}{\sqrt{2}}\delta_{\sigma_a,-}\delta_{k,k_{a}}\pm\delta_{\sigma_a,+}\frac{2^{3/2}}{ \ell}\frac{kk_{a}}{(k^2-k_{a}^2)\sqrt{g^2+k_{a}^2+\frac{2g}{ \ell}}}\right)\nn\\
&\times \left(\frac{1}{\sqrt{2}}\delta_{\sigma_b,-}\delta_{k,k_{b}}\pm\delta_{\sigma_b,+}\frac{2^{3/2}}{ \ell}\frac{kk_{b}}{(k^2-k_{b}^2)\sqrt{g^2+k_{b}^2+\frac{2g}{ \ell}}}\right) \;,
\end{align}
where the only difference with \eqref{KernelLR} is the introduction of the Fermi factors $f_{L/R}(k)$
and the fact that the sum over $k$ extends over the entire lattice $\Lambda_-$. Expanding the terms
in parenthesis one finds an expression similar to \eqref{exact_K}. The obtained expression defines the terms $A,B,C,D$ as in \eqref{K_ABCD}.

\begin{figure}[t] 
\includegraphics[width=\linewidth]{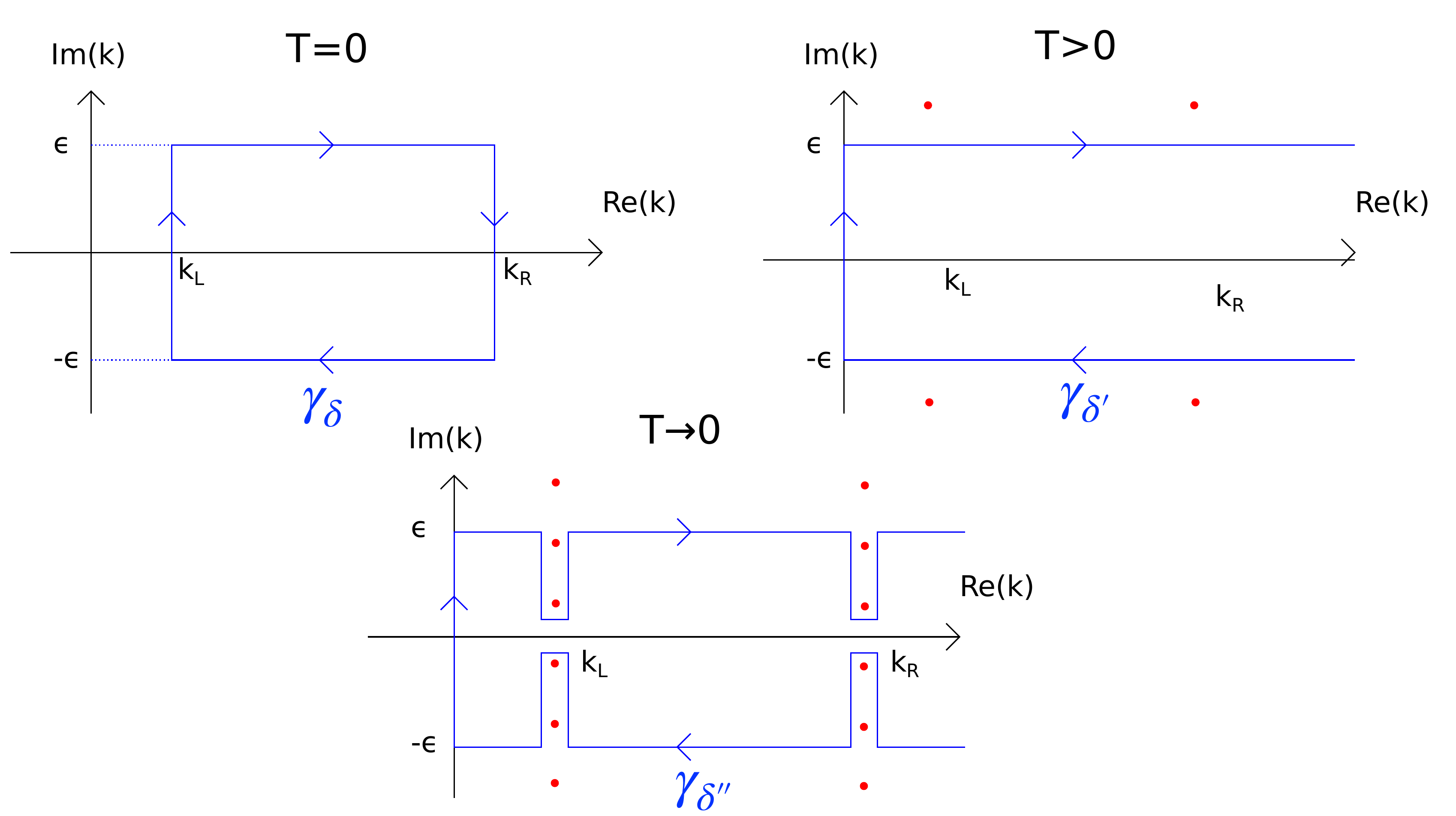}
\caption{Modification of the contour of integration for nonzero temperature. The top left part shows the zero temperature contour $\gamma_{\delta}$, while the top right panel shows the contour $\gamma_{\delta'}$ used at finite temperature. The red dots are the poles of the Fermi functions which, at low temperature, are located approximately along two vertical lines going through $k_L$ and $k_R$. As explained in the text the residues at these poles can be used to recover the results in the $T\to 0$ limit, using the contour $\gamma_{\delta ''}$ shown on the lower panel.}\label{FiniteTcontour}
\end{figure}

\subsection{Large $\ell$ limit} 

$\quad$ \\ 

The large $\ell$ limit is easy to perform on the term $A$ and one
obtains
\bea \label{AT} 
&&A_{L/R}(x,x')=\int_0^{\infty}\frac{dk}{2 \pi}f_{L/R}(k) \sin(kx)\sin(kx')\nn
\eea
which is the finite temperature generalization of \eqref{ref_sine}.

 For the term $C$ one needs again to use the contour integral trick. The contour $\gamma_\delta$ in Fig. \ref{Fig_contour_exact} 
 is now replaced by the semi-infinite rectangular contour $\gamma_{\delta'}$ with horizontal width $2 \epsilon$ shown in Fig. \ref{FiniteTcontour} (top right panel). 
 A formula analogous to
 \eqref{Cexact} can be written where the integrand in the first term now contains the additional factor 
 $f_R(k)-f_L(k)$ while the sum in the second term contains the additional factor 
 $f_R(k_b)-f_L(k_b)$ and $k_b$ is now summed over the whole lattice $\Lambda_+$. This formula is valid however only provided the contour $\gamma_{\delta'}$
 does not enclose a pole of the Fermi factors. Recall that the Fermi factor $f_{L/R}(k)$ has 
 poles at $k = \pm k_n$ with 
\bea \label{def_kn} 
k_n^{L/R} = \sqrt{ k_{L/R}^2 + 2 (2n+1) i \pi T} \quad, \quad n \in \mathbb{Z} \;. 
\eea
Hence we need to choose 
 \be \label{epsbound} 
 \epsilon < \sqrt{ k_{L/R}^2 + 2 i \pi T} \;,
 \ee 
which we will assume from now on. The limit $\ell \to +\infty$ can now be performed and as before (see Section \ref{sec:thermo})
the contribution of the lower half of $\gamma_\delta'$ (i.e., for ${\rm Im}(k)<0$) vanishes in that limit. This leads to
the finite temperature formula for the term $C$ at infinite $\ell$
\bea \label{CTT} 
&& \lim_{\ell\to\infty}  C(x,x',t)  = \int_0^{+\infty} \frac{dk_b}{\pi} \bigg[ 
\int_{\eta_{\infty}}\frac{dk}{\pi} \left(f_R(k)-f_L(k)\right)\frac{h_{x,x',t}(k,k_b)}{k-k_b}  \bigg] \nonumber \\
&& \hspace*{2.5cm}  + \int_{0}^{\infty} \frac{dk_b}{\pi}\left(f_R(k_b)-f_L(k_b)\right) \left( i + \frac{g}{k_b} \right) h_{x,x',t}(k_b,k_b) 
\eea 
where we denote $\gamma_\epsilon$ the integration contour $\int_{\gamma_\epsilon}= \int_{0}^{+i \epsilon} + \int_{+i \epsilon}^{\infty+i \epsilon}$,
which is the finite temperature analog of \eqref{res33} and $h_{xx't}$ is defined in \eqref{ff}. 
\\

The analysis of the term $B$ proceeds along the same line as in the zero-temperature case (see Section \ref{App_largel}).
The term $B$ is the sum of two parts as in \eqref{Bsum}. The starting formula for $B_R^{\text{off-diag}}(x,x',t)$
(and similarly for $B_L$) is \eqref{BR_offdiag_1} where the sum over $k$ is over the whole lattice $\Lambda_-$
and the Fermi factor $f_R(k)$ has been inserted. One then uses the contour integral trick 
with the same modifications of the contour as explained above (see Fig. \ref{FiniteTcontour}). 
This leads to the finite temperature analog of the formula \eqref{BBB} which reads 
\bea \label{BBB2} 
 && B_R^{\text{off-diag}}(x,x',t)= \int_0^{\infty}\int_0^{\infty}\frac{dk_a}{\pi}\frac{dk_b}{\pi} F_{x,x'}(k_a,k_b) e^{i(E(k_{a})-E(k_{b}))t}\nn\\
&& 
+ \left( \int_{0}^{i\epsilon} \frac{dk}{\pi} + \int_{+i\epsilon}^{+\infty+i\epsilon}\frac{dk}{\pi}  \right) f_R(k) \frac{k^2}{(k_a^2-k^2)(k_b^2-k^2)} \\
&& -2i \bigg[  (-\frac{1}{2}+\frac{i}{2}\frac{g}{k_a}){\rm Res}_{k=k_a}
+ (-\frac{1}{2}+\frac{i}{2}\frac{g}{k_b})Res_{k=k_b} \bigg] 
f_R(k) \frac{k^2}{(k_a^2-k^2)(k_b^2-k^2)} \nn
\eea
where $F_{x,x'}(k_a,k_b)$ is defined in \eqref{def_Fkakb} and $B_L^{\rm off-diag}(x,x') $ is obtained by changing $f_R \to f_L$.
The starting formula for $B_R^{\text{diag}}(x,x',t)$
is \eqref{BRR} where the sum over $k$ is over the whole lattice $\Lambda_-$
and the Fermi factor $f_R(k)$ has been inserted. The same argument shows, as in Appendix \ref{App_largel}, that the first part 
$B_R^{\text{diag},1}(x,x',t)$ vanishes at large $\ell$. The second part 
$B_R^{\text{diag},2}(x,x',t)$ is given by formula \eqref{BR_v} where 
the factor $f_R(k)$ has been inserted in the contour integral which is now over $\Gamma_1'$ 
which is the union of small circles oriented clockwise centered around $k=k_a$ with $0\leq k_a < +\infty$.
When computing the residues of the double pole one must be careful that the Fermi factor has been
inserted. Hence Eq. \eqref{double} is replaced by 
\bea 
 \int_{\Gamma'_1} \frac{d k}{2 \pi} \frac{\ell}{e^{i k \ell}-1} \frac{f_R(k) v(k,k_a)}{(k-k_a)^2}
&=& - i \frac{d}{dk} \left( f_R(k) v(k,k_a) \frac{\ell}{e^{i k \ell}-1} \right) \Bigg|_{k=k_a} \nn \\
&=& \frac{k_a^2 + g^2}{16 k_a^2} f_R(k_a) \ell^2 + O(\ell)
\eea 
This finally leads to the large $\ell$ limit of the $B_R^{\rm diag}$ term as 
\be \label{BdiagTT} 
B_R^{\rm diag}(x,x')  = B_R^{\rm diag,2}(x,x')  =  \int_0^{+\infty} \frac{dk_a}{\pi} f_R(k_a) F_{xx'}(k_a,k_a) \frac{k_a^2 + g^2}{4 k_a^2} \;,
\ee 
and $B_L^{\rm diag}(x,x') $ is obtained by changing $f_R \to f_L$.
\\

To summarize, the infinite $\ell$ limit of the kernel at finite temperature is equal to the following sum
\bea \label{KTsum} 
&& K(x,x',t) = A_L(x,x')+ A_R(x,x') + C(x,x',t)  + C(x',x,t)^*  \\
&& + B_R^{\text{off-diag}}(x,x',t) + B_L^{\text{off-diag}}(x,x',t) + B_R^{\rm diag}(x,x') +
B_L^{\rm diag}(x,x') \nn
\eea 
where $A_{L/R}$ is given in \eqref{AT}, $C$ is given in \eqref{CTT}, $B_{R/L}^{\text{off-diag}}$ is given in \eqref{BBB2} and
$B_{R/L}^{\rm diag}$ is given in \eqref{BdiagTT}. Note that $A_{L/R}$ and $B_{L/R}^{\rm diag}$ are time independent.
\\

{\bf Remark}. In the above formula we have assumed that $\epsilon$ satisfies the bound in \eqref{epsbound} such that the contour $\gamma'_\delta$ does not enclose any pole of the Fermi factors. However these poles get
closer to the real axis when temperature goes to zero. In other words the bound \eqref{epsbound}
becomes $\epsilon < \frac{\pi T}{k_R}$ at low $T$. So one can ask how is the $T=0$ recovered.
The answer is illustrated in the Fig. \ref{FiniteTcontour}. The contour $\gamma_{\delta'}$ can be deformed into the contour 
$\gamma_{\delta''}$
as shown in the third panel in Fig. \ref{FiniteTcontour}. Consider for instance the term $C$. One notes that as $T\to 0$ the contribution of the part of the
contour to the left of $k_L$ and the part of the contour to the right of $k_R$ vanishes. Hence the result is indentical in that
limit to the one obtained from the previously considered
contour $\gamma_\delta$ (see first panel in Fig. \ref{FiniteTcontour}), and to the $T=0$ result.
 
\subsection{Large time limit: regime of the NESS} 

$\quad$ \\

In the large time limit (once the infinite $\ell$ limit has been taken) the terms $B_{R/L}^{\text{off-diag}}(x,x',t)$ in \eqref{KTsum},
as well as the contour integral part of $C(x,x',t)$ are found to vanish, by similar arguments as for $T=0$. One is thus left 
with 
\be \label{KTsum2} 
 K_\infty(x,x') = A_L(x,x')+ A_R(x,x') + C_\infty(x,x')  + C_\infty(x',x)^* + B_R^{\rm diag}(x,x') +
B_L^{\rm diag}(x,x')
\ee
with 
\begin{align} 
C_\infty(x,x')=\frac{1}{2}\int_{0}^{\infty}\frac{dk}{\pi}\left(f_R(k)-f_L(k)\right)(g+ik)\sin(kx)\frac{k\cos(kx')+ g\sin(k |x'|)}{g^2+k^2} \;,
\end{align}
given by the residue part in \eqref{CTT}.

Putting all terms together, using also \eqref{AT} and \eqref{BdiagTT}
we obtain the kernel in the NESS at finite temperature for $g>0$ as
\bea \label{kernelfiniteT} 
&&K_\infty(x,x') =\int_0^{\infty}\frac{dk}{2\pi}(f_R(k)+f_L(k))\sin(k x)\sin(k x')\\
&&\quad\quad\quad\quad\quad + (f_R(k)+f_L(k))\frac{(k\cos(kx)+g\sin(k|x|))(k\cos(kx')+g\sin(k|x'|))}{g^2+k^2} \nn \\
&&\quad\quad\quad\quad\quad +(f_R(k)-f_L(k))(g+ik)\sin(k x)\frac{k\cos(kx')+g\sin(k|x'|)}{g^2+k^2} \nn \\
&&\quad\quad\quad\quad\quad +(f_R(k)-f_L(k))(g-ik)\sin(k x')\frac{k\cos(kx)+g\sin(k|x|)}{g^2+k^2} \;. \nn
\eea
We have checked that, in the limit $T\to 0$, this formula (\ref{kernelfiniteT}) coincides with the expression obtained for the zero-temperature kernel in the NESS given in Eqs. (\ref{Kpospos_intro}) and (\ref{Kposneg_intro}). 
From the kernel \eqref{kernelfiniteT}, 
we can recover the result for the density and the current at finite temperature given in \eqref{densitysteadystate>0T} and \eqref{currentsteadystate>0T}.
\\

{\bf Large time decay to the NESS}. The decay in time to the NESS is different at finite $T$. We will not present the analysis for the full kernel
but only show the decay of the term $C(x,x',t)$. For fixed $T$ let us compute the large time behaviour of $\Delta C=C-C\infty$. One has, from
\eqref{CTT}, where the time independent residue part is cancelled, 
\be
\!\! \Delta C(x,x',t)=\int_0^{\infty}\frac{dk_b}{\pi}\int_{\gamma_{\epsilon}} \frac{dk}{\pi} \left(f_R(k)-f_L(k)\right) \sin(k x)\frac{k_b\cos(k_b x')+g\sin(k_b |x'|)}{g^2+k_b^2} \frac{k k_b}{k^2-k_b^2}e^{i(k^2-k_b^2)\frac{t}{2}}
\ee
Taking the time derivative leads to the following decoupling of integrals
\bea
\partial_t \Delta C(x,x',t) =\frac{i}{2}G_b(t)G(t)
\eea
where $G_b(t)$ is defined in \eqref{Gb} and
\bea
G(t)=\int_{\gamma_\epsilon} \frac{dk}{\pi} \left(f_R(k)-f_L(k)\right)k \sin(k x)  e^{\frac{i}{2} k^2  t} \;.
\eea
The large time behavior of $G_b(t)$ was obtained before, and now the behavior of 
$G(t)$ is dominated by the vicinity of $k= i p = 0$ on the vertical part of the contour, leading to
\bea
&& G_b(t) \simeq \frac{1}{g^2 \sqrt{2 \pi}} (1 + g |x'|) 
\frac{1}{(i t)^{3/2}}\\
&& G(t) \simeq - i \left(f_R(0)-f_L(0)\right) \frac{x}{t^{3/2}} \int_0^{+\infty} \frac{dp}{\pi} p^2  e^{- \frac{i}{2} p^2} 
= (-1+ i) \left(f_R(0)-f_L(0)\right) \frac{x}{2 \sqrt{\pi}  t^{3/2}} \;. \nn \\
\eea
In summary we obtain the decay (provided $\beta_R \mu_R \neq \beta_L \mu_L$)
\bea
\Delta C(x,x',t) \sim t^{-2} \;.
\eea
It is slower that the result obtained at zero temperature, where we found an oscillating $t^{-5/2}$ decay dominated
by $k=k_{L/R}$ (see Eq. (\ref{DeltaC})).
\\

It is also interesting to check how to recover the decay in the zero temperature limit. To this aim we first send $\epsilon$ to infinity in the contour $\gamma_{\epsilon}$. When moving the contour it will cross the poles from the Fermi function (in the upper half-plane) as in Fig. \eqref{FiniteTcontour}.
Taking into account the series of residues at these poles we obtain
\bea
G(t)&=&\int_{0}^{+i\infty}\frac{dk}{\pi} \left(f_R(k)-f_L(k)\right)k \sin(k x)  e^{\frac{i}{2} k^2  t} +\sum_{n=0}^{+\infty} 2iT\sin(k_n^L x)e^{i\frac{(k_n^L)^2}{2}t}\nn \\
&-& \sum_{n=0}^{+\infty} 2iT\sin(k_n^R x)e^{i\frac{(k_n^R)^2}{2}t}
\eea
where we recall that $k_n^{L/R}$'s are defined in Eq. (\ref{def_kn}). In the zero temperature limit the integral part vanishes since $f_R(k)-f_L(k)$ decays to zero. 
Using that in that limit $\frac{dk_n}{dn} \simeq \frac{2  i \pi T}{k_n}$ the two series 
converges towards the already known integrals $G_L(t)$ and $G_R(t)$ defined in \eqref{Cdecay}, leading to
\bea
&& G(t)\underset{T\to 0}{\simeq} \int_{k_L}^{k_L+i\infty}\frac{dk}{\pi}k\sin(k x) e^{i\frac{k^2}{2}t}- \int_{k_R}^{k_R+i\infty}\frac{dk}{\pi}k\sin(k x) e^{i\frac{k^2}{2}t} \nn \\
&&\hspace*{1cm} =G_L(t)- G_R(t)
\eea
which leads to the zero temperature result $\Delta C(T)\underset{T\to 0}{\to} \Delta C(T=0)$ and allows to match
the finite temperature decay to the zero temperature decay.
\\

\subsection{Large time limit: ray regime at fixed $\xi=\frac{x}{t}$} 

$\quad$ \\

In the ray regime the Fermi factor do not change the arguments
about the contours in Section \ref{sec:light}. Hence Eqs.\eqref{Kp_xixi} and \eqref{Kxi_m} turn into 
\bea 
 K^{+}_{\xi}(y,y') &=&  \int_0^{\infty} \frac{dk}{\pi}f_R(k) \cos(k(y-y')) \Theta(\xi) + \int_0^{\infty} \frac{dk}{\pi}f_L(k) \cos(k(y-y')) \Theta(-\xi) 
\nn \\
& - & {\rm sign}(\xi) \int_{0}^{\infty} \frac{dk}{2 \pi}(f_R(k)-f_L(k)) T(k) 
e^{- i {\rm sign}(\xi) k (y-y') } \Theta(k - |\xi|)  \\
K^{-}_{\xi}(y,y') &=&
i \,  {\rm sign}(\xi) \int_{0}^{\infty} \frac{dk}{2 \pi}(f_R(k)-f_L(k)) \frac{g k}{g^2 + k^2} e^{- i {\rm sign}(\xi) k (y+y') }  \Theta(k - |\xi|) \;.
\eea  
As a consequence we obtain the expressions for the density \eqref{tilderhoT} and the current \eqref{tildeJT} in the ray-regime 
at finite temperature.

%As one can see on Fig. (\ref{FiniteTcontour}) $\epsilon$ needs to be small enough so that no pole of the Fermi function $f_{L/R}$ is inside the contour. It is also possible to recover the zero temperature result as $T$ goes to zero. In this case the Fermi functions poles get closer to each other forming an almost vertical line above $k_L$ and $k_R$. The contour has to be deformed as in the figure so that the poles don't cross the contour. Finally the part of the contour such that $Re(k)<k_L$ or $Re(k)>k_R$ decays to zero leaving only the interior contour that was here at zero temperature.

\setcounter{equation}{0}
\setcounter{figure}{0}
\renewcommand{\theequation}{E\arabic{equation}}
\renewcommand{\thefigure}{E\arabic{figure}}

\section{Energy current} \label{sec:heat}

In this section we give some details about the calculation of the energy current. 
We recall its definition
\bea \label{def_Ecurrent2}
J_Q^{L/R}(x,t) =  \sum_{n=1}^{\infty}f_{L/R}(k_n) 
{\rm Im} \left( 
 (\hat H_g \psi_{L/R}^n)(x,t)^* \partial_x \psi_{L/R}^n(x,t) \right) \;,
\eea
where $k_n=\frac{2 \pi n}{\ell}$ and $\hat H_g$ is the single-particle Hamiltonian with a delta-impurity in Eq. (\ref{def_H}). 
We recall that the evolution of a given state is given by
\bea
\psi_{R}^n(x,t)=\int_0^{\ell/2}dy \sum_{\sigma, k\in\Lambda_{\sigma}}\phi_{\sigma,k}(x)\phi^*_{\sigma,k}(y)e^{-iE(k)t}\psi_n(y,0) \;,
\eea
and similarly for $\psi_{L}^n(x,t)$. 
Substituting this expression in Eq. (\ref{def_Ecurrent2}), we get the time evolution of the heat current 
\bea
J_Q^R(x,t)&={\rm Im}[\sum_{k_n}f_R(k_n)\sum_{\sigma_E,k_E}\sum_{\sigma_j,k_j}\phi^*_{\sigma_E,k_E}(x)E(k_E)\partial_x\phi_{\sigma_j,k_j}(x) e^{i(E(k_E)-E(k_j))t}\\
& \times \int_0^{\ell/2}dy\phi_{\sigma_E,k_E}(y)\psi_n(y,0)\int_0^{\ell/2}dy\phi^*_{\sigma_j,k_j}(y)\psi_n(y,0)] \;. \nn
\eea
where, here and below, $k_n \in \Lambda_-$ and $k_E,k_j \in \Lambda_+$.
Injecting the overlaps as in the calculation of the kernel (see Eq. \ref{KR_bis}) we obtain
\bea
J_Q^{L/R}(x,t)&={\rm Im}[\sum_{k_n}f_{L/R}(k_n)\sum_{\sigma_E,k_E}\sum_{\sigma_j,k_j}\phi^*_{\sigma_E,k_E}(x)E(k_E)\partial_x\phi_{\sigma_j,k_j}(x) e^{i(E(k_E)-E(k_j))t} \nn \\
&(\frac{1}{\sqrt{2}}\delta_{\sigma_E,-1}\delta_{k_E,k_n}\pm\delta_{\sigma_E,1}\frac{2^{3/2}}{\ell}\frac{k_E k_n}{(k_n^2-k_E^2)\sqrt{g^2+k_E^2+\frac{2g}{\ell}}})  \\
&(\frac{1}{\sqrt{2}}\delta_{\sigma_j,-1}\delta_{k_j,k_n}\pm\delta_{\sigma_j,1}\frac{2^{3/2}}{\ell}\frac{k_j k_n}{(k_n^2-k_j^2)\sqrt{g^2+k_j^2+\frac{2g}{\ell}}})] \;. \nn
\eea
As before for the computation of the kernel, we decompose $J_Q(x,t) = J_Q^{L}(x,t) + J_Q^{R}(x,t)$ in four different contributions
\bea
&& J_Q(x,t)=\underbrace{{\rm Im}[\sum_{k_n}\frac{1}{\ell}\left(f_R(k_n)+f_L(k_n)\right)E(k_n)\sin(k_n x) k_n \cos(k_n x)]}_{A^Q}\\
&&+{\rm Im}[2 \sum_{k_n,k_E,k_j}(\frac{2}{\ell})^3\left(f_R(k_n)+f_L(k_n)\right) E(k_E)\frac{k_E\cos(k_E x)+g\sin(k_E|x|)}{g^2+k_E^2+\frac{2g}{\ell}}\nn \\
&&\underbrace{\times k_j\frac{-k_j\sin(k_j x)+g \, {\rm sgn}(x)\cos(k_j x)}{g^2+k_j^2+\frac{2g}{\ell}}\frac{k_Ek_jk_n^2}{(k_n^2-k_E^2)(k_n^2-k_j^2)}e^{i(E(k_E)-E(k_j) t}]}_{B^Q} \nn \\
&&+{\rm Im}[\sum_{k_n,k_j}(\frac{2}{\ell})^2\left(f_R(k_n)-f_L(k_n)\right) E(k_n) \sin(k_n x) \nn \\
&& \times\underbrace{k_j \frac{-k_j\sin(k_j x)+g \, {\rm sgn}(x)\cos(k_j x)}{g^2+k_j^2+\frac{2g}{\ell}} e^{i(E(k_n)-E(k_j))t}\frac{k_j k_n}{k_n^2-k_j^2}]}_{C^Q} \nn \\
&&+{\rm Im}[\sum_{k_n,k_E}(\frac{2}{\ell})^2\left(f_R(k_n)-f_L(k_n)\right)k_n \cos(k_n x) \nn \\
&&\times \underbrace{ E(k_E) \frac{k_E\cos(k_E x)+g \, \sin(k_E|x|)}{g^2+k_E^2+\frac{2g}{\ell}}e^{i(E(k_E)-E(k_n))t}\frac{k_E k_n}{k_n^2-k_E^2}]}_{D^Q} \;. \nn
\eea
The term $A_Q$ is trivially zero since the summand is real. The term $B_Q$ can again be split into off-diagonal and diagonal part.
The off-diagonal part vanishes in the large time limit by analogy with the calculation of the kernel (see Section \ref{sec:Blargetime}). The diagonal part, i.e.
keeping only $k_E=k_j$ in the sum is non zero but becomes zero when taking the imaginary part.

Let us now consider the crossed terms ($C^Q= C^Q_L + C^Q_R$ and $D^Q=D^Q_L + D^Q_R$):
\bea
&&C^Q_{R/L}= \pm {\rm Im}\sum_{k_n,k_j}(\frac{2}{\ell})^2f_{R/L}(k_n)  k_j E(k_n) \sin(k_n x)\frac{-k_j\sin(k_j x)+g{\rm sgn}(x)\cos(k_j x)}{g^2+k_j^2+\frac{2g}{\ell}}\\
&& \quad\quad\quad\quad\quad\quad\quad\times e^{i(E(k_n)-E(k_j))t}\frac{k_j k_n}{k_n^2-k_j^2}]\nn \\
&&\underset{\ell\to\infty}{\simeq} \pm {\rm Im} [\int_0^{\infty}\frac{dk_j}{\pi}[\int_{\gamma_{\epsilon}}\frac{dk_n}{\pi}f_{R/L}(k_n)-2i(-\frac{1}{2}+\frac{i}{2}\frac{g}{k_j})f_{R/L}(k_j){\rm Res}_{k_j=k_n}]\\
&&k_jE(k_n) \sin(k_n x)\frac{-k_j\sin(k_j x)+g{\rm sgn}(x)\cos(k_j x)}{g^2+k_j^2}e^{i(E(k_n)-E(k_j))t}\frac{k_j k_n}{k_n^2-k_j^2}] \nn
\eea
Here $\gamma_{\epsilon}$ is the contour such that $\int_{\gamma_{\epsilon}}=\int_{0}^{i\epsilon}+\int_{i\epsilon}^{+\infty}$. 
In the large time limit the contour integral vanishes and 
\be
\lim_{t\to\infty} C_{R/L}^Q % \pm % Re[-i\int_0^{\infty}\frac{dk}{2\pi}f_{L/R}(k)(g+ik)E(k) k\sin(kx)\frac{-k\sin(k x)+g{\rm sgn}(x)\cos(k x)}{g^2+k^2}]\\
%&&
=\pm \int_0^{\infty}\frac{dk}{2\pi}f_{L/R}(k)kE(k) k\sin(kx)\frac{-k\sin(k x)+g{\rm sgn}(x)\cos(k x)}{g^2+k^2}]
\ee
A similar method gives $D_{L/R}^Q$.
\bea
&&D_{L/R}^Q
%\pm Re[-i\sum_{k_n,k_E}(\frac{2}{\ell})^2f_{L/R}(k_n) E(k_E) \frac{k_E\cos(k_E x)+g\sin(k_E|x|)}{g^2+k_E^2+\frac{2g}{\ell}}k_n \cos(k_n x)\\
%&&\quad\quad\quad\quad\quad\quad\quad\times e^{i(E(k_E)-E(k_n))t}\frac{k_E k_n}{k_n^2-k_E^2}]\\
\underset{\ell\to\infty}{\simeq}\pm {\rm Im}[\int_0^{\infty}\frac{dk_E}{\pi}[\int_{\eta_{L/R}}\frac{dk_n}{\pi}f_{L/R}(k_n)+2i(-\frac{1}{2}-\frac{i}{2}\frac{g}{k_E})f_{L/R}(k_E){\rm Res}_{k_E=k_n}]\\
&&E(k_E) \frac{k_E\cos(k_E x)+g\sin(k_E|x|)}{g^2+k_E^2+\frac{2g}{\ell}}k_n \cos(k_n x)e^{i(E(k_E)-E(k_n))t}\frac{k_E k_n}{k_n^2-k_E^2}]\nn \\
&&\underset{t\to\infty}{\simeq}
%\pm Re[-i\int_0^{\infty}\frac{dk}{2\pi}f_{L/R}(k)(g-ik)E(k)\frac{k\cos(kx)+g\sin(k|x|)}{g^2+k^2} k\cos(k x)]\\
%&&=
\mp \int_0^{\infty}\frac{dk}{2\pi}f_{L/R}(k)kE(k) k\cos(k x)\frac{k\cos(kx)+g\sin(k|x|)}{g^2+k^2}
\eea
After summing these contributions, $J^Q_\infty = C^Q_R+C^Q_L+D^Q_R+D^Q_L$ we obtain the result for the asymptotic energy current
in the NESS \eqref{def_JQ}.
%\bea
%J^Q_\infty =-\int_0^{\infty}\frac{dk}{2\pi}(f_{R}(k)-f_{L}(k))k E(k)\frac{k^2}{g^2+k^2}
%\eea
%\\

\setcounter{equation}{0}
\setcounter{figure}{0}
\renewcommand{\theequation}{F\arabic{equation}}
\renewcommand{\thefigure}{F\arabic{figure}}

\section{Low temperature expansions} \label{App_lowT}

In this appendix we perform the low temperature expansions for the energy and particle currents.

\subsection{Energy current} 

$\quad$ \\

Let us compute the energy current $J_{Q, \infty}$ as given by formula \eqref{def_JQ}, recalling that here $E(k)=k^2/2$. 
Using the variable $\epsilon=\frac{k^2}{2}$ it takes the form
$J_{Q, \infty}= J_{Q, \infty}^L - J_{Q, \infty}^R$ with
\be 
J_{Q, \infty}^{L/R} = \int_{0}^{+\infty} d\epsilon  \frac{h(\epsilon)}{1+e^{\beta_{L/R}(\epsilon-\mu_{L/R})}}
\ee 
where $h(\epsilon)=\frac{\epsilon^2}{\pi(g^2+2\epsilon)}$. At low temperature one one can use the Sommerfeld expansion
\bea \label{sommerfeld} 
\int_{0}^{+\infty}d\epsilon \frac{h(\epsilon)}{1+e^{\beta(\epsilon-\mu)}}\underset{\beta\to\infty}{\simeq}\int_{0}^{\mu}d\epsilon h(\epsilon)+\frac{\pi^2 h'(\mu)}{6\beta^2} + \frac{7 \pi^4}{360 \beta^4} h'''(\mu) + O(\beta^6) 
\eea
and one obtains
\be 
J_{Q, \infty} = - \int_{\mu_L}^{\mu_R} d\epsilon \frac{\epsilon^2}{\pi(g^2+2\epsilon)} +  \frac{\pi}{6} \left( T_L^2 \frac{2 \mu_L (g^2 + \mu_L)}{(g^2 + 2 \mu_L)^2}
- T_R^2 \frac{2 \mu_R (g^2 + \mu_R)}{(g^2 + 2 \mu_R)^2} \right) + O(T_R^4,T_L^{4}) \;.
\ee

In the case $\mu_L=\mu_R= \frac{k_F^2}{2}$ it simplifies (using \eqref{sommerfeld} up to order $\beta^{-4}$) into the formula \eqref{JQ_lowT}
given in the text.
%\be 
%J_Q = \frac{\pi \cos^2(\alpha)}{12}(T_L^2-T_R^2) - \frac{7 \pi^3}{30} \frac{g^4}{(g^2 + k_F^2)^4} (T_L^4-T_R^4)
%+ O(T_R^6,T_L^{6}) \quad , \quad \cos^2(\alpha)=\frac{k_F^2(2g^2+k_F^2)}{(g^2+k_F^2)^2} 
%\ee 
%$h'(\frac{k_F^2}{2})=\frac{k_F^2(2g^2+k_F^2)}{2\pi(g^2+k_F^2)^2}$
In the absence of impurity, for $g=0$, one has $h(\epsilon)= \frac{\epsilon}{2 \pi}$ and all terms beyond $O(\beta^{-2})$ in the Sommerfeld expansion vanish.
To obtain the subleading terms which are exponential in $\beta$ one uses the low temperature expansion
\bea \label{sommerfeld2} 
\int_{0}^{+\infty}d\epsilon \frac{\epsilon}{2 \pi( 1+e^{\beta(\epsilon-\mu)})}
= - \frac{1}{2 \pi \beta^2} {\rm Li_2}(- e^{\beta \mu}) = 
\frac{\mu^2}{4 \pi} + \frac{\pi}{12 \beta^2}- \frac{1}{2 \pi \beta^2} e^{- \beta \mu} + O(e^{-2 \beta \mu}) \;,
\eea
where ${\rm Li_2}(z)= \sum_{k\geq 1}z^k/k^2$ is the di-logarithm function. This expansion (\ref{sommerfeld2}) leads to the formula \eqref{JQ_geq0} in the text.
%\be 
%J_Q|_{g=0} = \frac{ \mu_L^2 - \mu_R^2}{4 \pi} + \frac{\pi}{12}(T_L^2-T_R^2) - \frac{1}{2 \pi} (T_L^2 e^{- \mu_L/T_L} - T_R^2 e^{- \mu_L/T_R} ) 
%\ee 

\subsection{Particle current}

$\quad$ \\

For the particule current one has
$J_{\infty}= J_{\infty}^L - J_{\infty}^R$ with
\be 
J_{\infty}^{L/R} = \int_{0}^{+\infty} d\epsilon  \frac{\tilde h(\epsilon)}{1+e^{\beta_{L/R}(\epsilon-\mu_{L/R})}}
\ee 
where $\tilde h(\epsilon)=\frac{\epsilon}{\pi(g^2+2\epsilon)}$. Using the Sommerfeld expansion one obtains
\be 
J_{\infty} = - \int_{\mu_L}^{\mu_R} d\epsilon \frac{\epsilon}{\pi(g^2+2\epsilon)} +  \frac{\pi}{6} \left( T_L^2 \frac{g^2}{(g^2 + 2 \mu_L)^2}
- T_R^2 \frac{g^2}{(g^2 + 2 \mu_R)^2} \right) + O(T_R^4,T_L^{4})
\ee 
In the case $\mu_L=\mu_R= \frac{k_F^2}{2}$ it simplifies (using \eqref{sommerfeld} up to order $\beta^{-4}$) into the formula \eqref{Jinf_g}
given in the text.
%\be 
%J = \frac{\pi}{6}(T_L^2 -T_R^2) \frac{g^2}{(g^2 + k_F^2)^2} 
%+ \frac{7 \pi^3}{15} \frac{g^2}{(g^2 + k_F^2)^4} (T_L^4-T_R^4)
%+ O(T_R^6,T_L^{6}) 
%\ee 

In the absence of impurity, for $g=0$, one uses
\bea \label{sommerfeld2} 
\int_{0}^{+\infty}d\epsilon \frac{1}{2 \pi( 1+e^{\beta(\epsilon-\mu)})}
=  \frac{1}{2 \pi \beta} \log(1 + e^{\beta \mu}) = 
\frac{\mu}{2 \pi}  + \frac{1}{2 \pi \beta} e^{- \beta \mu} + O(e^{-2 \beta \mu}) \;,
\eea
which leads to the formula \eqref{Jinf_g0} given in the text. 
%\be 
%J|_{g=0} = \frac{ \mu_L - \mu_R}{2 \pi}  + \frac{1}{2 \pi} (T_L e^{- \mu_L/T_L} - T_R e^{- \mu_L/T_R} ) 
%\ee 

\section{Remarks on NESS and GGE}
\label{sec:GGE} 

For non-interacting fermions the prediction from the GGE takes the form for the density matrix
\be 
\hat D_{\rm GGE} = \frac{1}{Z_{\rm GGE} } e^{ \sum_\ell f_\ell c^\dagger_{\ell} c_{\ell}} \;,
\ee 
where here $\ell$ labels the eigenstates $|\varphi_\ell \rangle$ of the single particle post-quench Hamiltonian $\hat H_g$
and the $c^\dagger_\ell$ are the corresponding creation operators. Since it has a Gaussian form  
it leads to the kernel
\be  \label{KGGE} 
K_{\rm GGE} = \sum_\ell \langle c^\dagger_\ell c_\ell \rangle_{\rm GGE}  | \varphi_\ell \rangle \langle \varphi_\ell | \quad , \quad  
\langle c^\dagger_\ell c_\ell \rangle_{\rm GGE} = \nu_\ell = \frac{1}{1+ e^{-f_\ell}}  \;.
\ee 
In the GGE the coefficients $f_\ell$ (which should not be confused with the Fermi factors $f_{L/R}$) are determined so that the occupation numbers $\nu_\ell$ are equal to those in the 
initial state (which has density matrix $\hat D_0$). Hence one has 
\be 
\langle c^\dagger_\ell c_\ell \rangle_{\rm GGE}=\langle c^\dagger_\ell c_\ell \rangle_{t=0}= \langle \varphi_\ell | \hat D_0 | \varphi_\ell \rangle  \;.
\ee

In the present problem, for any finite size $\ell$ (i.e. before taking the thermodynamic limit),
the post-quench eigenstates are denoted $|\varphi_\ell \rangle= |\phi_{\sigma_a,k_a} \rangle$,
where $\sigma_a=\pm 1$ for respectively even and odd eigenstates given explicitly in \eqref{def_pm} and \eqref{def_lm}, 
and where $k_a$ belongs to either the even or odd lattices respectively, $k_a \in \Lambda^\pm$, see \eqref{def_lp} and \eqref{def_lm} respectively. 
The initial density matrix reads
\be 
\hat D_0 = \sum_{k_n}  \left[f_L(k_n) | \phi_n^L \rangle \langle \phi_n^L | + f_R(k_n) | \phi_n^R \rangle \langle \phi_n^R |   \right] \;,
\ee 
where the $| \phi_n^{L/R} \rangle$ are defined in \eqref{def_eigen} and $f_{L/R}$ are the Fermi factors for the left and
right sides of the system. Hence the GGE prediction (for any $\ell$) takes the form
\bea \label{KGGEprediction} 
K_{\rm GGE}(x,x') &=& \sum_{\sigma_a = \pm 1, k_a \in \Lambda_{\sigma_a}} 
\sum_{k_n} \Big( f_L(k_n) \langle \phi_{\sigma_a,k_a} | \phi_n^L \rangle \langle \phi_n^L | \phi_{\sigma_a,k_a} \rangle \\
&& + \; f_R(k_n) \langle \phi_{\sigma_a,k_a}  | \phi_n^R \rangle \langle \phi_n^R | \phi_{\sigma_a,k_a} \rangle \Big ) \phi_{\sigma_a,k_a}^*(x) \phi_{\sigma_a,k_a}(x')\;. \nn
\eea 

We can now compare \eqref{KGGEprediction} to the exact formula for $K(x,x',t)=K_R(x,x',t)+K_L(x,x',t)$
at finite $\ell$ given in \eqref{KR_bis} in the particular case of an initial condition at zero temperature. 
Clearly $K_{\rm GGE}(x,x')$ is obtained by retaining only the terms $k_a=k_b$ and $\sigma_a=\sigma_b$
in the double sum over post-quench eigenstates. This is also called the {\it diagonal approximation}
and sometimes denoted $K_d$ (the fact that the two coincide quite generally for a finite size system
or for trapped fermions was discussed in \cite{DLSM2019}). In the expression \eqref{exact_K} it corresponds to
keeping only the terms $A(x,x')$ (which correspond to $\sigma_a=\sigma_b=-1$) and 
the term denoted $B^{\rm diag}(x,x')$ in \eqref{Bsum} (which correspond to $k_a=k_b$ in the sum $\sigma_a=\sigma_b=1$). 
Hence one has, for any $\ell$,
\be  \label{KGGEAB} 
K_{\rm GGE}(x,x')  = A(x,x') + B^{\rm diag}(x,x')  \;,
\ee 
which is time independent by construction. This has a well defined $\ell=+\infty$ limit, which
is studied in this paper.

The important remark is that the GGE prediction \eqref{KGGEAB} in the $\ell=+\infty$ limit is
different from the result that we obtained for the NESS, i.e. $K_{\rm GGE}(x,x') \neq K_\infty(x,x')$. 
Indeed one has 
\be  \label{KGGEABC} 
K_{\infty}(x,x') =  K_{\rm GGE}(x,x') + C(x,x',t=+\infty) + C(x',x,t=+\infty)^*  \;.
\ee 
While $K_{\rm GGE}(x,x')$ is real and does not carry any current, the additional terms $C$ lead to a non zero current 
in the NESS. Although they are not strictly diagonal for finite $\ell$ they contain "almost diagonal" oscillating terms 
of the form $e^{i t ( E(k)-E(k_a))}$ where $k$ and $k_a$ do not belong to the same lattice.
Since $E(k)-E(k_a)$ can be of order $1/\ell^2$ at large $\ell$ (for some couples $(k, k_a)$) they do lead to a finite contribution in the NESS
where one takes $\ell \to +\infty$ first. 

The above result also implies that $\hat D_{\rm NESS} \neq \hat D_{\rm GGE}$. 
One can in principle obtain $\hat D_{\rm NESS}$ from our result for the kernel $K_{\infty}(x,x')$. Let us just 
indicate it in the simpler case $g=0$. In that case one has from \eqref{kernelfiniteT}
\be  \label{g00} 
K_{\infty}(x,x') = \int_{-\infty}^{+\infty} \frac{dk}{2 \pi} ( f_L(k) \theta(k) + f_R(k) \theta(-k) ) e^{- i k(x-x')} \;.
\ee 
Using that $c^\dagger_x c_{x'} =    \int \frac{dk}{2 \pi} c^\dagger_k c_k e^{- i k (x-x')}$
we obtain 
\be 
\hat D_{\rm NESS} = \frac{1}{Z_{\rm NESS} } e^{  \int_{-\infty}^{+\infty} \frac{dk}{2 \pi} h_k c^\dagger_{k} c_{k}} 
\quad , \quad \frac{1}{1 + e^{-h_k}} = f_R(k) \theta(-k) + f_L(k) \theta(k)  \;.
\ee
Hence $h_k = \beta_L (\frac{k^2}{2} - \mu_L) \theta(k) +  \beta_R (\frac{k^2}{2} - \mu_R) \theta(-k)$, 
i.e. the fermions with positive momentum as those coming from the left and reciprocally.

%\be 
%|x >  = \int \frac{dk}{2 \pi} e^{- i  k x} |k> = \int \frac{dk}{2 \pi} e^{- i  k x} \int dy e^{i k y} |y>  = |x > 
%\ee 
%\be 
%|k> = \int dy e^{i k y} |y> 
%\ee 

%\be 
%c_x^\dagger = |x > < 0| 
%\ee 

%\be 
%< x | k > = e^{i k x}
%\ee 
%\be 
%c^\dagger_x |0> = |x> \quad , \quad <k | c^\dagger_x |0> = e^{- i k x}
%\ee 
%\bea 
%c^\dagger_x c_{x'} = ? =    \int \frac{dk}{2 \pi} c^\dagger_k c_k e^{- i k (x-x')} \\
%<0|c^\dagger_x c_{x'}|0> = \int \frac{dk}{2\pi}  <0|c^\dagger_x |k> <k| c_{x'}|0>
%\eea 

%\be 
%<k| c_{x'}|0> = ? = e^{i k x'} 
%\ee 

\end{appendix}

\bibliographystyle{iopart-num}

\end{document}